%% file: main.tex
\documentclass[twocolumn]{aastex631}
\usepackage{xspace}
\usepackage{amsmath}
\usepackage{multirow}

\newcommand{\hess}{H.E.S.S.\xspace}
\newcommand{\fermiLAT}{\emph{Fermi}-LAT\xspace}
\newcommand{\FermiLAT}{\emph{Fermi}~LAT\xspace}
\newcommand{\fermi}{\emph{Fermi}\xspace}
\newcommand{\crpropa}{\texttt{CRPropa}\xspace}
\newcommand{\gammapy}{\texttt{gammapy}\xspace}
\newcommand{\gray}{gamma-ray\xspace}

\newcommand{\Grays}{gamma~rays\xspace}

\shorttitle{Constraints on the IGMF with \hess and \FermiLAT}
\shortauthors{\hess and \fermiLAT Collaborations}
\graphicspath{{./}{figures/}}

\begin{document}

\title{Constraints on the intergalactic magnetic field using \fermiLAT and \hess blazar observations}

\newif\ifshortauthorlist
\shortauthorlistfalse

\ifshortauthorlist
\author[0000-0002-0738-7581]{Manuel Meyer}
\affiliation{Institute for Experimental Physics, University of Hamburg, 
Luruper Chaussee 149, 
22761 Hamburg, Germany}
\email{manuel.meyer@uni-hamburg.de}

\author{David A. Sanchez}
\affiliation{Universit\'e Savoie Mont Blanc, CNRS, Laboratoire d’Annecy de
Physique des Particules – IN2P3, 74000 Annecy, France}

\author{Mathieu de Bony de Lavergne}
\affiliation{Universit\'e Savoie Mont Blanc, CNRS, Laboratoire d’Annecy de
Physique des Particules – IN2P3, 74000 Annecy, France}

\author{Tomas Bylund}
\affiliation{Department of Physics and Electrical Engineering, Linnaeus University, 351 95 V\"axj\"o, Sweden}
\collaboration{300}{\hess Collaboration}
\collaboration{1}{\fermiLAT Collaboration}
\else
\input{authors_list.tex}
\fi





\begin{abstract}

Magnetic fields in galaxies and galaxy clusters are believed to be the result of the amplification of intergalactic seed fields during the formation of large-scale structures in the universe. However, the origin, strength, and morphology of this intergalactic magnetic field (IGMF) remain unknown.
Lower limits on (or indirect detection of) the IGMF can be obtained from observations of high-energy gamma rays from distant blazars.
Gamma~rays interact with the extragalactic background light to produce electron-positron pairs, which can subsequently initiate electromagnetic cascades. 
The \gray signature of the cascade depends on the IGMF since it deflects the pairs.
Here we report on a new search for this cascade emission using a combined data set from the \fermi Large Area Telescope and the High Energy Stereoscopic System. Using state-of-the-art Monte Carlo predictions for the cascade signal, our results 
place a lower limit on the IGMF of $B > 7.1\times10^{-16}\,$G
for a coherence length of 1\,Mpc even when blazar duty cycles as short as 10~yr are assumed. This improves on previous lower
limits by a factor of 2. 
For longer duty cycles of $10^4$ ($10^7$) yr, 
IGMF strengths below 
$1.8\times10^{-14}$\,G ($3.9\times10^{-14}$\,G)
are excluded,
which rules out specific models for IGMF generation in the early universe. 

\end{abstract}

\keywords{BL Lacertae objects: general; galaxies: active; gamma rays: galaxies; magnetic fields}


\section{Introduction} \label{sec:intro}


Astrophysical magnetic fields in galaxies and galaxy clusters are ubiquitously observed and believed to have been seeded by a turbulent intergalactic magnetic field (IGMF), which should still be present at its seed field value and coherence length in cosmic voids \citep[see, e.g.,][for reviews]{2013A&ARv..21...62D,2021Univ....7..223A}. 
Such a seed field should have a small field strength in the voids, making it 
extremely difficult to observe directly. Currently, only upper limits of the order of $10^{-9}$\,G are known from
Faraday rotation measurements
\citep{2016PhRvL.116s1302P}.

Detecting or constraining the IGMF would provide insight into many processes in the early universe;
it might be possible to determine whether the IGMF was created during inflation or during phase transitions 
and to probe contributions of beyond-the-standard-model physics. 
Alternatively, the IGMF could be found to have an
astrophysical origin. 

Observations of blazars (active galactic nuclei (AGNs) with relativistic bulk outflows closely aligned with the line of sight (LOS)) at \gray energies offer a unique alternative probe of the IGMF,
with different systematics from other established methods \citep[e.g.][]{2013A&ARv..21...62D}   
During propagation, gamma rays interact with photons of background radiation fields, most notably the extragalactic background light (EBL), and produce electron-positron pairs.
These pairs can inverse Compton (IC) scatter cosmic microwave background (CMB) photons up to \gray energies, thereby initiating a cascade as these upscattered photons again pair-produce.
The deflection of the pairs in the IGMF delays the arrival of the \Grays produced in the cascades (compared to the primary gamma rays; \citealt{1995Natur.374..430P}), which should appear as an extended halo around the blazar \citep{1994ApJ...423L...5A} and enhance its apparent brightness at lower \gray energies \citep{2009PhRvD..80l3012N}.

The higher the value of the IGMF, the stronger the deflection of the pairs and, in turn, the larger the pair halos (with lower surface brightness), the lower the GeV excess of cascade photons, and the longer the time delays. 
The non-observation of an excess of GeV photons in individual blazar observations with the \fermi Large Area Telescope (LAT) has thus been used to place lower limits on the IGMF of the order of $B \gtrsim10^{-15}\,\mathrm{G}~$\citep[e.g.,]{2010Sci...328...73N}. 
Similarly, searches for a halo with imaging air Cherenkov telescopes (IACTs) have ruled out an IGMF strength between 
$\sim10^{-16.5}\,\mathrm{G}$ and $\sim10^{-15.5}\,\mathrm{G}$ \citep{2014A&A...562A.145H} as well as between 
 $\sim10^{-14.5}\,\mathrm{G}$ and $\sim 10^{-13.5}\,\mathrm{G}$ \citep{2017ApJ...835..288A}.
The \gray constraints are considerably relaxed when short activity times of blazars are considered \citep{2011ApJ...733L..21D}.\footnote{
We define the activity time (duty cycle) as the length of the time during which the blazar has been emitting \Grays. Throughout this work, we assume that a blazar has emitted a flux that is constant in time, which can be described with the average spectrum observed today. 
}
However, even if blazars have only been active for the past 10\,yrs, a joint analysis of \fermiLAT and archival IACT spectra from several sources is still able to exclude $B\lesssim3\times 10^{-16}\,\mathrm{G}$ for a coherence length of $\ell_B \gtrsim 10\,\mathrm{kpc}$ \citep{2018ApJS..237...32A}.
A comprehensive overview of current constraints is provided by \citet{2021Univ....7..223A}.

Here we perform an updated search for the cascade emission of hard-spectrum blazars. 
In contrast to previous cascade searches,
we use a combination of \fermiLAT data and observations carried out with the High Energy Stereoscopic System (\hess) 
based on the full Poisson likelihood for gamma-ray event statistics
and calculate pair halo templates with state-of-the-art Monte Carlo simulations.

\section{Observations}
\label{sec:obs}

\subsection{Source Selection}
Among the blazars observed with both \hess and the LAT, we aim to select those for which the cascade emission is expected to be maximal. 
Extreme high-frequency-peaked BL Lac-type objects (EHBLs) with an observed synchrotron peak $\gtrsim 1\,$keV are the most promising for such searches. 
Their hard spectra are measured up to TeV energies without finding evidence for a source-intrinsic spectral cutoff. 
Therefore, a lot of energy is potentially available for reprocessing into the cascade. 
To select EHBL sources, we start from the fourth catalog of AGNs observed with the LAT \citep[4LAC;][]{2020arXiv201008406L}.
We select blazars that (i) follow a power-law spectrum in energy $E$, $dN/dE = N_0 (E / E_0)^{-\Gamma}$ with spectral index $\Gamma < 2$ within statistical uncertainties; 
(ii) have a known spectroscopic redshift $z$; 
and (iii) have a synchrotron peak $\nu_\mathrm{peak} > 10^{17}\,$Hz. 
Additionally, we demand that the flux have a chance probability of being variable at less than $99\,\%$ in the LAT energy band on monthly time scales.
This choice excludes sources that exhibit strong variability.
Otherwise, the past temporal flux evolution would have to be taken into account to predict the time-delayed cascade emission. 
This selection results in five sources that have been previously detected with \hess~\citep{2006Natur.440.1018A,
2007A&A...473L..25A,
2007A&A...470..475A,
2010A&A...521A..69A}. 
These are listed in Table~\ref{tab:sources}.
Except for PKS~0548$-$322, all of the selected sources are also included in the analysis of \citet{2018ApJS..237...32A}.
In addition, \citet{2018ApJS..237...32A} considered the \hess-detected blazar 1ES~0414+009; however, this source fails our cuts on the synchrotron peak frequency and the probability for variability. 

\begin{deluxetable*}{l|c|ccc|cc}
\tablecaption{Selected sources and observation summaries.\label{tab:sources}}
\tablewidth{0pt}
\tablehead{
\colhead{}& \colhead{} & \multicolumn3c{\hess} & \multicolumn2c{\FermiLAT} \\
\colhead{Source} & \colhead{$z$} & \colhead{Live Time} & \colhead{Significance} &
\colhead{$\Gamma_\mathrm{int}$} & \colhead{$\sqrt{\mathrm{TS}}$} & \colhead{$\Gamma_\mathrm{int}$} \\
\colhead{} & \colhead{} & \colhead{(hr)} & \colhead{($\sigma)$} &
\colhead{} &  \colhead{} & \colhead{}
} 
\startdata
1ES~0229+200 &  0.140 &  144.1 & 16.5 & $1.76\pm0.12$  & 14.2 & $1.73 \pm 0.12$ \\
1ES~0347$-$121 & 0.188 & 59.2 & 16.1 & $2.12\pm0.15$ & 18.3 & $1.81\pm0.11$ \\
PKS~0548$-$322 & 0.069 & 53.9 & 10.2 & $1.92\pm0.12$ & 11.7 & $2.06\pm0.17$ \\
1ES~1101$-$232 & 0.186 & 71.9 & 18.7 & $1.66\pm0.09$ & 19.5 & $1.65\pm0.09$ \\
H~2356$-$309 & 0.165 & 150.5 & 23.4 & $2.10\pm0.09$ & 21.7 & $1.81\pm0.09$ \\
\enddata
\tablecomments{
Second column: spectroscopic redshift, $z$, taken from 
 (in order of increasing R.A.): 
\citet{2021AJ....161..196P,2005ApJ...631..762W,1976ApJ...207L..75F,1989ApJ...345..140R,2009MNRAS.399..683J}.
Third column: total observation time taking data quality cuts and dead time into account. 
Fourth and sixth columns: detection significance in units of $\sigma$. 
Fifth and seventh columns: power-law indices for EBL-corrected spectra. 
}
\end{deluxetable*}

\subsection{\fermi-LAT Observations}
\label{sec:lat-obs}

\begin{figure}
    \centering
    \includegraphics[width=.99\linewidth]{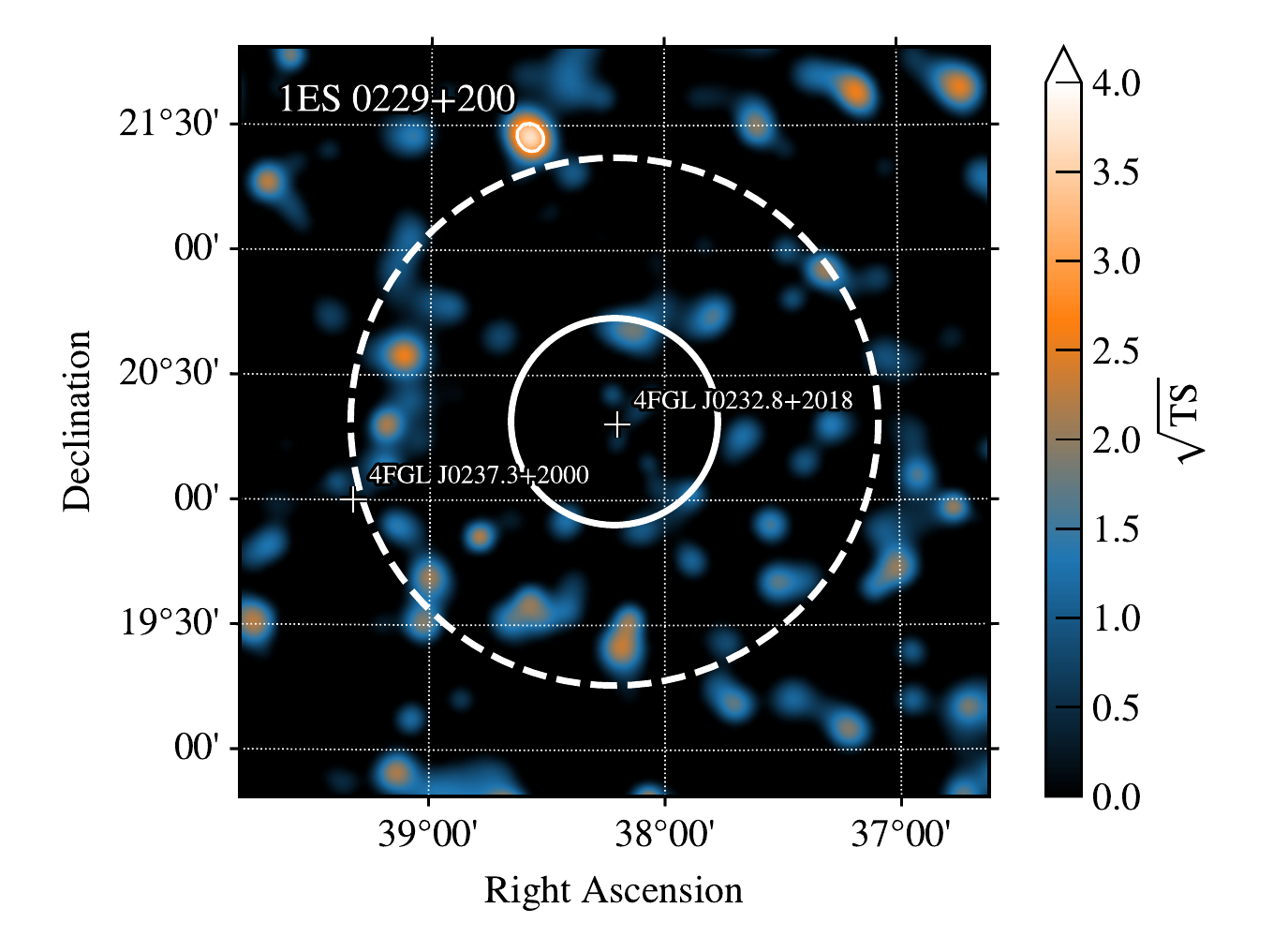}
    \caption{
    A TS map for the LAT analysis of the ROI around 1ES~0229+200.
    Point sources within the ROI are marked with plus signs.
    The 68\,\% containment radii at 1\,GeV for the LAT point spread function for the events with the worst and best angular reconstruction are shown as a dashed and solid circle, respectively. 
    The thin white contour indicates a region where 
    $\sqrt{\mathrm{TS}} = 3$.
    }
    \label{fig:tsmap}
\end{figure}

The \fermi LAT is a pair-conversion telescope that detects \Grays from 20\,MeV to beyond 300\,GeV \citep{2009ApJ...697.1071A}.
For our five sources, we analyze 11.5\,yr of LAT data between 1\,GeV and 3\,TeV extracted from a $6^\circ \times 6^\circ$ region of interest (ROI) centered on the nominal source position. The data selection closely follows \citet{2018ApJS..237...32A}.
Using events and instrumental response functions of the \texttt{P8R3\_SOURCE\_V3} class, 
we find the optimal model for the ROI that includes all point sources listed in the second data release of the fourth LAT point-source catalog \citep[4FGL-DR2;][]{2020arXiv200511208B}, as well as templates for the diffuse Galactic and isotropic backgrounds (see Appendix \ref{app:fermi} for further details). 
All of our blazars are modeled with power laws in our initial ROI model. 
After an initial optimization of the model, we change the spectral model to a power law with index $\Gamma_\mathrm{int}$ that includes absorption on the EBL, given by $\exp(-\tau_{\gamma\gamma})$, where $\tau_{\gamma\gamma}$ is the optical depth. 
Throughout this article, the EBL absorption is taken from the model of \citet{2011MNRAS.410.2556D}.
The obtained spectral indices and source significances, expressed as the square root of the test statistic (TS)\footnote{The test statistic is defined as $\mathrm{TS} = -2(\ln\mathcal{L}_1 - \ln\mathcal{L}_0)$, where $\mathcal{L}_1$ ($\mathcal{L}_0$) is the likelihood of the model including (excluding) a source.} 
are listed in Table~\ref{tab:sources}.
In Figure~\ref{fig:tsmap}, we show a sky map with TS
values for our optimized ROI model for 1ES~0229+200.
In each pixel, the significance of a putative source with a power-law index equal to 2 is calculated. 
A potential halo could reveal itself as a TS excess around the central source.
This is, however, not observed for any of the studied sources (see also Appendix~\ref{app:fermi}). 

\subsection{\hess Observations}
\label{sec:hess-obs}

\hess is an array of five IACTs located in the Khomas highlands in Namibia. 
The array is composed of four telescopes with an effective mirror diameter of 12\,m called CT1-CT4, and one with an effective mirror diameter of 28 meters called CT5. 
New cameras were installed between 2015 and 2016  for CT1-4 
and in 2019 for CT5. 
\hess can detect very high energy \Grays between $\approx50$\,GeV and $\approx100\,$TeV.
All sources considered here have been detected with \hess. 
For all EHBLs except PKS~0548$-$322, additional data have been taken since the original publications during different phases of the experiment (with and without CT5, before and after the camera exchange campaigns).
In order to have a data set with a 
hardware setup as homogeneous as possible, 
we restrict ourselves to runs\footnote{One run is a consecutive observation of a source with a duration of approximately 28\,minutes.} with good data quality in which three or four of the smaller telescopes participated.
For the same reason, we only use data taken with the cameras originally mounted on the 12\,m telescopes.
The energies of the recorded \gray-like events are reconstructed using the \texttt{ImPACT} method and standard selection cuts \citep{2014APh....56...26P}.
The spectral analysis is performed using \texttt{gammapy} version 0.18.2 \citep{gammapy:2017}. 
Source counts are extracted from an ``On'' region of radius of 
$0.07^\circ$
centered on the source location, 
which corresponds to the 68\,\% confidence radius of the point spread function.
The observed spectra are modeled with power laws with EBL attenuation; models with spectral curvature 
are not preferred.  
The resulting spectral indices 
are provided in Table~\ref{tab:sources} together with  observation live times and detection significances. 
Further details on the analysis are provided in Appendix \ref{app:hess}.

\section{Search for the pair halo and constraints on the IGMF}

\subsection{Cascade predictions}
\begin{figure}
    \centering
    \includegraphics[width=.99\linewidth]{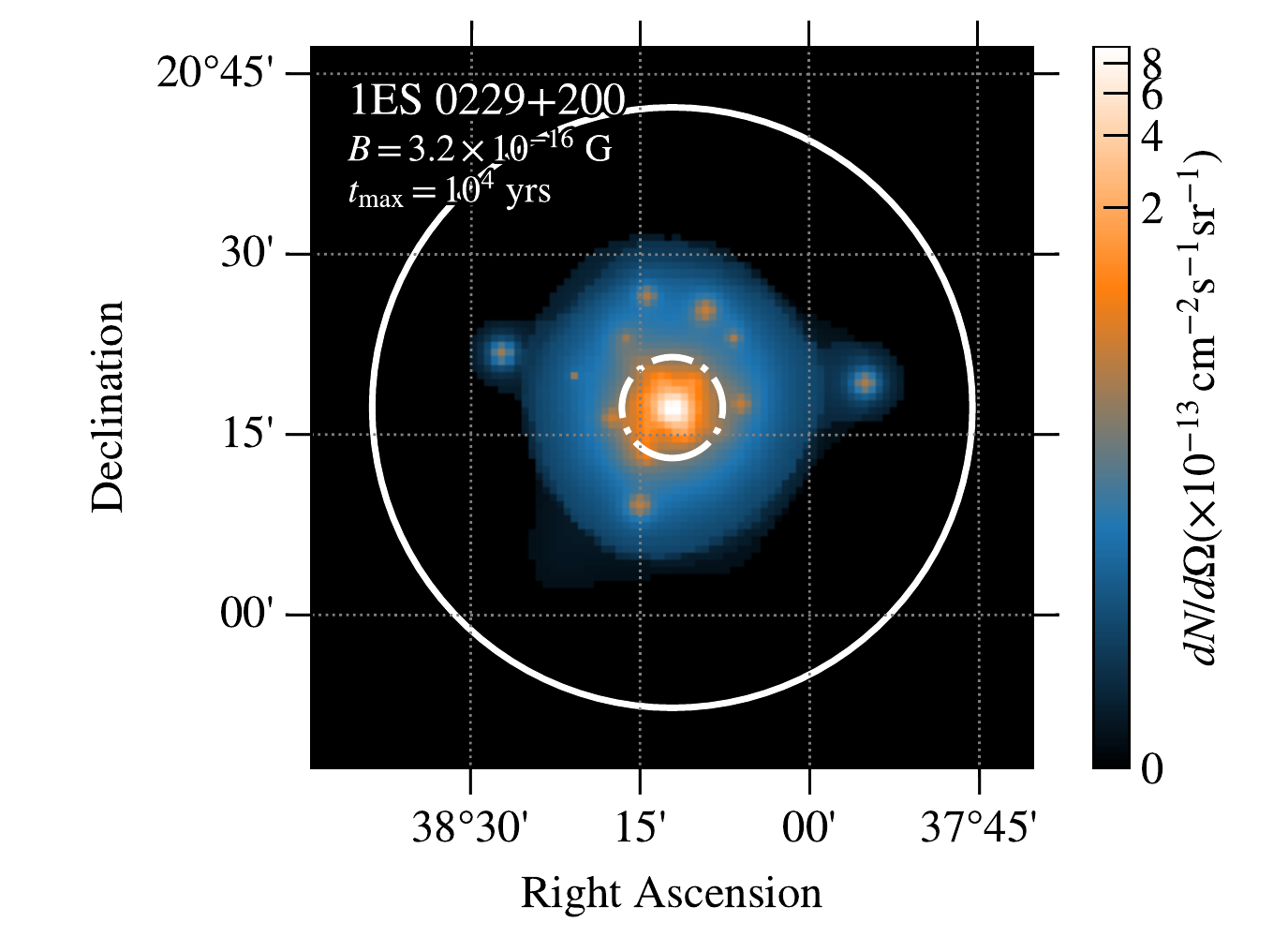}
    \caption{
    Simulated halo photon flux per solid angle, $dN/d\Omega$, integrated above 1\,GeV for the best-fit spectrum (see Section~\ref{sec:combined}) for 1ES~0229+200.
    An activity time of $t_\mathrm{max} = 10^{4}\,$yr and a field strength of $B = 3.2\times10^{-16}\,$G are assumed.
    The same 68\,\% containment radius as in Figure~\ref{fig:tsmap} is shown for comparison, along with the On region for the \hess observations (dotted-dashed circle).
    The hot spots correspond to individual simulated particles.
    }
    \label{fig:cascmap}
\end{figure}

We generate templates for the pair halo for each redshift using the \crpropa Monte Carlo simulation package version 3.1.6\footnote{\url{https://github.com/CRPropa/CRPropa3}} \citep{2016JCAP...05..038A}. 
In the simulation, we include all relevant particle interactions and consider 
cascade photons arriving with a time delay shorter than a maximum blazar activity time $t_\mathrm{max}$.
Using a spectral reweighting \citep{2018ApJS..237...32A} and parallel transport \citep{2016PhRvD..94h3005A}, the simulations are converted into energy-dependent sky maps for arbitrary input spectra and angles $\theta_\mathrm{obs}$ between the jet axis and the LOS (see Appendix \ref{app:crpropa} for further details on the simulation).\footnote{A Python package performing these calculations is publicly available at \url{https://github.com/me-manu/simCRPropa}.} 
These serve as input templates for our data analysis.
For simplicity, $\theta_\mathrm{obs}=0^\circ$ is assumed throughout. 
For $0 < \theta_\mathrm{obs} < \theta_\mathrm{jet} / 2$, the templates will be asymmetric. 
However, in the considered energy range, the asymmetry is small and our results do not depend strongly on $\theta_\mathrm{obs}$ and the orientation of the halo.
The IGMF is modeled as cell-like; each cell has a side length of $\ell_B = 1$\,Mpc and the $B$-field orientation changes randomly from one cell to the next. 
The templates are generated for each source and for seven values of the field strength, $B = 10^{-16}\,\mathrm{G},10^{-15.5}\,\mathrm{G},\ldots,10^{-13}$\,G. 
For higher values of $B$, the pairs are quickly isotropized and the cascade emission would appear as an additional component to the isotropic \gray background in the LAT energy band. 
An example of the simulated energy-integrated sky map is shown in Figure~\ref{fig:cascmap} and an output spectrum is shown in Figure~\ref{fig:sed}.
Spectra for all other sources and different choices of the IGMF are presented in Figure~\ref{fig:spectra-tmax1e7} in Appendix~\ref{app:halo-fit}.

\subsection{Combined LAT and \hess analysis}
\label{sec:combined}

We test for the presence of a halo for a given 
field strength using a likelihood ratio test. 
Different fixed values of $t_\mathrm{max} = 10\,\mathrm{yr}, 10^4\,\mathrm{yr}$, and $10^7\,\mathrm{yr}$ are assumed, which correspond to a conservative, intermediate, and optimistic scenario, respectively, in terms of the amount of cascade emission.
Our model is the sum of the point-source flux $\phi_{\mathrm{src}}(E, \boldsymbol{\theta}_i)$ for central source $i$ and spectral parameters $\boldsymbol{\theta}_i$ and the diffuse halo flux $\phi_\mathrm{halo}(E, \mathbf{p},\boldsymbol{\theta}_i, B)$ which also depends on the sky position $\mathbf{p}$ and the IGMF. 
In contrast to Section~\ref{sec:obs}, the point-source spectra are now modeled with a power law with a cutoff at energy $E_\mathrm{cut}$,
\begin{equation}
    \phi_\mathrm{src} = N_0 (E / E_0)^{-\Gamma_\mathrm{int}} \exp(-E/E_\mathrm{cut}-\tau_{\gamma\gamma}),\label{eq:src-spec}
\end{equation}
 i.e.,  $\boldsymbol{\theta} = (N_0, \Gamma_\mathrm{int}, E_\mathrm{cut}, E_0)$. 
With this choice, 
we obtain plausible constraints on the IGMF since the high-energy cutoff reduces the number of gamma rays that can initiate the cascade.
Assuming instead a simple power law as the intrinsic spectrum would bear the risk of overestimating the flux of primary gamma rays and thus of the cascade. 

The likelihood of observing the data $\mathcal{D}_i$ given the model parameters
is the product of the individual Poisson likelihoods for LAT and \hess data, $\mathcal{L}_i \equiv\mathcal{L}(B, \boldsymbol{\theta}_i | \mathcal{D}_i) = \mathcal{L}(B, \boldsymbol{\theta}_i | \mathcal{D}_{\mathrm{LAT}, i})\mathcal{L}(B, \boldsymbol{\theta}_i | \mathcal{D}_{\mathrm{H.E.S.S.}, i})$.
The log-likelihood ratio test can then be written as
\begin{equation}
\label{eq:likelihood}
    \lambda(B) = -2 \sum\limits_i\ln\left(\frac{\mathcal{L}(B, \hat{\hat{\boldsymbol{\theta}}}_i | \mathcal{D}_i)}{\mathcal{L}(\hat{B}, \hat{\boldsymbol{\theta}}_i | \mathcal{D}_i)}\right),
\end{equation}
where $\hat{\hat{\boldsymbol{\theta}}}_i$ denotes the spectral parameters maximizing $\mathcal{L}_i$ for a fixed field strength, whereas $\hat{\boldsymbol{\theta}}_i$ and $\hat{B}$ are the parameters that maximize $\mathcal{L}_i$ unconditionally.
We first extract $\ln\mathcal{L}(B, \boldsymbol{\theta}_i | \mathcal{D}_{\mathrm{LAT}, i})$ over a grid of IGMF values and spectral parameters by adding the reweighted cascade template to the ROI  model (see Section~\ref{sec:lat-obs}) and reoptimizing the spectral parameters of the background sources (see Appendix \ref{app:fermi-halo} for further details).
In a second step, we determine the spectral parameters that maximize the total $\mathcal{L}_i$ for a given value of $B$ using the \texttt{MINUIT} fit routine in \gammapy.
For the \hess data, the model includes the point source and the pair halo intensity within the On region.
In each step of the parameter optimization, the cascade template is reweighted to account for the changing spectrum, and the extracted value of $\ln\mathcal{L}(B, \boldsymbol{\theta}_i | \mathcal{D}_{\mathrm{LAT}, i})$ is added to $\ln\mathcal{L}(B, \boldsymbol{\theta}_i | \mathcal{D}_{\mathrm{H.E.S.S.}, i})$.
In this way, for a given set of IGMF parameters, the best-fit spectrum of each source for both the LAT and \hess energy ranges is determined.
An example of the fit for 1ES~0229+200 is shown in Figure~\ref{fig:sed}.
Figures for other sources are shown in  Appendix~\ref{app:all-fits} where we also report the best-fit parameters.\footnote{
We provide the \fermiLAT and \hess spectra, best-fit values, as well as the cascade simulation output electronically at \url{https://zenodo.org/record/8014311}.
}

\begin{figure}
    \includegraphics[width=.99\linewidth]{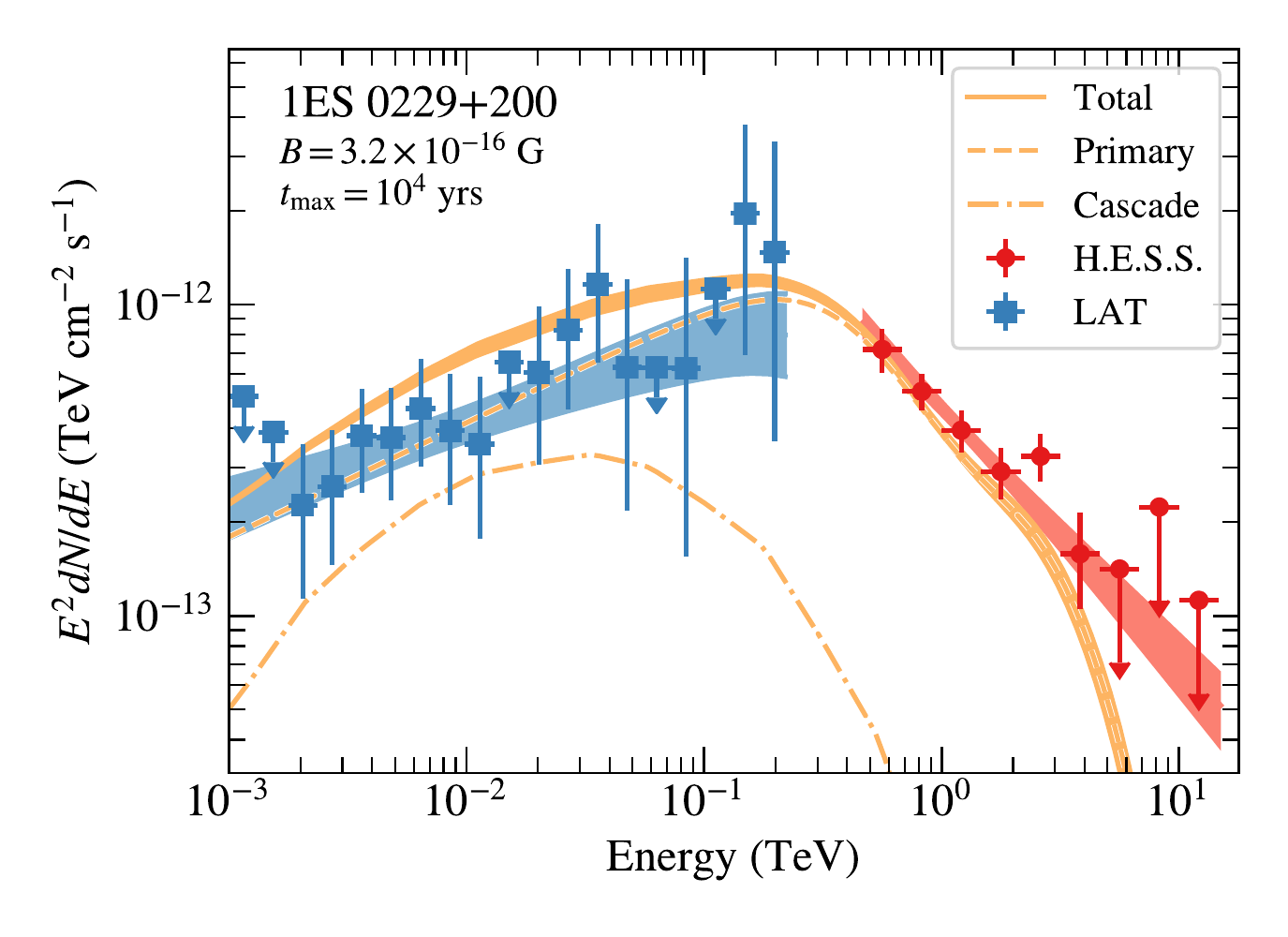}
    \caption{
    Observed LAT and H.E.S.S. spectra of 1ES~0229+200 (blue and red markers and lines, respectively), together with the total best-fit spectrum (orange solid band, including uncertainties) for a point-source model of the primary \gray flux (dashed orange line) and the halo flux (dash-dotted orange line).
    Spectral points are shown as upper limits if the detection significance is below $2\,\sigma$.
    The same values for $t_\mathrm{max}$ and  $B$ are assumed as in Figure~\ref{fig:cascmap}.
    }
    \label{fig:sed}
\end{figure}

\begin{figure*}
    \centering
    \includegraphics[width=.85\linewidth]{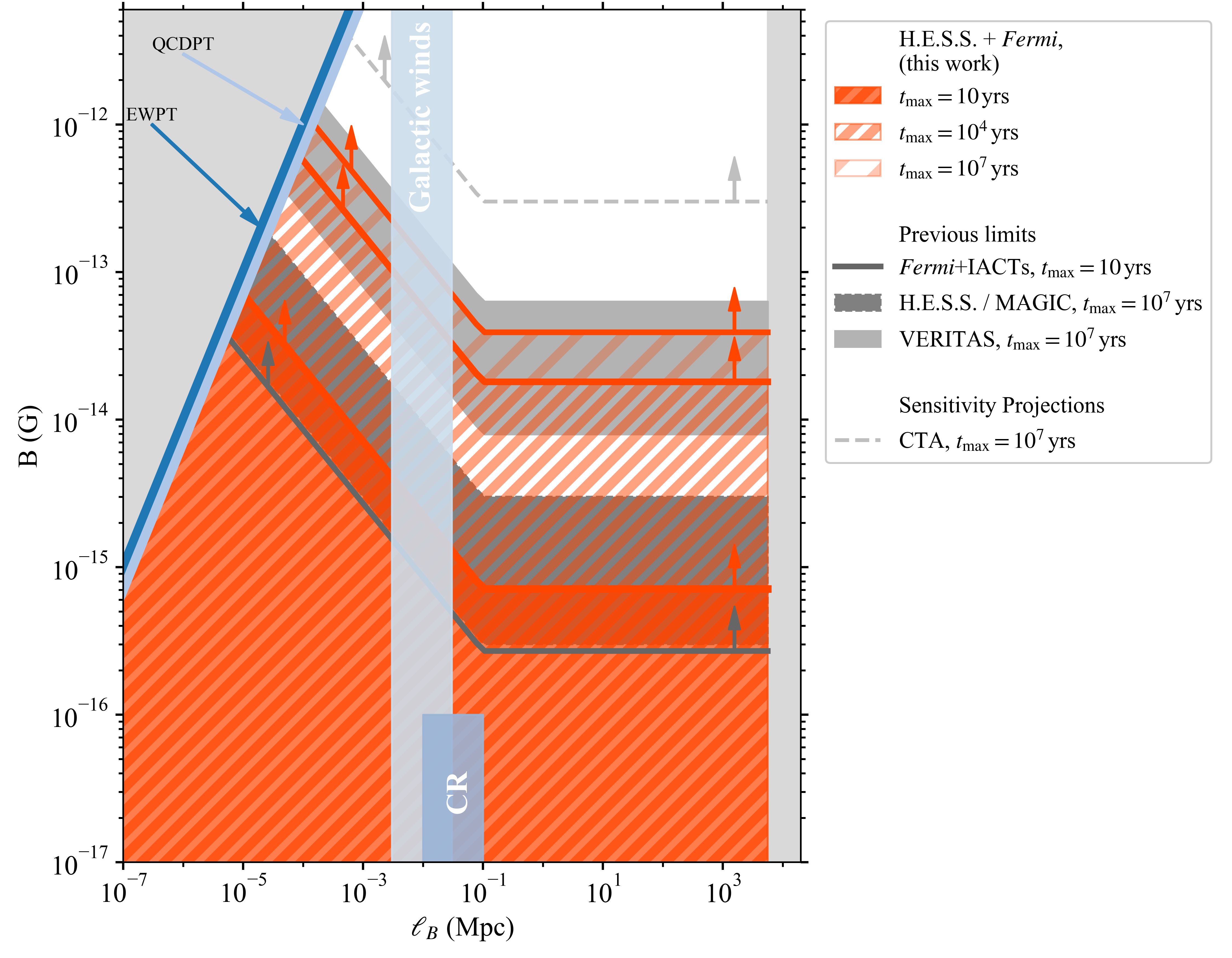}
    \caption{The IGMF parameter space probed with \gray instruments.
    Lower limits at 95\,\% confidence level on the IGMF derived from the  combined \fermiLAT and \hess analysis for different assumed blazar activity times are shown as orange filled and hatched regions.
    A scaling of the limits with the coherence length as $\ell_B^{-1/2}$ is assumed.
    Exclusion regions from 
    H.E.S.S., MAGIC, and VERITAS
    are shown as gray filled regions~\citep[][]{2010A&A...524A..77A,2014A&A...562A.145H, 2017ApJ...835..288A}. 
    The CTA could exclude field strengths below the dashed gray line \citep{2021JCAP...02..048A}.
    The dark-gray line shows combined \emph{Fermi}-LAT and IACT lower limits for $t_\mathrm{max} = 10\,$yr \citep{2018ApJS..237...32A}. 
    Light-gray shaded regions are disfavored by theory. 
    Blue lines and regions indicate theoretically favored regions for the generation of the IGMF during either electro-weak (EWPT) or QCD phase transitions (QCDPT), 
    by Galactic winds 
    or induced by cosmic-ray (CR) streaming \citep[see][and references therein]{2013A&ARv..21...62D}. 
    }
    \label{fig:igmf}
\end{figure*}

In general, high values of $B$ provide a better fit to the data than low IGMF values, implying halos with large extensions and a low surface brightness. 
For low values of $B$, the fit prefers low values of $E_\mathrm{cut}$ to minimize the cascade contribution.
For instance, for $t_\mathrm{max} \geqslant 10^4\,$yr and $B=10^{-16}\,$G the cutoff energy is below $E_\mathrm{cut}\lesssim3\,$TeV for all sources except PKS\,0548$-$322.
However, even the fits assuming high IGMF values are not significantly preferred over the fit without cascade emission.
The largest value for the log-likelihood ratio between the fit without and with a halo is found to be $\mathrm{TS}\sim0.9$ for $B=10^{-13}\,\mathrm{G}$ and $t_\mathrm{max} = 10^4\,$yrs. 
Thus, we do not find any evidence for the presence of a halo and proceed to set lower limits on the IGMF strength.
As the field strength constitutes one additional degree of freedom, we set 95\,\% confidence limits when $\lambda > 2.71$.
We find that we can rule out an IGMF with a field strength $B < 3.9\times10^{-14}\,\mathrm{G}$ for $t_\mathrm{max} = 10^7$\,yrs. For a more conservative choice of the blazar activity time, we exclude
$B < 1.8\times10^{-14}\,\mathrm{G}$ for $t_\mathrm{max} = 10^4$\,yrs and $B < 7.1\times10^{-16}\,\mathrm{G}$ for $t_\mathrm{max} = 10$\,yrs. The strongest constraints are provided by 1ES~0229+200 and 1ES~1101$-$232 (see Appendix \ref{app:likelihoods} for the full likelihood profiles).
These limits, together with previous constraints and theoretically motivated parameters, are shown in Figure~\ref{fig:igmf}. 

\section{Discussion}
The analysis presented here improves on previous analyses of a combination of IACT and LAT data in two ways.
Firstly, we use the \crpropa code to generate predictions for the spectral and spatial shape of the halo. As the code traces the full trajectories of the particles, we are not limited by the assumption of, e.g., small deflections of the pairs in the IGMF, which limited the validity range of the results of \citet{2018ApJS..237...32A}.
Secondly, we combine LAT and \hess data on the likelihood level and self-consistently include the halo contribution.  
As a result, we are able to improve on the constraints found by \citet{2018ApJS..237...32A} by a factor of two for blazar activity times as short as 10\,yr.
To our knowledge, these are the tightest constraints on the IGMF from \gray observations to date (compare Figure~\ref{fig:igmf}).  

The assumed activity time of the sources poses the largest systematic uncertainty in our study.
Changing $t_\mathrm{max}$ from 10 to $10^4\,$yr strengthens the constraint by almost a factor of 30
(activity times of $10^{7}$-$10^{8}$\,yr could be realistic; \citealt{2002NewAR..46..313P}).
For $t_\mathrm{max} = 10^{7}$, we cannot rule out as high IGMF values as \citet{2017ApJ...835..288A}.
The likely reason is that we do not include the spatial halo extension in the \hess energy range, as we only consider the cascade photons that arrive within the On region. 
We are instead able to probe low IGMF values thanks to the combination of LAT and \hess data.
We have deliberately selected sources that do not experience strong variability in the LAT data.
However, at least H\,2356$-$309 
and 1ES\,0229+200 
show flux variations by factors $\sim$2-3
above 100\,GeV 
on monthly and yearly time scales, respectively
\citep{2010A&A...516A..56H,2023A&A...670A.145A}. 
Our reweighting framework developed in Appendix~\ref{app:crpropa} could in principle be extended to study how long-term variability affects the predictions for the cascade. This is left for future study.
Additional systematic uncertainties with a smaller effect on the results compared to the change in activity time,
such as the choice of the EBL model, the choice of $\theta_\mathrm{jet}$, 
and the different absolute energy scales of the LAT and \hess 
(which can be treated self-consistently in our likelihood framework), 
are discussed in Appendix~\ref{app:sys}.

Under the assumptions that (a) the sources have been active 
for  $10\,\mathrm{yrs}\lesssim t_\mathrm{max}\lesssim10^4\,\mathrm{yrs}$ and (b) our constraints scale as $\ell_B^{-1/2}$ in the case when $\ell_B$ is shorter than the mean free path of the pairs for IC scattering (\citealt{2009PhRvD..80l3012N}; see, however, \citealt{2015PhRvD..91l3514C}), our results challenge particular models for the IGMF generation. 
For example, we can rule out models where the seed fields are created through axion-driven magnetogenesis (which predicts 
a present-day IGMF with $B\sim10^{-13}$\,G and $\ell_B \sim 20\,$pc; see \citealt{2018PhRvL.121b1301M}) and  
the low end of allowed $B$-field values generated during or shortly after inflation with values $B\gtrsim10^{-15}$\,G on megaparsec scales \citep{2014JCAP...05..040K}.

It could be possible that the pairs  predominantly lose their energy by heating the intergalactic medium through plasma instabilities, which would suppress the cascade \citep{2012ApJ...752...22B}. 
Whether such instabilities develop is, however, still debated \citep[see, e.g., ][]{2021Univ....7..223A}. 
We can introduce an additional normalization for our halo templates, $ 0 \leqslant N_\mathrm{halo} \leqslant 1$ to mimic the possibility that the pairs lose part of their energy through this process ($N_\mathrm{halo} = 1$  corresponds to the case discussed so far).
The fit tends to select $N_\mathrm{halo} \sim 0$ as the halo is not present in the data, i.e., no energy is dissipated through IC scattering. In this case, no limits on the IGMF are possible.
We note, however, that our constraints for $t_\mathrm{max} = 10\,$yr are not affected, as the energy loss due to beam instabilities would only dominate over IC cooling after hundreds of years of blazar activity~\citep{2012ApJ...752...22B}.

In the future,
observations with the upcoming Cherenkov Telescope Array (CTA) should be able to detect an IGMF as strong as $10^{-13}\,$G for sufficiently long blazar activity times (see Figure~\ref{fig:igmf}) thanks to its improved angular resolution, point-source sensitivity, 
and covered energy range \citep{2016ApJ...827..147M,2021JCAP...02..048A}.

\begin{acknowledgments}

The support of the Namibian authorities and of the University of
Namibia in facilitating the construction and operation of H.E.S.S.
is gratefully acknowledged, as is the support by the German
Ministry for Education and Research (BMBF), the Max Planck Society,
the German Research Foundation (DFG), the Helmholtz Association,
the Alexander von Humboldt Foundation, the French Ministry of
Higher Education, Research and Innovation, the Centre National de
la Recherche Scientifique (CNRS/IN2P3 and CNRS/INSU), the
Commissariat \`a  l’\'energie atomique et aux \'energies alternatives 
(CEA), the U.K. Science and Technology Facilities Council (STFC),
the Irish Research Council (IRC) and the Science Foundation Ireland
(SFI), the Knut and Alice Wallenberg Foundation, the Polish
Ministry of Education and Science, agreement No. 2021/WK/06, the
South African Department of Science and Technology and National
Research Foundation, the University of Namibia, the National
Commission on Research, Science \& Technology of Namibia (NCRST),
the Austrian Federal Ministry of Education, Science and Research
and the Austrian Science Fund (FWF), the Australian Research
Council (ARC), the Japan Society for the Promotion of Science, the
University of Amsterdam, and the Science Committee of Armenia grant
21AG-1C085. We appreciate the excellent work of the technical
support staff in Berlin, Zeuthen, Heidelberg, Palaiseau, Paris,
Saclay, T\"ubingen, as well as in Namibia in the construction and operation
of the equipment. This work benefited from services provided by the
H.E.S.S. Virtual Organisation, supported by the national resource
providers of the EGI Federation.

The \textit{Fermi}-LAT Collaboration acknowledges support for LAT development, operation, and data analysis from NASA and DOE (United States), CEA/Irfu and IN2P3/CNRS (France), ASI and INFN (Italy), MEXT, KEK, and JAXA (Japan), and the K.A.~Wallenberg Foundation, the Swedish Research Council, and the National Space Board (Sweden). Science analysis support in the operations phase from INAF (Italy) and CNES (France) is also gratefully acknowledged. This work performed in part under DOE contract DE-AC02-76SF00515.
\end{acknowledgments}

%

\vspace{5mm}
\facilities{\FermiLAT, \hess}


\software{astropy \citep{2013A&A...558A..33A},
          numpy \citep{harris2020array},
          scipy \citep{2020SciPy-NMeth}, 
          CRPropa3 \citep{2016JCAP...05..038A}, 
          fermipy \citep{2017ICRC...35..824W},
          fermitools \citep{2019ascl.soft05011F},
          gammapy \citep{gammapy:2017}
          }



\appendix

\section{Details on the \fermiLAT data analysis}
\label{app:fermi}

\subsection{Data selection and initial model optimization}

In this appendix, we provide further details on the \fermiLAT data selection and analysis choices. 
Events of the \texttt{P8R3\_SOURCE\_V3} class with arrival times between 2008 August 4 and 2020 January 4 are extracted from a $6^\circ \times 6^\circ$ sky region in the energy range from 1\,GeV to 3\,TeV that arrived with a zenith angle $\leqslant 100^\circ$ in order to minimize contamination of our sample with gamma rays from the Earth limb. 
The standard data filter \texttt{DATA\_QUAL$>$0 \&\& LAT\_CONFIG==1} is applied.
Following \citet{2018ApJS..237...32A}, the events are then binned using a spatial binning of $0.025^\circ\,\mathrm{pixel}^{-1}$ and a logarithmic binning in energy with 8 bins per decade. 
Furthermore, we split the data sample into two subsamples using the quality of the angular reconstruction (the so-called PSF event classes; see \citealt{2013arXiv1303.3514A}). The first subsample contains all events belonging to the PSF event classes 0-2 (event type 28) and the second sample contains the events with the best reconstructed arrival 
directions, PSF event class 3 (event type 32). A combined likelihood analysis is performed on these two subsamples using \texttt{fermipy}\footnote{\url{https://fermipy.readthedocs.io/en/latest/}} \citep{2017ICRC...35..824W} version 0.19.0 and the \texttt{fermitools} version 1.2.23\footnote{\url{https://fermi.gsfc.nasa.gov/ssc/data/analysis/}} 
with the \texttt{MINUIT} optimizer. 

\begin{figure*}
    \centering
        \gridline{
          \fig{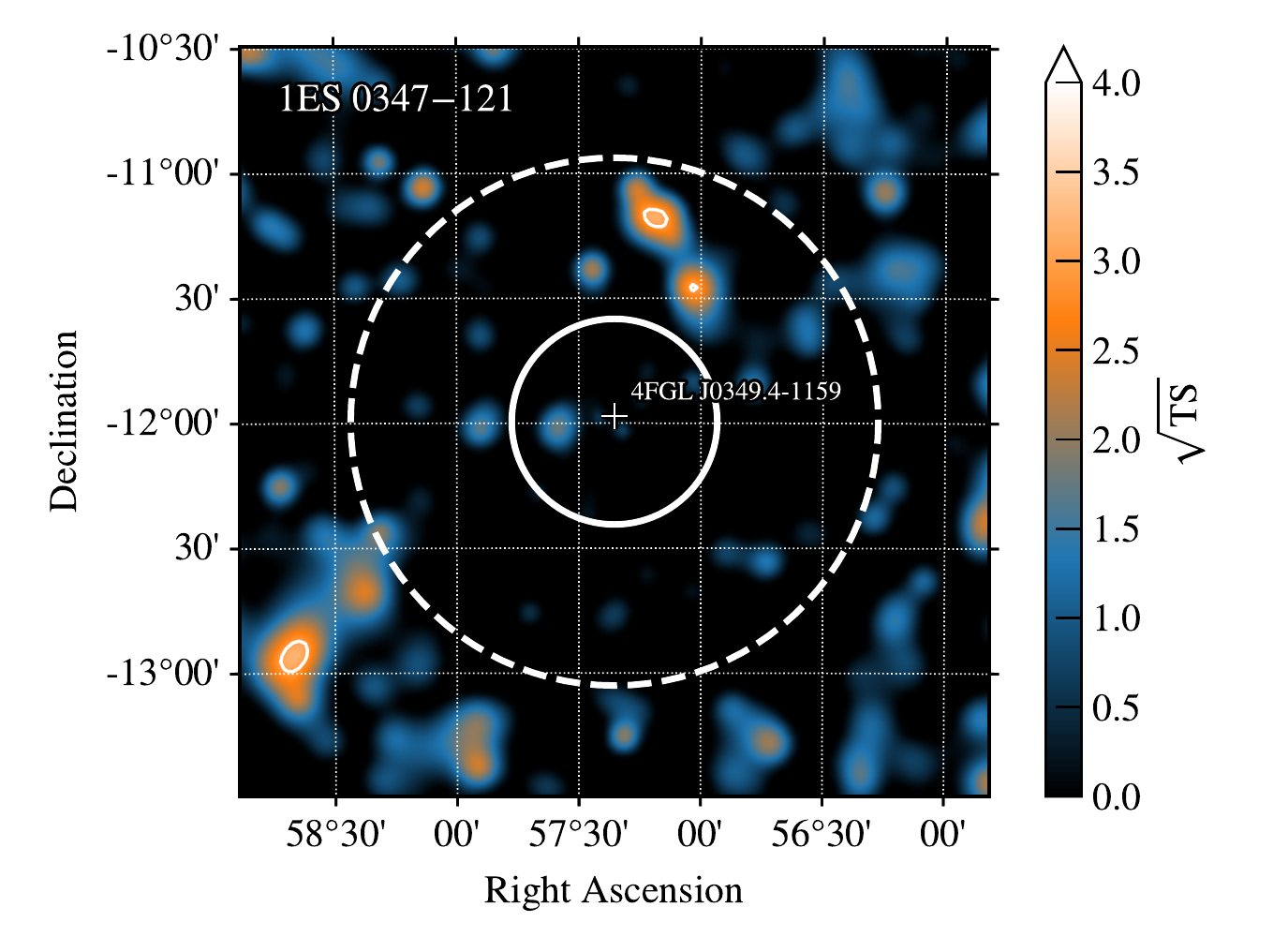}{0.5\textwidth}{}
          \fig{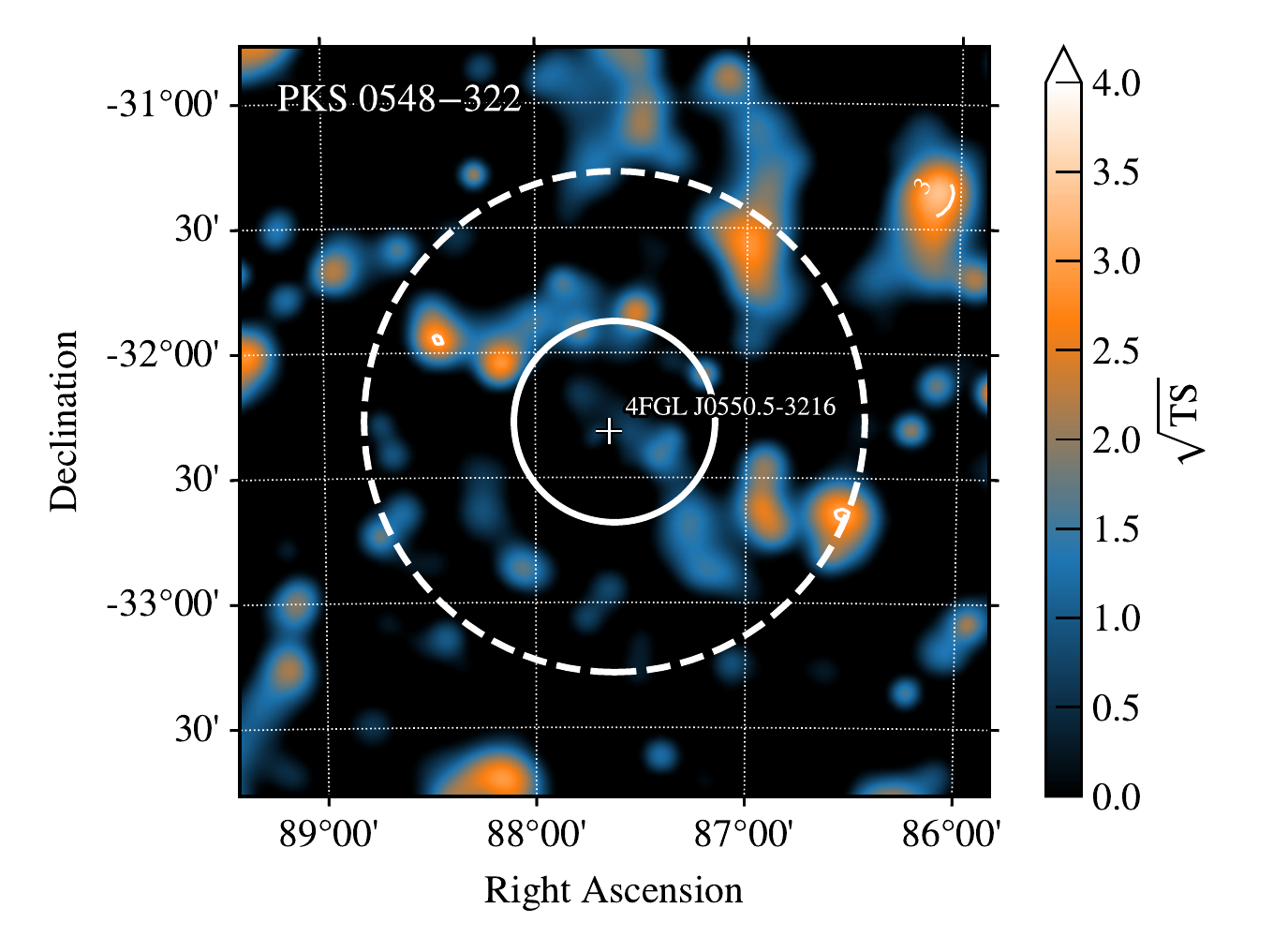}{0.5\textwidth}{}
        }
        \gridline{
          \fig{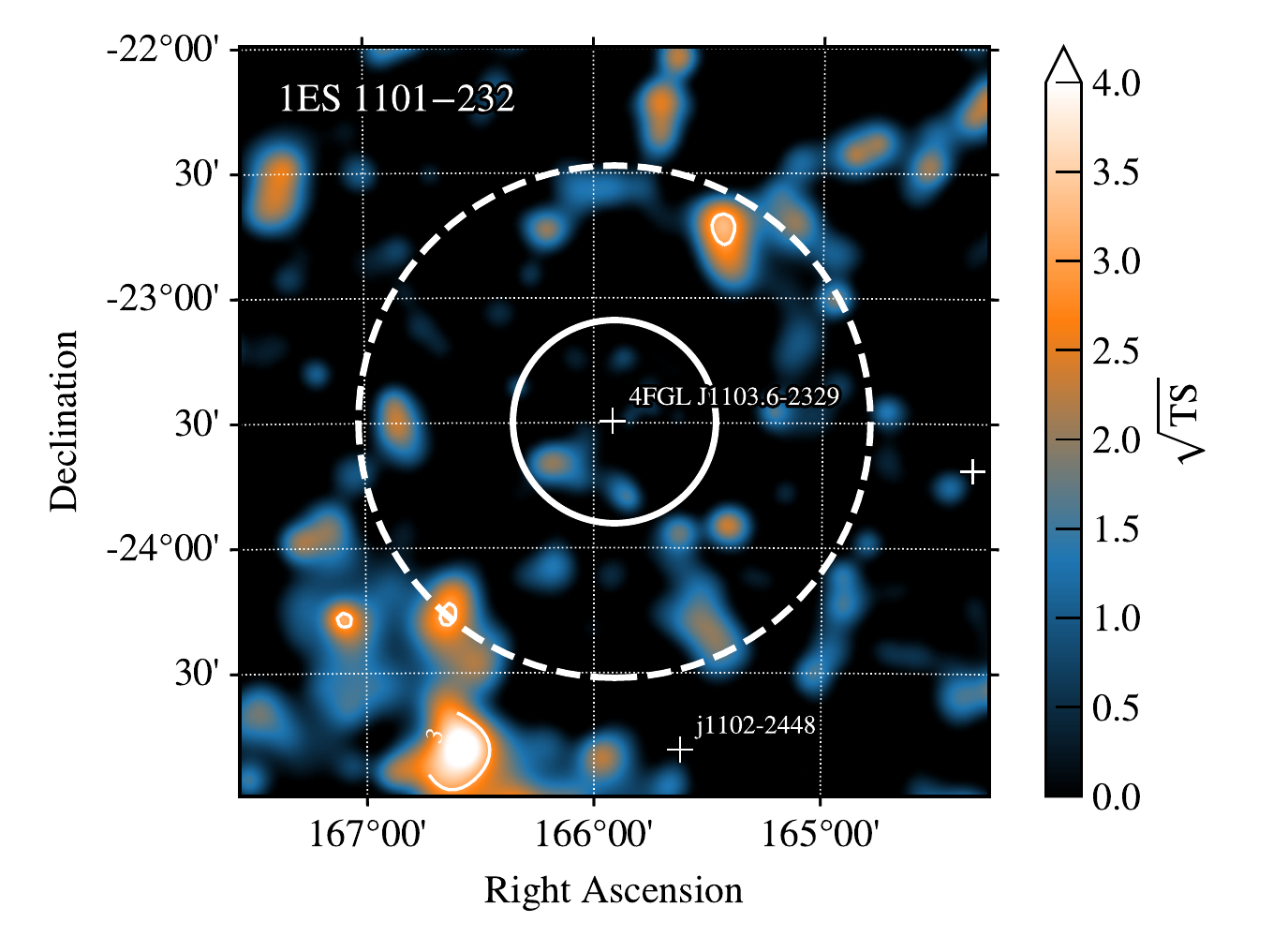}{0.5\textwidth}{}
          \fig{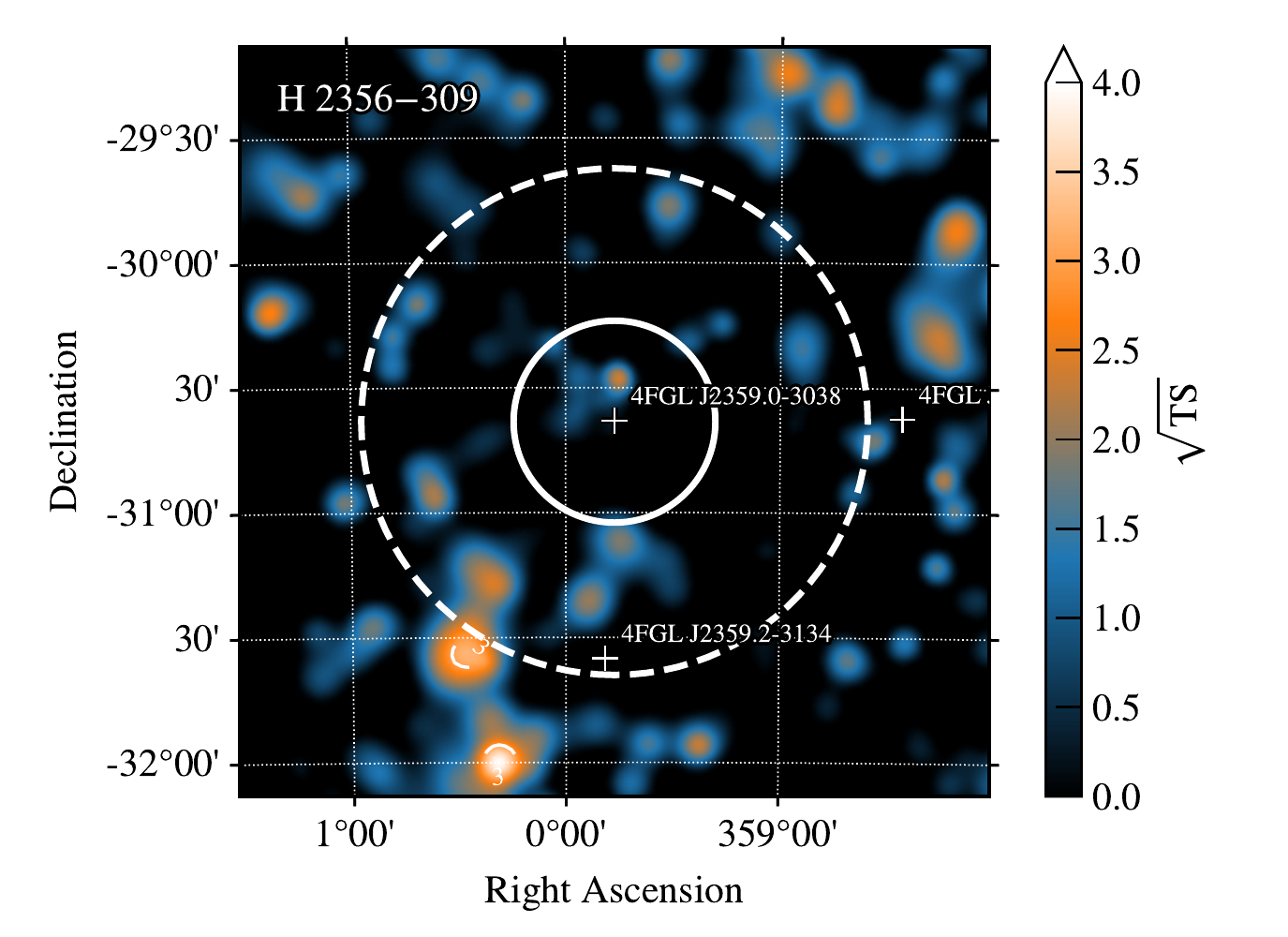}{0.5\textwidth}{}
          }

    \caption{TS maps as in Figure~\ref{fig:tsmap} for all other considered sources. Again, no significant excess is observed close to any considered source. The additional source in the ROI of 1ES~1101$-$232 is marked with a lowercase j and its coordinates. }
    \label{fig:fermi-tsmaps}
\end{figure*}

For the initial optimization of the ROI model we include the point sources listed in the 4FGL-DR2 up to an angular distance of $10^\circ$ from the ROI center. Templates for the galactic diffuse (\texttt{gll\_iem\_v07}) and isotropic diffuse (\texttt{iso\_P8R3\_SOURCE\_V3\_v1}) emission are added to the model as well\footnote{See \url{https://fermi.gsfc.nasa.gov/ssc/data/access/lat/BackgroundModels.html}. For the isotropic diffuse emission, we calculate an average of the templates for the PSF classes 0-2.}. 
All spectral parameters of sources up to a radius of $6^\circ$ are left free to vary, and only spectral normalizations are free parameters for sources at larger distances. After an initial optimization using the \texttt{fermipy} \texttt{optimize} routine, we fix all spectral source parameters (spectral shape parameters, i.e., all parameters but the spectral normalization) 
for sources that are detected with a $\mathrm{TS} < 5$ ($\mathrm{TS} < 9$) or that have a predicted number of photons less than $10^{-2}$. The flux normalization parameter of sources with $\mathrm{TS} > 50$ that was originally frozen is also left free to vary during the fit. 
After a first fit to the data, a $\mathrm{TS}$ map is computed. Putative point sources (modeled with a power law) are added consecutively to the ROI to hot spots with $\mathrm{TS} > 25$. 
This procedure could tentatively mask any additional unaccounted flux from the cascade halo.
Only one additional source had to be added in the vicinity of 1ES~1101$-$232 at 
a distance of larger than $1^\circ$, which should not have an effect on the halo search.
From this it appears unlikely that our analysis procedure could mask an additional halo contribution. 
The TS maps for 1ES~0229+200 are shown in Figure~\ref{fig:tsmap} and for all other sources in Figure~\ref{fig:fermi-tsmaps}.
The above analysis choices result in generally flat spectral and spatial residuals.

The best-fit spectral parameters for a point-source analysis with and without EBL absorption are provided in Table~\ref{tab:fermi_only}. 
For the fit with EBL, we only provide the spectral index, $\Gamma_\mathrm{int}$, which, for all sources, is compatible with the fit without absorption (with index $\Gamma$) within $1\,\sigma$ uncertainties. 
The reason is the limited number of observed \Grays by the LAT in the energy range where the optical depth $\tau_{\gamma\gamma} \gtrsim 1$.
The obtained point-source spectra for all sources are shown as blue squares and solid lines in Figure~\ref{fig:spectra-tmax1e7} in Appendix~\ref{app:all-fits}.
We note that the LAT spectrum of 1ES~0347-121 exhibits evidence for a  detection ($\sqrt{\mathrm{TS}} \gtrsim 4$) of the source above 1\,TeV.
This result is due to two photons detected above 1\,TeV in the vicinity of the source.

\begin{table*}[htb]
    \centering
    \caption{Best-fit parameters for the LAT spectra of the considered EHBL sources without a halo component.\label{tab:lat-spectra}}
    \begin{tabular}{lcccc}
        \hline\hline
         Source Name &  $N_0$ ($10^{-14}$ MeV$^{-1}$ cm$^{-2}$ s$^{-1}$) & $E_0$ (GeV) & $\Gamma$ & $\Gamma_\mathrm{int}$ \\
         \hline
         1ES\,0229+200 &  1.02             $\pm$ 0.14  & 5.80 & 1.77 $\pm$              0.09 & 1.73 $\pm$ 0.12  \\
         1ES\,0347-121  &  1.90             $\pm$ 0.21  & 4.82 & 1.81 $\pm$              0.08 & 1.81 $\pm$ 0.11  \\
         PKS\,0548-322  &  2.17             $\pm$ 0.32  & 3.29 & 1.99 $\pm$              0.12 & 2.06 $\pm$ 0.17  \\
         1ES\,1101-232  &  2.09             $\pm$ 0.23  & 4.84 & 1.68 $\pm$              0.07 & 1.65 $\pm$ 0.09 \\
         H\,2356-309  &  3.49             $\pm$ 0.35  & 3.70 & 1.78 $\pm$              0.07 & 1.81 $\pm$ 0.09  \\
    \hline
    \end{tabular}
    \label{tab:fermi_only}
\end{table*}

\subsection{Extracting Likelihood Values in the Presence of the Halo}
\label{app:fermi-halo}

In the presence of a halo, it is nontrivial to extract the likelihood values, $\mathcal{L}(B, \boldsymbol{\theta}_i| \mathcal{D}_{\mathrm{LAT}, i})$, using the built-in functionality of \texttt{fermipy}. 
The reason is that the cascade flux depends on the intrinsic spectral parameters of the blazar. 
Therefore, we loop over the spectral parameters of the central blazar and fix them at each step.
Then, we add the corresponding reweighted halo template and extract the likelihood values. 
More specifically, we start from the optimized ROI model of the LAT data for each source $i$ as described in Section \ref{sec:lat-obs}. We loop through values of the source-intrinsic spectral index, $\Gamma_\mathrm{int} = \Gamma_{\mathrm{int},0} + \Delta\Gamma_\mathrm{int}$, where $\Gamma_{\mathrm{int},0}$ denotes the best-fit value from LAT data (see Table \ref{tab:sources}) and $\Delta\Gamma_\mathrm{int}$ ranges from $-0.75$ to $0.75$ with a step size of $0.075$. We further sample the exponential cutoff energy of the spectrum, $E_\mathrm{cut} = 1, 3, \ldots, 13\,$TeV (which itself is not constrained from LAT data) and the spectral normalization $N_0$.
The choice is motivated by the energy range of the \hess spectra. 
For each combination of the spectral parameters, we reweight the cascade template, add it to the ROI, and extract the multidimensional profile likelihood cube for the model with the halo emission to describe the data.
The normalizations of the other (diffuse) sources in the ROI are optimized in each step of the loop. 

\section{Details on the \hess data analysis}
\label{app:hess}
In this appendix, details on the analysis choices for the \hess observations are given in addition to the main selection criteria described in Section~\ref{sec:hess-obs}.
For each source, the runs are stacked onto one data set per source for runs that pass our selection given in the main text in combination with an offset between telescope pointing and source direction of $\leqslant 2^\circ$ and a zenith angle of $Z \leqslant 60^\circ$. 
The background in the On region is estimated from multiple reflected
``Off'' regions of the same size that have a minimum separation to the source location of $0.2^\circ$~\citep{2006A&A...457..899A}.
The energy threshold $E_\mathrm{thr}$ of this stacked data set is computed from the condition that the energy bias of the energy dispersion matrix is below 10\,\%. 
The median zenith angles $\langle Z \rangle$ and values of $E_\mathrm{thr}$ are provided in Table~\ref{tab:hess-only}.
We choose an energy binning of the recorded events of 24 bins per decade. 
Using the \texttt{MINUIT} back end in \gammapy, we fit a simple power law without (with) EBL absorption to the observations. 
The best-fit parameters, $N_{0}$ ($N_{0,\mathrm{int}}$) and $\Gamma$ ($\Gamma_\mathrm{int}$) are also provided in Table~\ref{tab:hess-only}.
The EBL absorption effect is clearly visible as $\Gamma_\mathrm{int} < \Gamma$ for all sources.
The energy normalization $E_0$ is determined from the fit without EBL absorption such that the off-diagonal terms of the covariance matrix are minimal. 
The resulting point-source spectra are shown in Figure~\ref{fig:spectra-tmax1e7} in Appendix~\ref{app:all-fits} as a red bow tie and red points. For the spectral points, we use 9 bins between $E_\mathrm{thr}$ and $10^{1.25}\,$TeV.
A cross-check analysis with an independent analysis chain \citep{2009APh....32..231D} yields consistent results. 

\begin{table*}[htb]
    \centering
    \caption{
    Mean zenith angle $\langle Z \rangle$, energy threshold $E_\mathrm{thr}$, 
    and best-fit parameters for the \hess spectra of the considered EHBL sources without a halo component.\label{tab:hess-only}}
    \begin{tabular}{l|cc|ccccc}
        \hline\hline
         Source &  $\langle Z \rangle$ & $E_\mathrm{thr}$ & $N_{0}$  &  $N_{0, \mathrm{int}}$ & $E_0$ & \multirow{2}{*}{$\Gamma$} & \multirow{2}{*}{$\Gamma_\mathrm{int}$} \\
         Name & (deg) & (TeV) & ($10^{-12}$ TeV$^{-1}$ cm$^{-2}$ s$^{-1}$) & ($10^{-12}$ TeV$^{-1}$ cm$^{-2}$ s$^{-1}$) & 
         (TeV) & & \\
         \hline
         1ES\,0229+200 & 44.4 & 0.46 & 0.32             $\pm$ 0.02  & 1.64 $\pm$ 0.11 & 1.14 & 2.81 $\pm$              0.11 & 1.76 $\pm$ 0.12 \\
         1ES\,0347-121  & 16.1 & 0.22 &  4.70             $\pm$ 0.35  & 11.94 $\pm$ 0.09  & 0.43 & 3.40 $\pm$              0.14 & 2.12 $\pm$ 0.15  \\
         PKS\,0548-322  &  11.0 & 0.22 & 0.44             $\pm$ 0.05  & 0.85 $\pm$ 0.09  & 0.89 & 2.44 $\pm$              0.11 & 1.92 $\pm$ 0.12  \\
         1ES\,1101-232  & 21.3 & 0.16 & 2.75             $\pm$ 0.18  & 10.13 $\pm$ 0.64 &  0.54 & 2.93 $\pm$              0.09 & 1.66 $\pm$ 0.09  \\
         H\,2356-309  &  13.5 & 0.22 & 3.60             $\pm$ 0.18  & 9.14 $\pm$ 0.46 &  0.46 & 3.21 $\pm$              0.09 & 2.10 $\pm$ 0.09  \\
    \hline
    \end{tabular}
\end{table*}

\section{Details on the cascade simulations with \crpropa}
\label{app:crpropa}

The \crpropa simulations are carried out in the large sphere observer framework, with the source placed at the origin of a sphere with radius equal to the comoving distance $D$ to the source. 
We use a standard $\Lambda$CDM cosmology with $H_0 = 67.3\,\mathrm{km}\,\mathrm{s}^{-1}\,\mathrm{Mpc}^{-1}$ and a matter density of $\Omega_m = 0.315$ (these are default \crpropa values). 
From the source, gamma rays are injected into a cone with opening angle  $\theta_\mathrm{jet} = 6^\circ$ (which is in the range of typically observed angles; e.g., \citealt{2017ApJ...846...98J}) with its central axis pointing along the negative $x$-axis in the simulation coordinate system. 
We include energy losses through pair production, IC scattering, synchrotron emission of electrons in the IGMF, and propagation of particles in an expanding universe. 
Within the propagation module of \crpropa, we set the tolerance parameter, which determines the numerical precision, to $10^{-9}$ and set the minimum step length of the simulation such that time delays $\tau$ down to 1~yr can be resolved. 
In order to reduce the computational time, we stop tracing particles if their energy drops below 0.1\,GeV or if their total propagating distance becomes larger than 4\,Gpc. 
We inject monochromatic gamma rays with energies $\epsilon$ equal to the central energies of bins $\Delta\epsilon$ between 1\,GeV and $\approx31.6$\,TeV using 4 bins per decade. 
The value of the maximum energy is limited by the available computational resources for the simulations.
As shown in Section~\ref{sec:combined}, the intrinsic blazar spectra will tend to have lower cutoff energies so that this choice does not affect  our results. 
The time delay $\tau$ of the cascade gamma rays is calculated through $c\tau \approx (d - D),$ where $c$ is the speed of light. 
To mitigate simulation noise introduced through a finite number of injected gamma rays, the sky maps are adaptively smoothed using the \texttt{ASMOOTH} algorithm  \citep{2006MNRAS.368...65E}.
From the \crpropa simulation, we obtain a multidimensional histogram of photon counts $\mathcal{N}$. 
We can convert this histogram into the number of counts arriving per injected energy $\epsilon$, observed energy $E$, solid angle $\Omega$, time delay interval $d\tau$, and injected particle in the interval $\Delta\epsilon$ as

\begin{equation}
\frac{d\mathcal{N}}{d\epsilon dE d\tau d\Omega} = \frac{1}{N_\mathrm{inj}(\Delta\epsilon)}
\frac{\mathcal{N}}{\Delta\epsilon \Delta E \Delta \tau \Delta \Omega},
\label{eq:casc-spec-initial}
\end{equation}
where $\Delta X$ is the chosen bin width of the \crpropa histogram for quantity $X$.
To speed up computations, the number of injected particles $N_\mathrm{inj}$ decreases for larger injected energies and thus depends on the injected energy.
For a single simulation, $N_\mathrm{inj} = 10$ for $\epsilon < 1$\,TeV, $N_\mathrm{inj} = 5$ for $1\leqslant\epsilon < 10$\,TeV, and $N_\mathrm{inj} = 3$ for $\epsilon \geqslant 10$\,TeV. 
For magnetic fields $\log_{10}(B/\mathrm{G}) \leqslant -15$, 1000 independent simulations are performed that are combined into a single spectrum. For higher $B$ fields we run more simulations as more particles are deflected at large angles and discarded (see Section~\ref{app:crpropa-rotations}), which increases 
the noise of the simulations. 
We increase the number of simulations by 200 for each tested IGMF, resulting in 1800 simulations for $\log_{10}(B/\mathrm{G}) = -13$.

To obtain the cascade flux that arrives within some maximum time delay $\tau_\mathrm{max}$ for an arbitrary injected (time-averaged) photon spectrum $dN/d\epsilon$ (e.g., in units $\mathrm{s}^{-1}\mathrm{cm}^{-2}\mathrm{eV}^{-1}$), we need to convolve the initial cascade spectrum in Eq.~\eqref{eq:casc-spec-initial} with the injected spectrum and integrate over the delay times, 
\begin{equation}
    \frac{d\mathcal{N}}{dE d\Omega} = 
    \int\limits_0^\infty d\epsilon \int\limits_0^\infty d\epsilon' \frac{dN}{d\epsilon'} \int\limits_{0}^{\tau_\mathrm{max}} d\tau \frac{d\mathcal{N}}{d\epsilon' dE d\tau d\Omega} \approx
    \sum_i \sum_j  \Delta\epsilon_i \Delta\tau_j w_i \left(\frac{d\mathcal{N}}{d\epsilon dE d\tau d\Omega}\right)_{ij}.
    \label{eqn:casc-flux-no-time}
\end{equation}

In the last step, we have discretized the two integrals and introduced the spectral weights $w_i$ for the $i$th energy bin of the injected spectrum $\Delta\epsilon_i$,
\begin{equation}
w_i = \int\limits_{\Delta\epsilon_i} \frac{dN}{d\epsilon} d\epsilon. 
\label{eqn:spec-weights}
\end{equation}

\subsection{Applied rotations in the large sphere observer framework}
\label{app:crpropa-rotations}

Each event is recorded when it hits the sphere at some position $\vec{x}$ in the global simulation coordinate system $S_\mathrm{sim}$ with Cartesian unit vectors $\hat{x}_\mathrm{sim}$, $\hat{y}_\mathrm{sim}$, and $\hat{z}_\mathrm{sim}$. 
For each event the final (initial) unit momentum vector $\vec{p}$ ($\vec{p}_0$), final (initial) energy $E$ ($\epsilon$), and propagation length $d$ are recorded.
The sphere has a radius equal to the comoving distance $D$ to the source, and the source is located at the center of the sphere with Cartesian coordinates $\vec{x}_0 = (0,0,0)$.
For the simulation, we assume a source that emits particles uniformly into a cone with aperture $\theta_\mathrm{jet}$ with the jet axis $\vec{j}$ pointing in the $(-1,0,0)$ direction.

We adopt the large sphere observer framework and follow the same method as in \citet{2016PhRvD..94h3005A}, placing an observer at each hit position $\vec{x}$ on the surface of the sphere and defining a local coordinate system $S_\mathrm{obs}$.
In $S_\mathrm{obs}$, the unit vector of the $z_\mathrm{obs}$-axis points along the connection between the hit location and the source, i.e., the LOS, defined by the unit vector $\vec{\xi} = \xi_x \hat{x}_\mathrm{sim} + \xi_y \hat{y}_\mathrm{sim} + \xi_z \hat{z}_\mathrm{sim} = (\vec{x} - \vec{x}_0) / |\vec{x} - \vec{x}_0|$. 
With this definition, we see that $\hat{z}_\mathrm{obs} = -\vec{\xi}$.
The $\hat{x}_\mathrm{obs}$ and $\hat{y}_\mathrm{obs}$ unit vectors are then defined in such a way that $S_\mathrm{obs}$ is a right-handed coordinate system.
Introducing the definitions, 
\begin{eqnarray}
 \rho &=& \sqrt{\xi_x^2 + \xi_y^2} \\
 r &=& \sqrt{\xi_x^2 + \xi_y^2 + \xi_z^2} \\
 \cos\phi &=& \frac{\xi_x}{\rho} \\
 \sin\phi &=& \frac{\xi_y}{\rho} \\
 \cos\vartheta &=& \frac{\xi_z}{r} \\
 \sin\vartheta & = & \frac{\rho}{r}, 
\end{eqnarray}
the unit vectors $\hat{x}_\mathrm{obs}$ and $\hat{y}_\mathrm{obs}$ in $S_\mathrm{sim}$ are found to be

\begin{eqnarray}
 \hat{x}_\mathrm{obs} & = & (-\sin\phi, \cos\phi, 0)^T, \\
 \hat{y}_\mathrm{obs} & = & (\cos\phi\cos\vartheta, \sin\phi\cos\vartheta, -\sin\vartheta)^T.
\end{eqnarray}

We proceed by projecting the momentum vectors $\vec{p}$ and $\vec{p_0}$ into the new basis, 
\begin{eqnarray}
 \vec{p}_\mathrm{obs} &=& \langle \vec{p}, \hat{x}_\mathrm{obs} \rangle \hat{x}_\mathrm{obs} + \langle \vec{p}, \hat{y}_\mathrm{obs} \rangle \hat{y}_\mathrm{obs} + \langle \vec{p}, \hat{z}_\mathrm{obs} \rangle \hat{z}_\mathrm{obs}, \\
  \vec{p}_{0,\mathrm{obs}} &=& \langle \vec{p}_0, \hat{x}_\mathrm{obs} \rangle \hat{x}_\mathrm{obs} + \langle \vec{p}_0, \hat{y}_\mathrm{obs} \rangle \hat{y}_\mathrm{obs} + \langle \vec{p}_0, \hat{z}_\mathrm{obs} \rangle \hat{z}_\mathrm{obs}.
\end{eqnarray}

We can now consider arbitrary orientations of the jet axis with respect to the LOS by defining the jet vector in $S_\mathrm{obs}$ to be 
\begin{equation}
    \vec{j}_\mathrm{obs} = \cos\phi_\mathrm{jet} \sin\vartheta_\mathrm{jet} \hat{x}_\mathrm{obs} + \sin\phi_\mathrm{jet} \sin\vartheta_\mathrm{jet} \hat{y}_\mathrm{obs} + \cos\vartheta_\mathrm{jet}\hat{z}_\mathrm{obs},
\end{equation}
with some values for $\phi_\mathrm{jet}$ and $\vartheta_\mathrm{jet}$. 
The $\vartheta_\mathrm{jet}$ angle is related to the angle $\theta_\mathrm{obs}$ between the LOS and the jet axis through $\theta_\mathrm{obs} = \pi / 2 - \vartheta_\mathrm{jet}$  (here we choose $\phi_\mathrm{jet} = \pi$ and $\vartheta_\mathrm{jet} = \pi/2$).

Only photons will be included in the cascade flux for which the angle $\alpha_\mathrm{obs}$ between the initial photon momentum $\vec{p}_{0, \mathrm{obs}}$ and the jet axis $\vec{j}_\mathrm{obs}$ is smaller than the assumed half-opening angle of the jet, $\theta_\mathrm{jet} / 2$,

\begin{equation}
    \cos\alpha_\mathrm{obs} = \langle -\vec{j}_\mathrm{obs}, \vec{p}_{0, \mathrm{obs}} \rangle \geqslant \cos(\theta_\mathrm{jet} / 2).
\end{equation}

In this way, we can use the same simulation for arbitrary angles between the jet axis and the LOS.

\section{Details on fits with halo components}
\label{app:halo-fit}

In this appendix, we provide additional details on the combined fits of LAT and \hess data in the presence of a halo.

\subsection{Best-fit parameters}
\label{app:all-fits}

\begin{figure}
    \centering
        \gridline{
          \fig{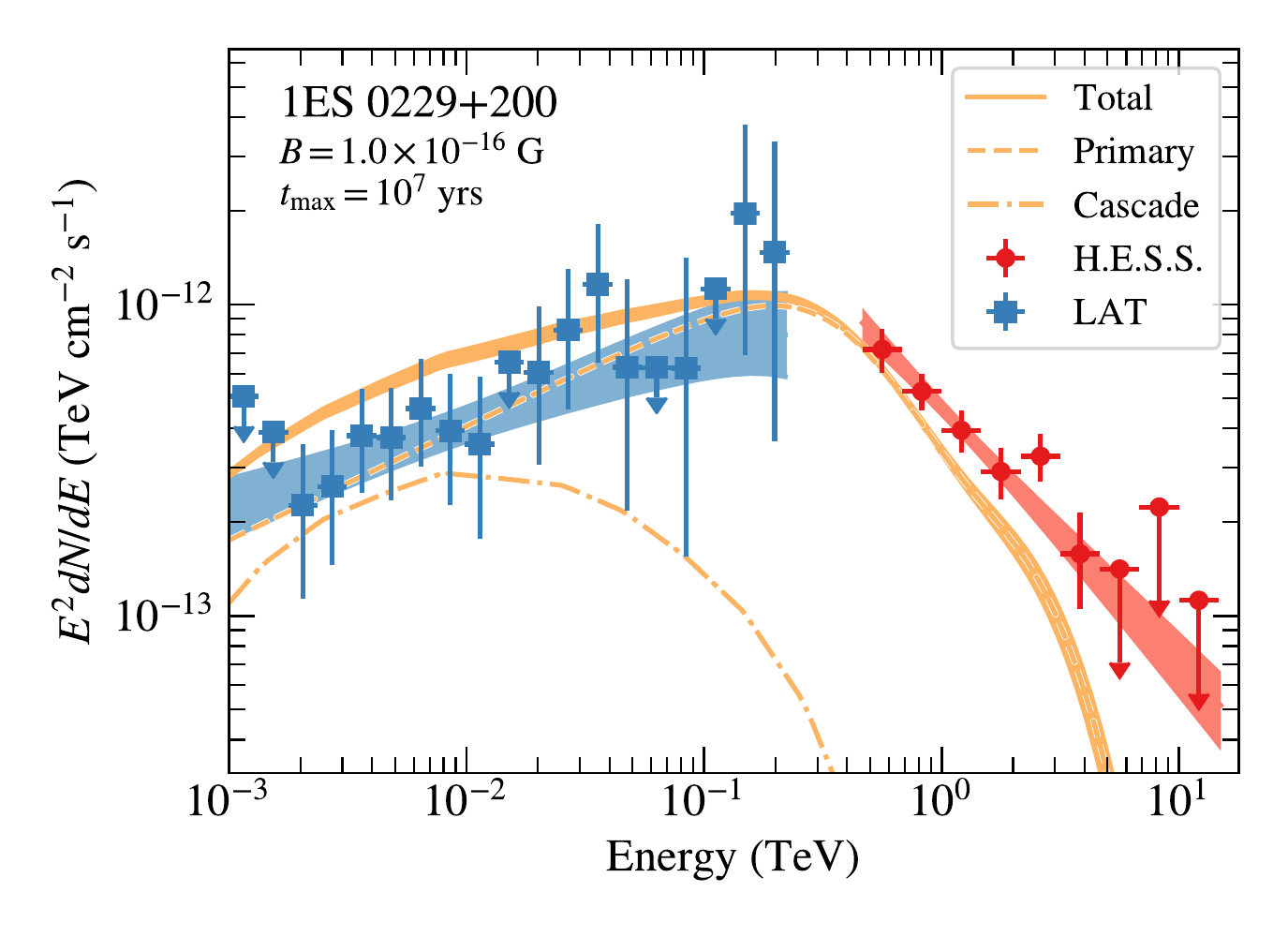}{0.33\textwidth}{}
          \fig{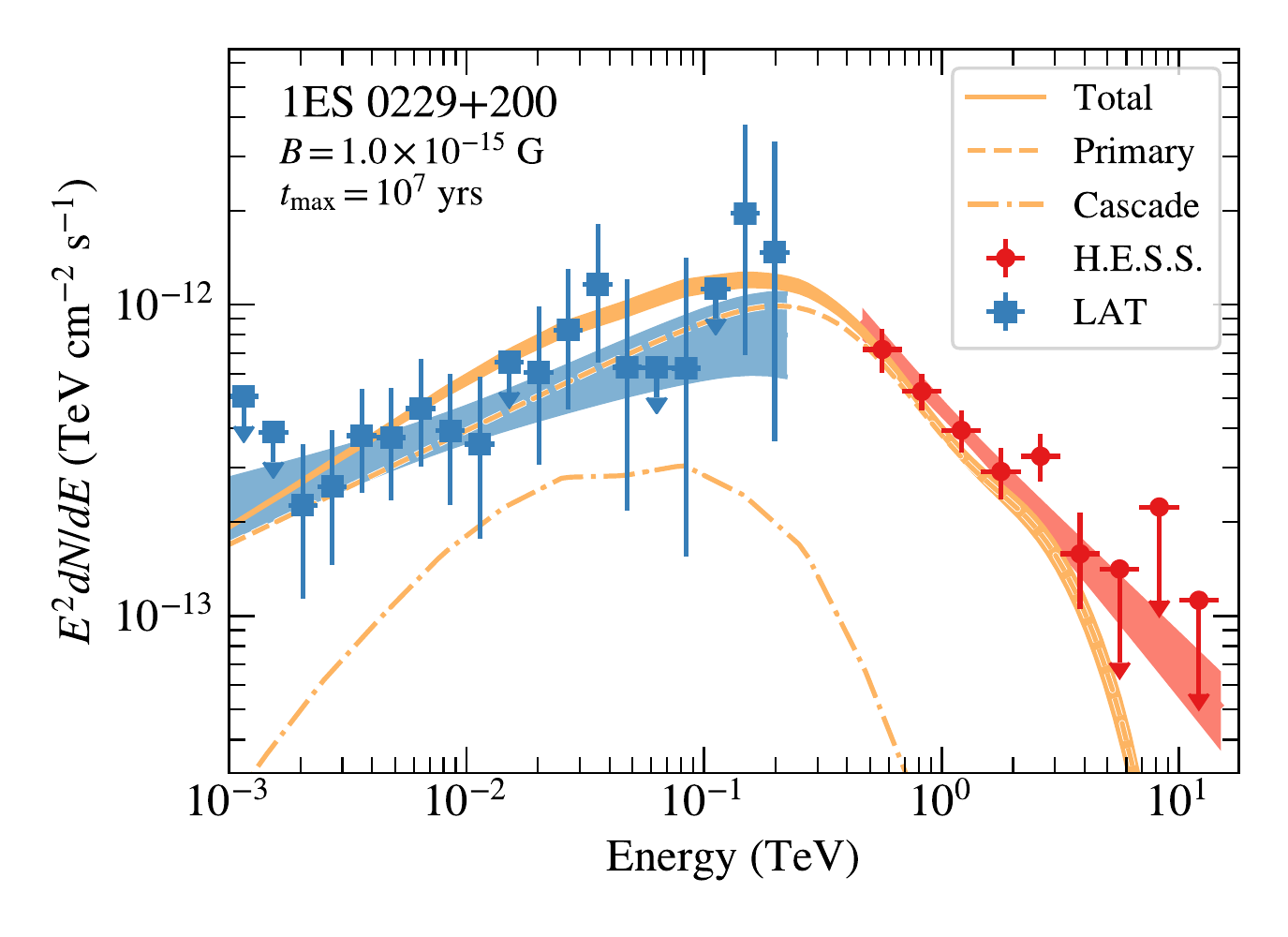}{0.33\textwidth}{}
          \fig{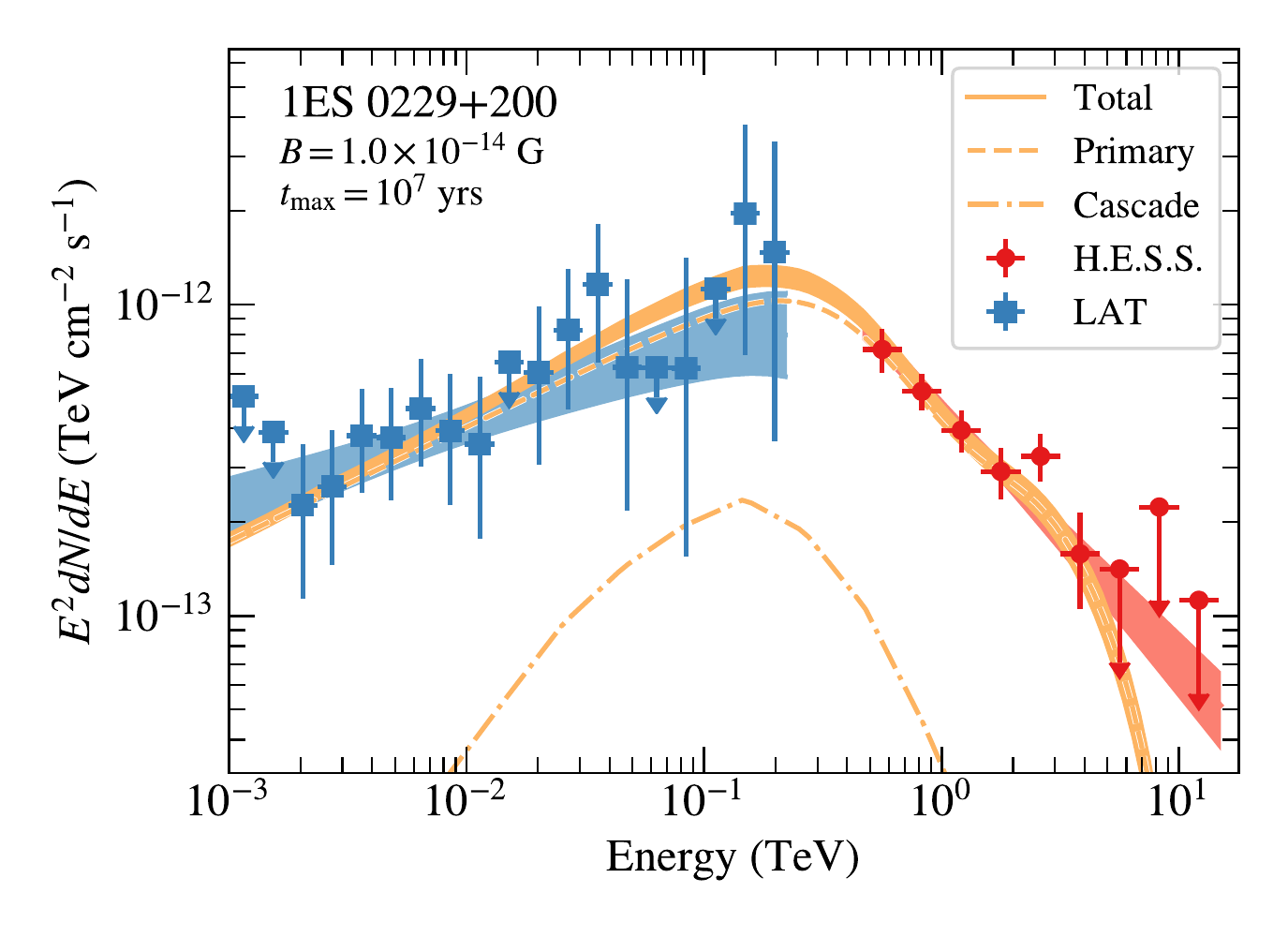}{0.33\textwidth}{}
          }\vspace{-30pt}
        \gridline{
          \fig{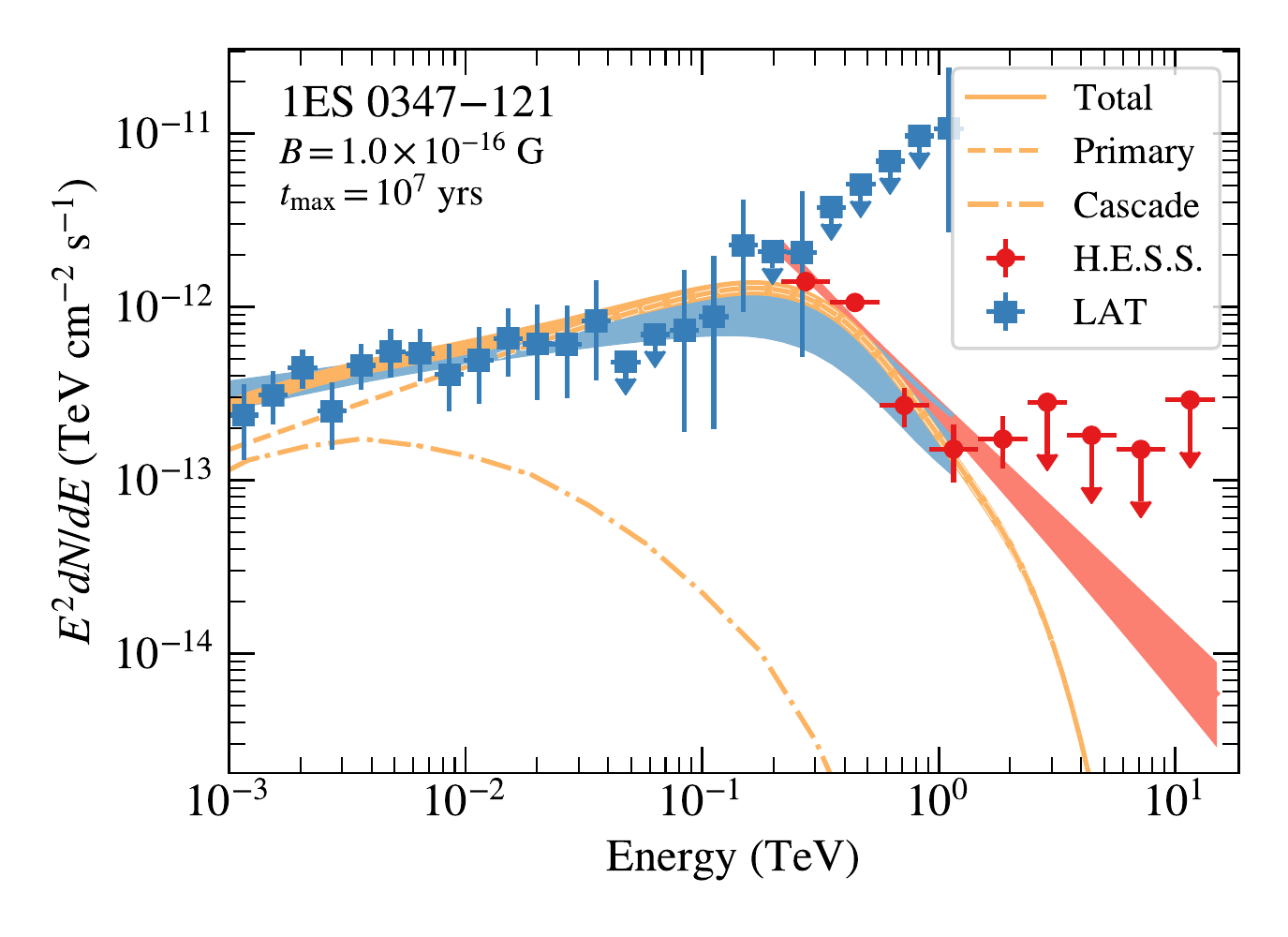}{0.33\textwidth}{}
          \fig{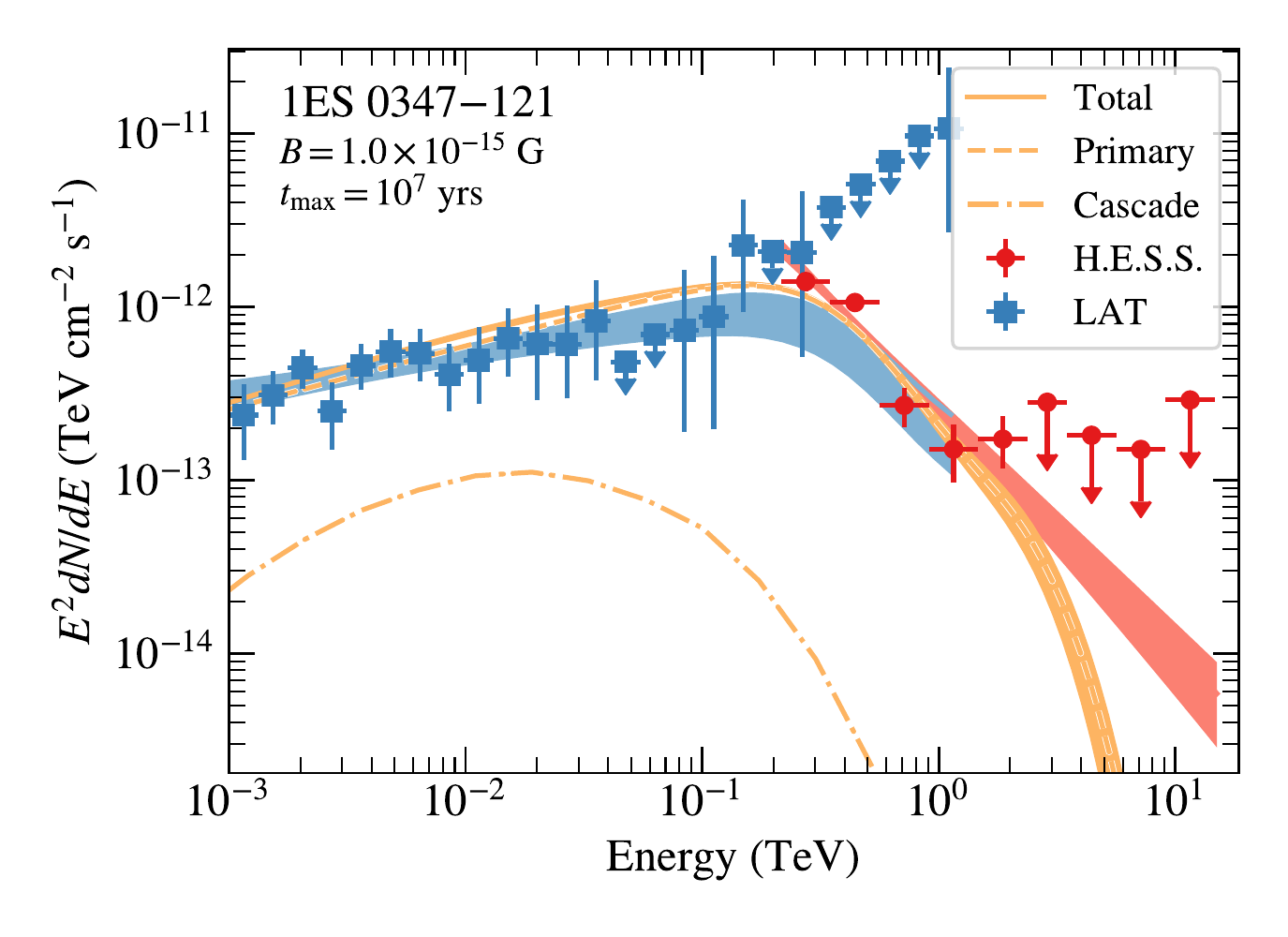}{0.33\textwidth}{}
          \fig{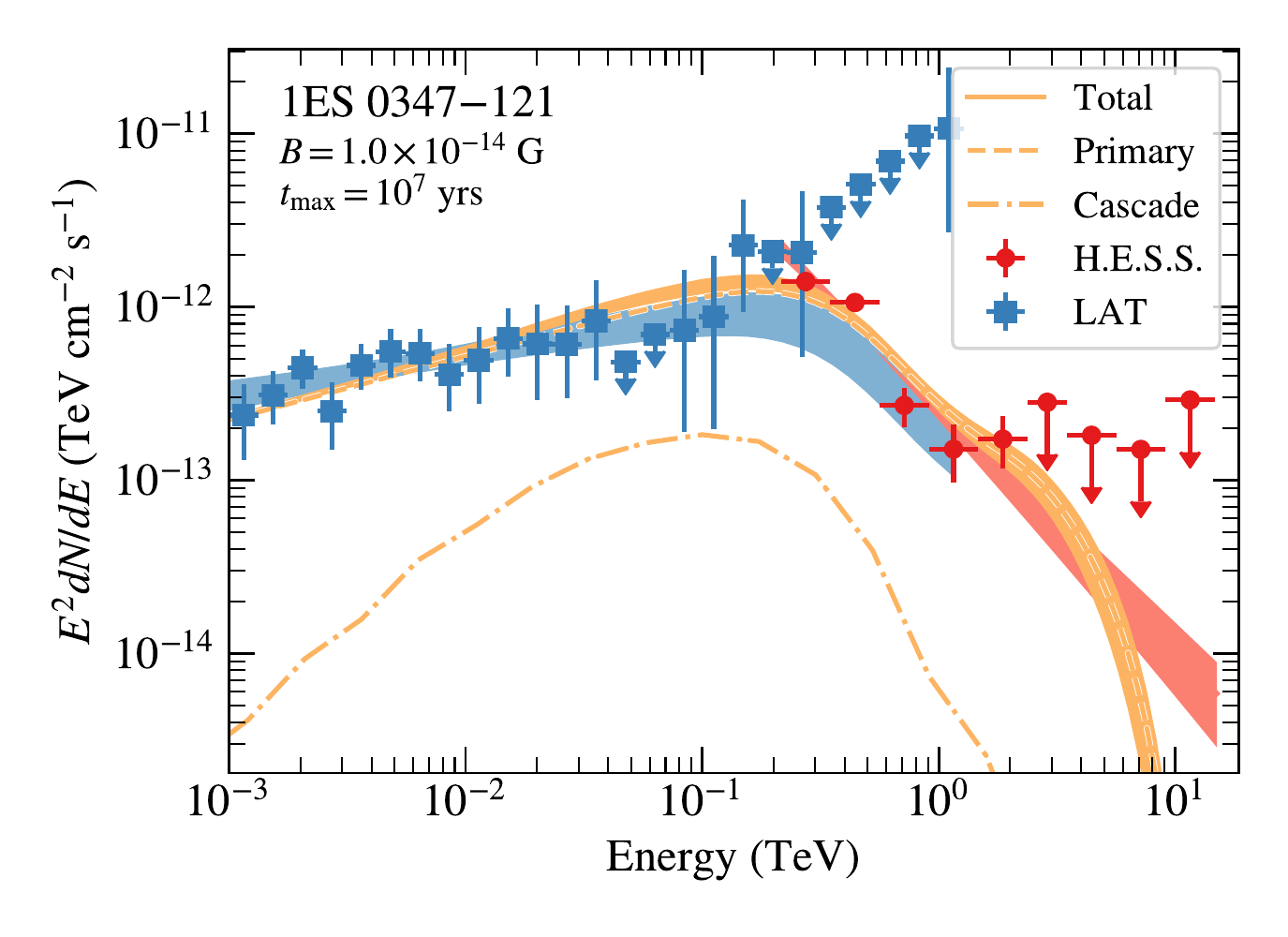}{0.33\textwidth}{}
          }\vspace{-30pt}
        \gridline{
          \fig{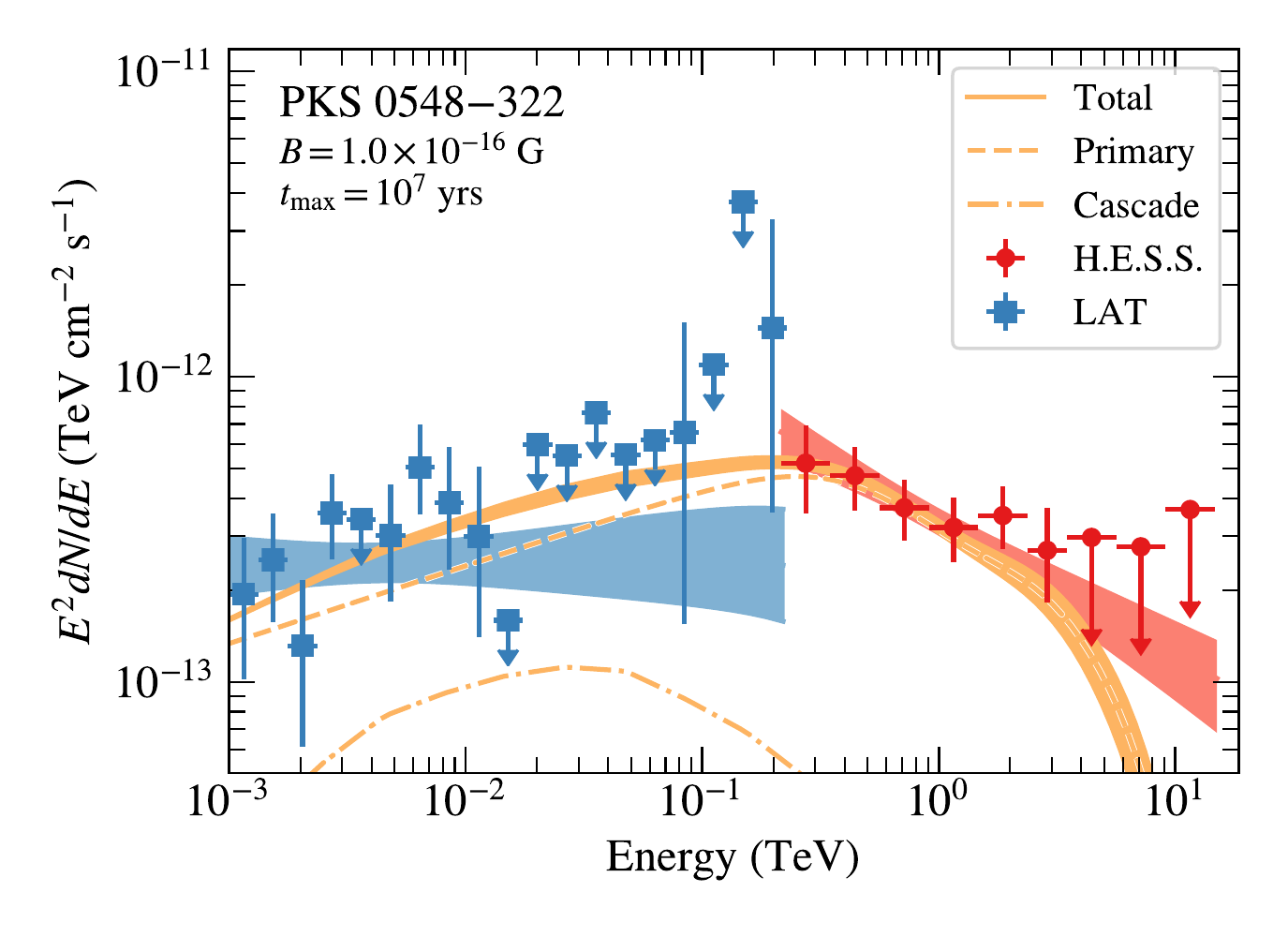}{0.33\textwidth}{}
          \fig{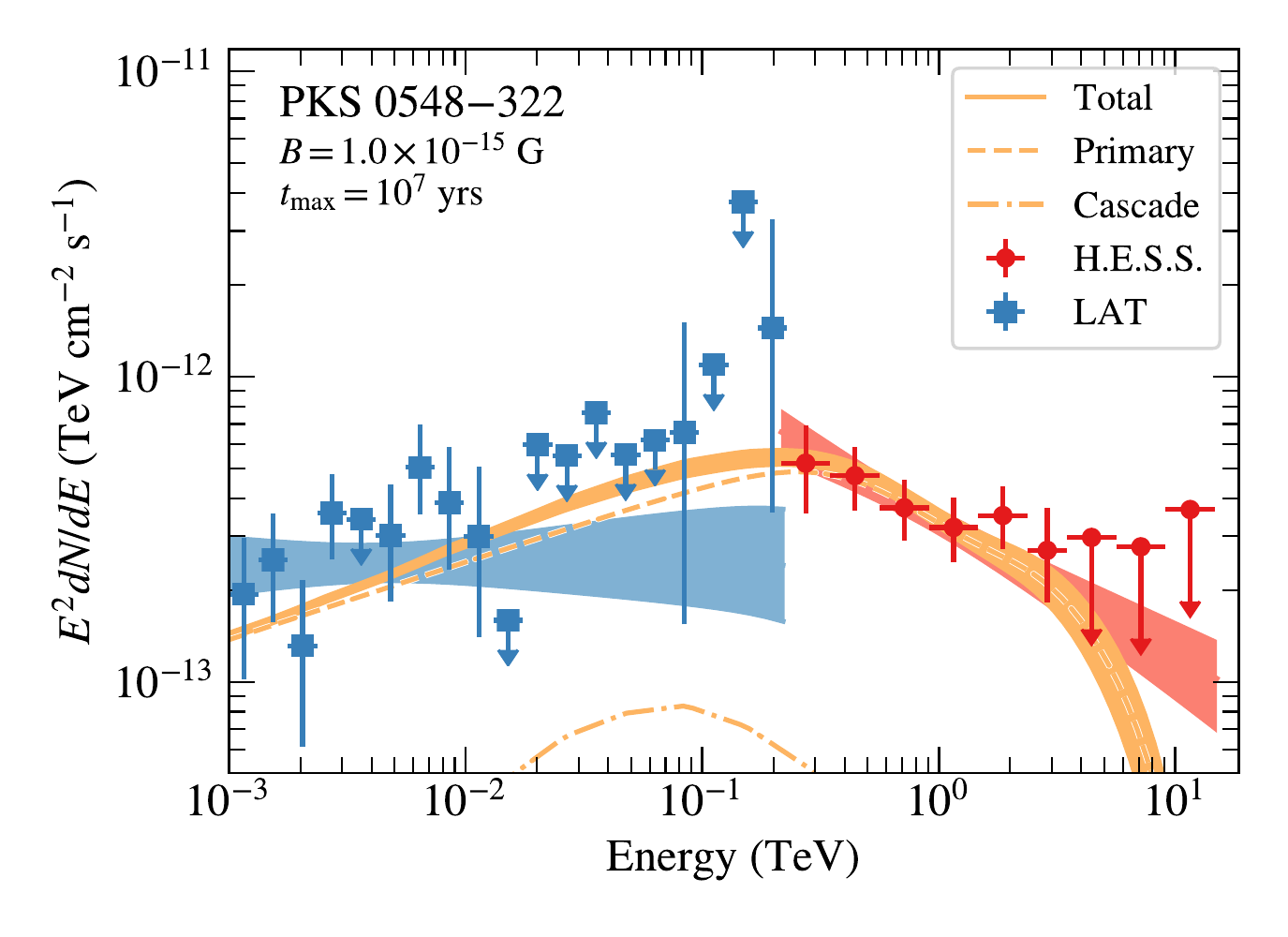}{0.33\textwidth}{}
          \fig{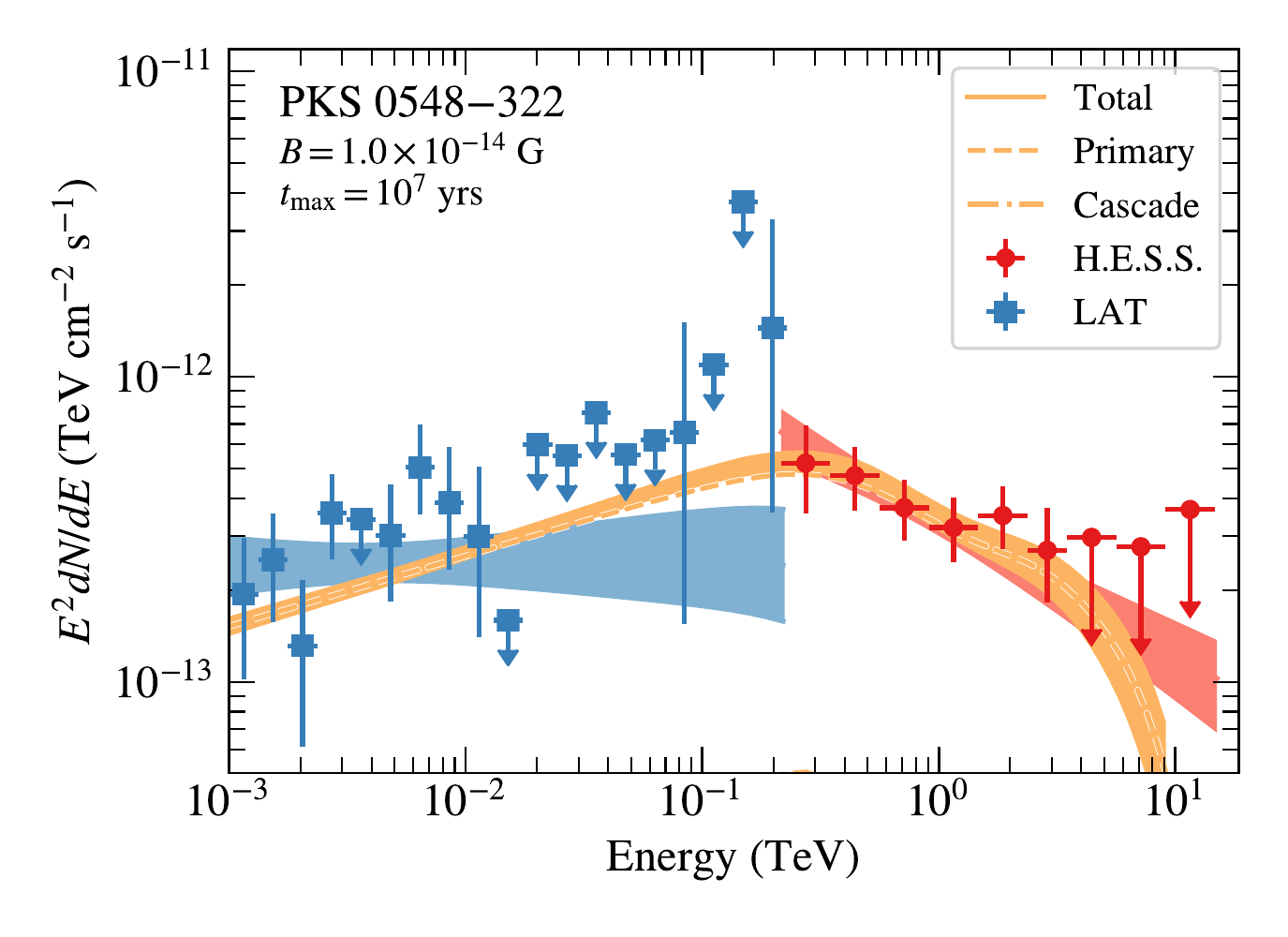}{0.33\textwidth}{}
          }\vspace{-30pt}
        \gridline{
          \fig{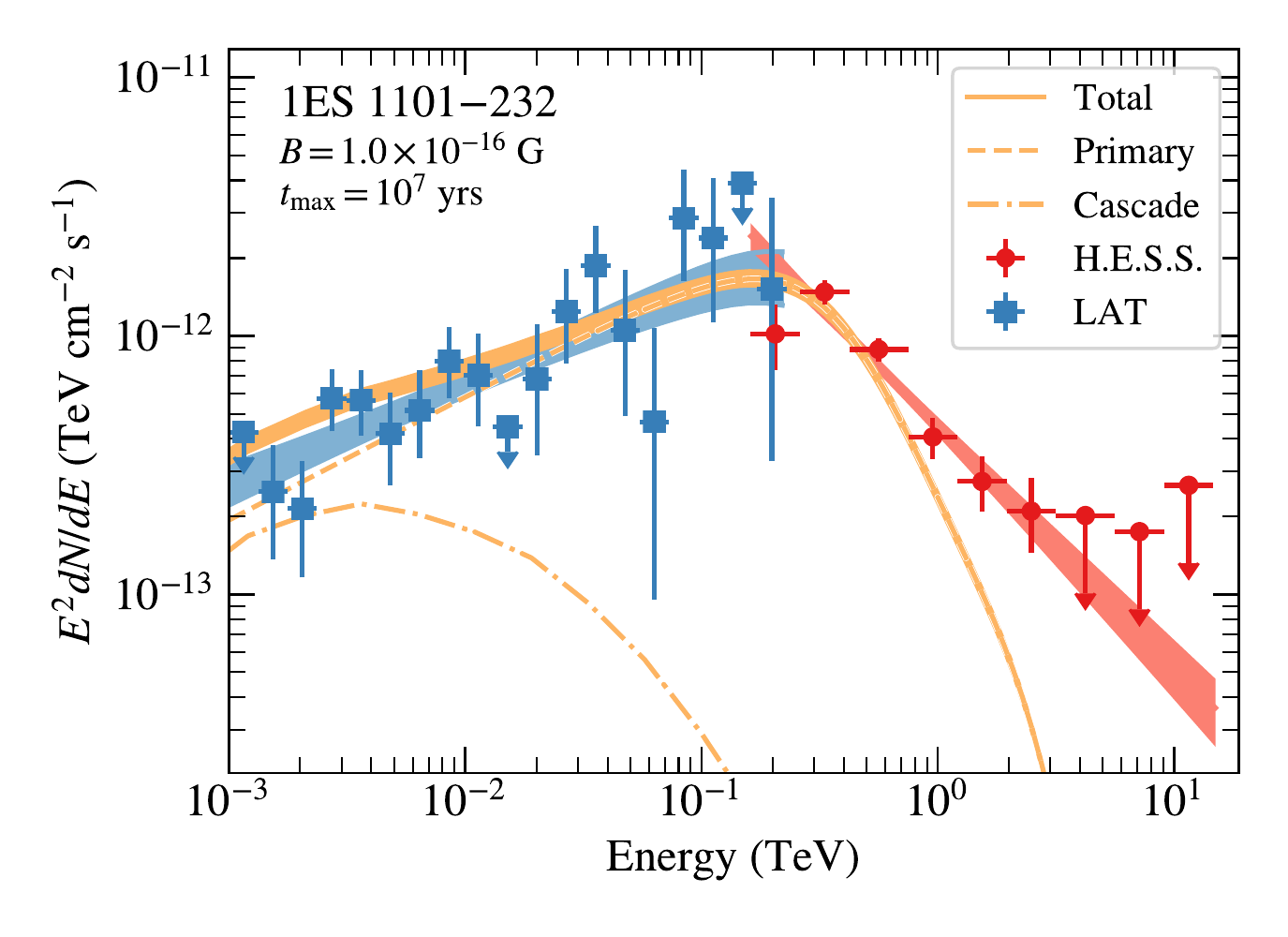}{0.33\textwidth}{}
          \fig{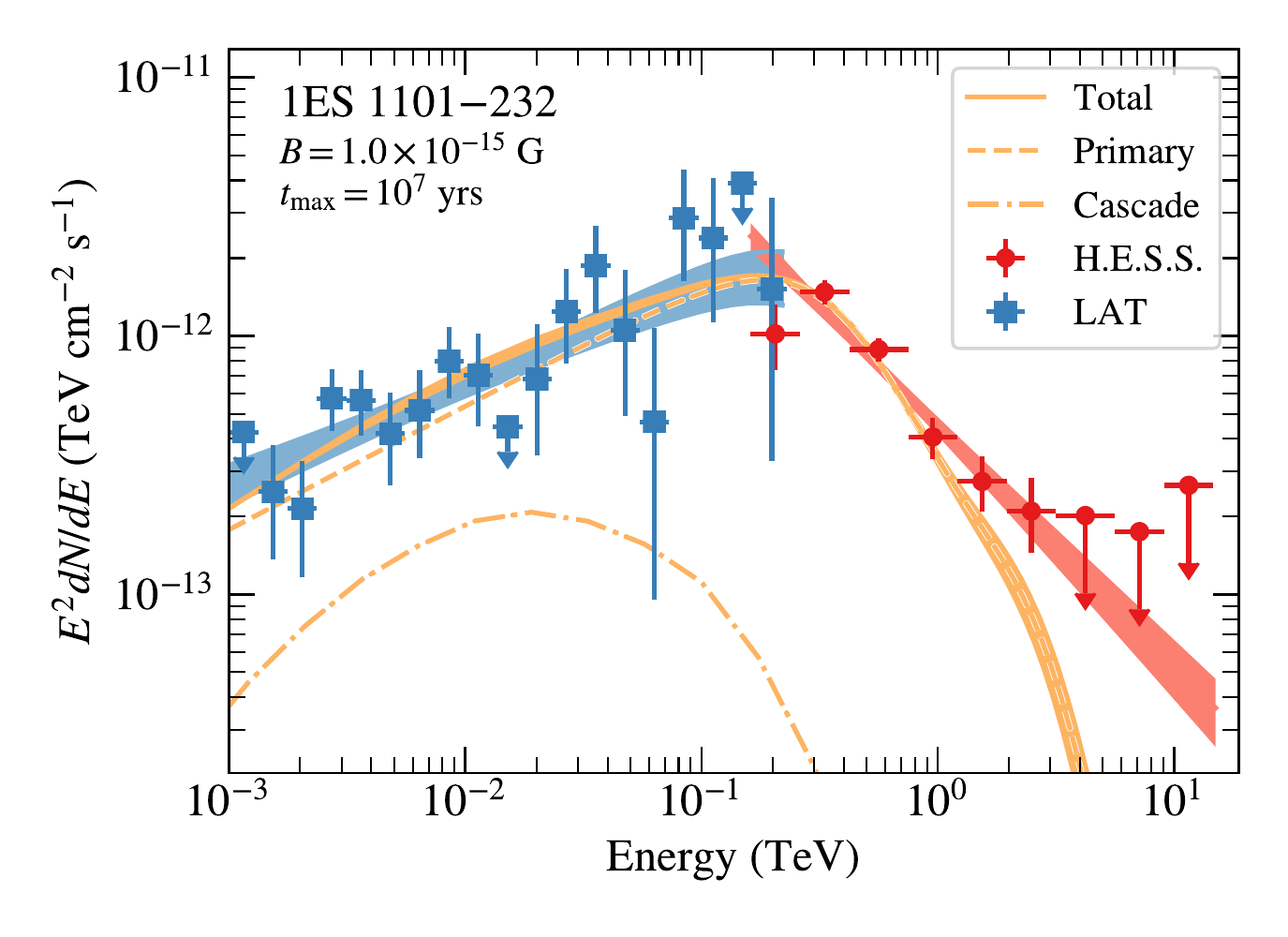}{0.33\textwidth}{}
          \fig{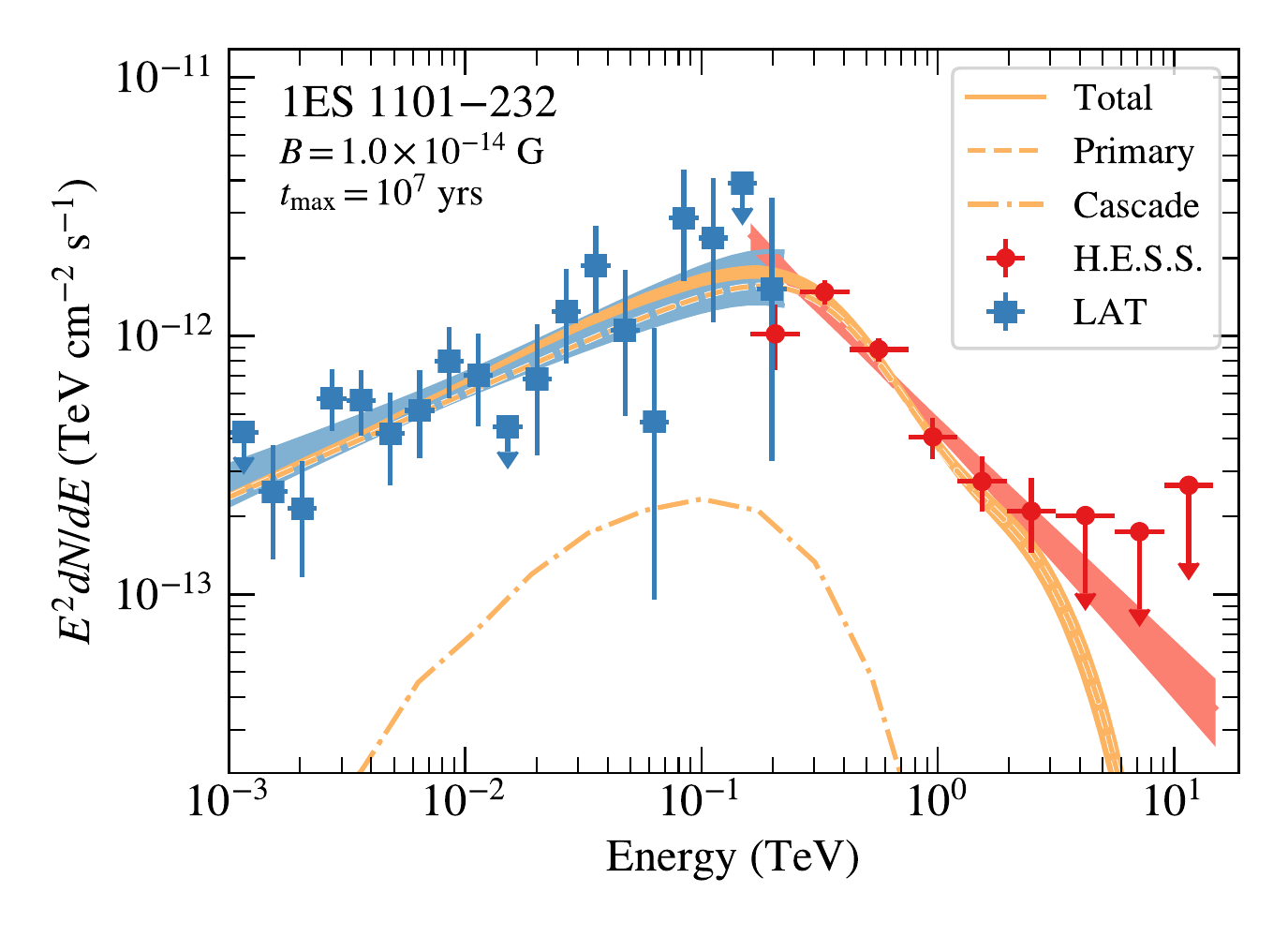}{0.33\textwidth}{}
          }\vspace{-30pt}
        \gridline{
          \fig{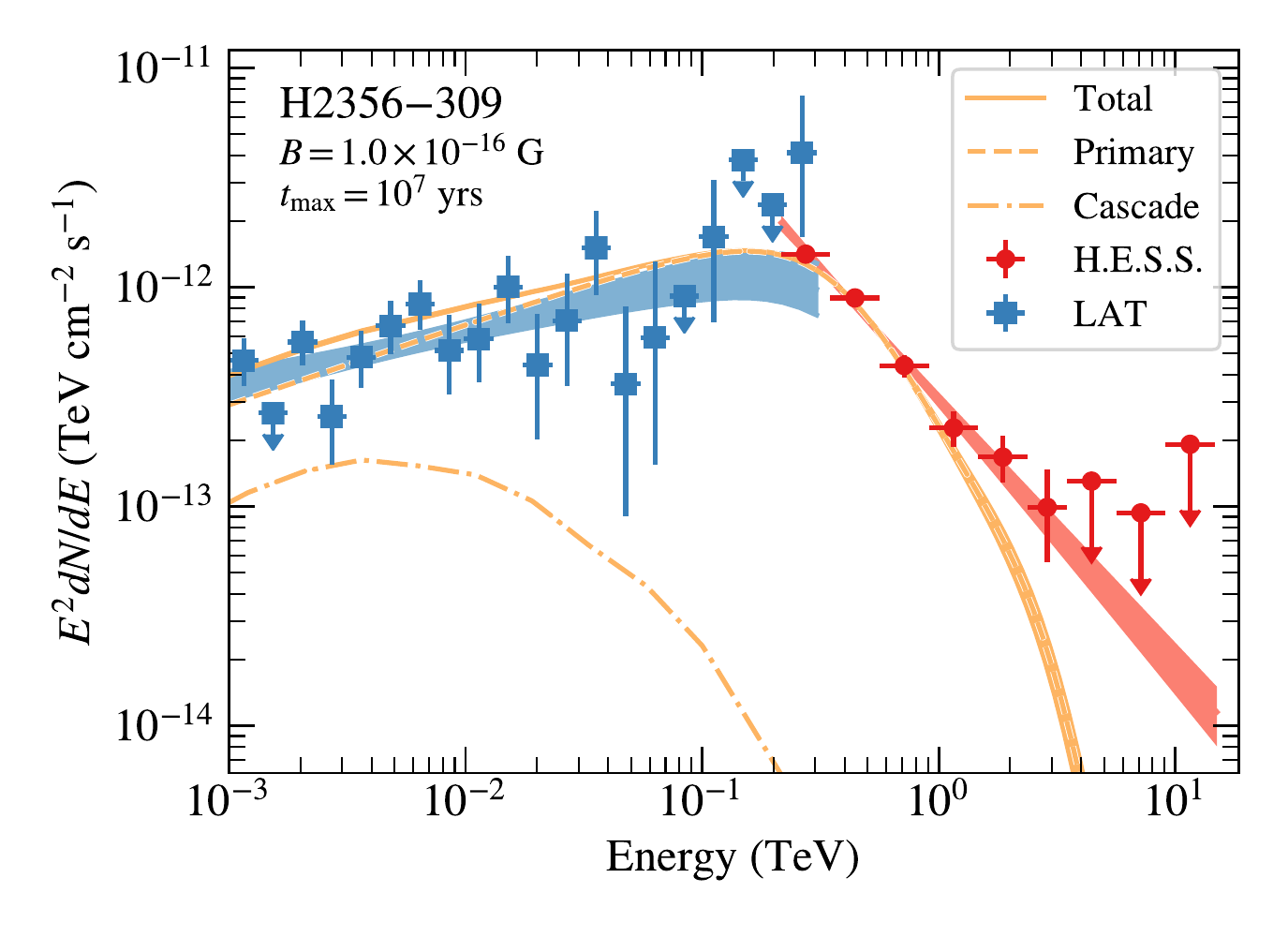}{0.33\textwidth}{}
          \fig{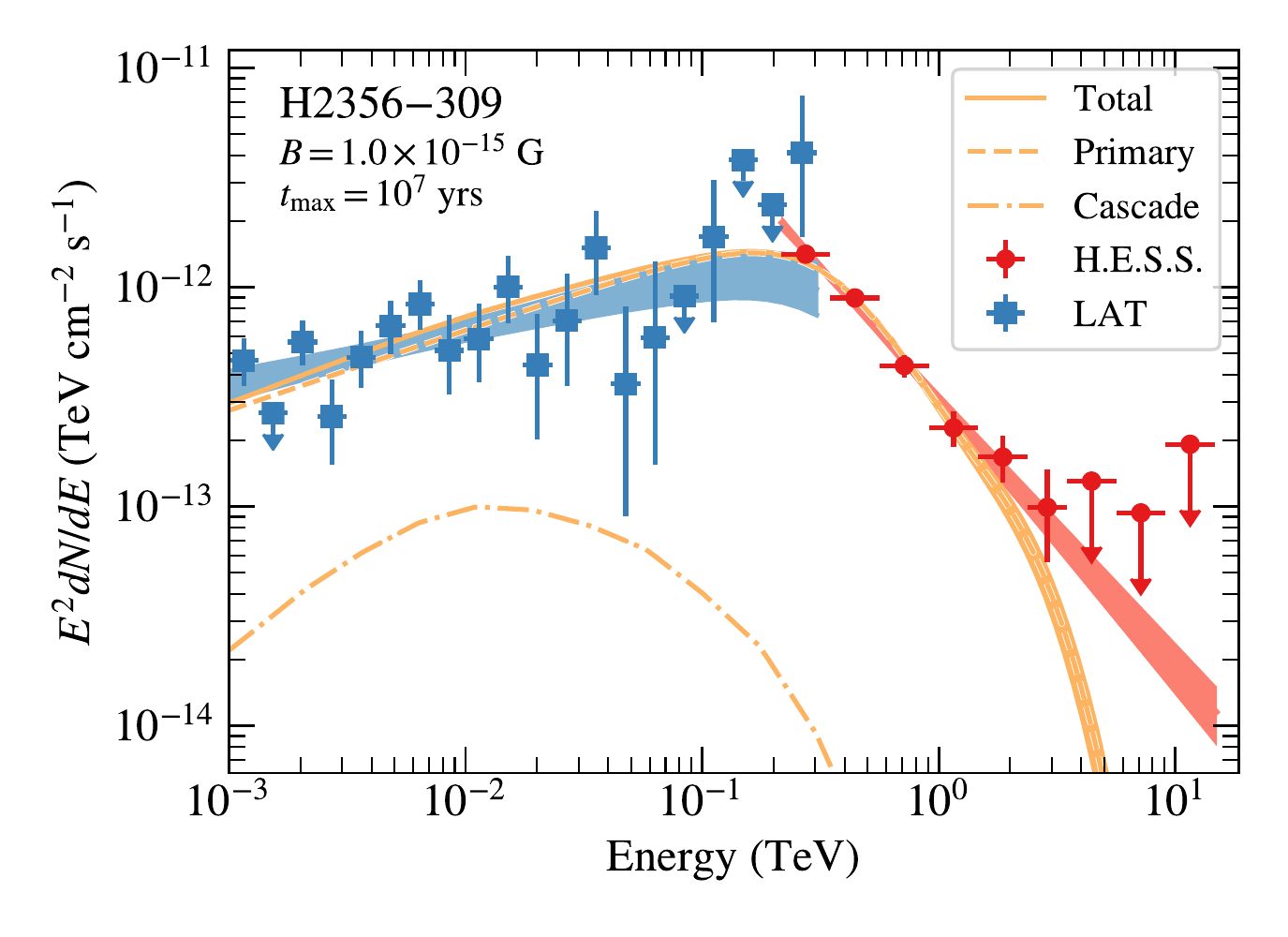}{0.33\textwidth}{}
          \fig{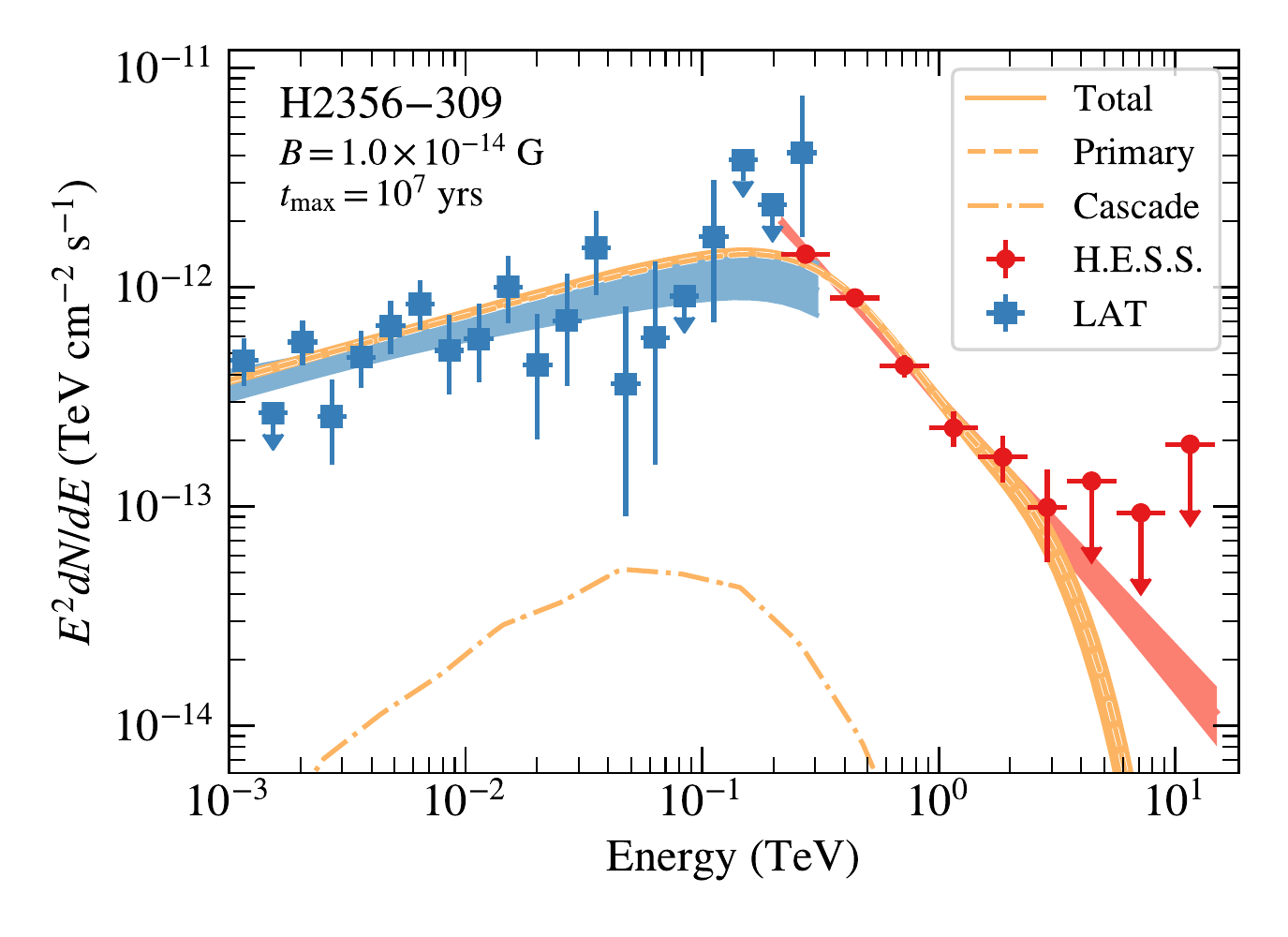}{0.33\textwidth}{}
          }\vspace{-30pt}
    \caption{Best-fit spectra assuming a point-source model in addition to the halo flux templates for a combined LAT and \hess analysis. An activity time of $t_\mathrm{max} = 10^{7}\,$yr is assumed for all sources. 
    The columns show different field strengths $B \in [10^{-16}, 10^{-15}, 10^{-14}]\,$G and each source is shown in one row. A coherence length of 1\,Mpc is assumed throughout, together with the EBL model of \citep{2011MNRAS.410.2556D}. Blue and red markers and lines show the point-source spectra measured with the LAT and \hess, respectively.}
    \label{fig:spectra-tmax1e7}
\end{figure}

The best-fit parameters for the spectrum in Eq.~\eqref{eq:src-spec} for all sources for all tested values of the IGMF and $t_\mathrm{max}$ are reported in Table~\ref{tab:best-fit-par}.
For simplicity, $E_0$ is fixed to 1\,TeV. 
Example spectra for all sources for three different values of the IGMF and $t_\mathrm{max} = 10^7\,$yr are shown in Figure~\ref{fig:spectra-tmax1e7} (same as Figure~\ref{fig:sed} in the main text).
From Figure~\ref{fig:spectra-tmax1e7} and the best-fit values of $\kappa$ in the last column of Table~\ref{tab:best-fit-par}, it becomes clear that exponential cutoffs at lower energies are preferred for low magnetic fields. This again shows the absence of a halo, as a lower-energy cutoff leads to fewer cascade photons. 
Figure~\ref{fig:spectra-tmax1e7} also illustrates that an increasing magnetic field leads to a suppression of the cascade at lower energies. 
Under the assumption that the coherence length of the magnetic field is much larger than the IC cooling length, the deflection of the pairs roughly scales as $E_e^{-2} B$, where $E_e$ is the electron energy~\citep{2009PhRvD..80l3012N}.
As $B$ increases, the low-energy pairs in particular will exhibit a large deflection and the upscattered cascade emission will not reach the observer. 
As a consequence, the peak of the cascade emission shifts toward higher energies as can be seen in each row of Figure~\ref{fig:spectra-tmax1e7}.

\startlongtable
\begin{deluxetable}{ccccc}
\tablehead{
\colhead{Source} & \colhead{IGMF Strength} & \colhead{$N_{0,\mathrm{int}}$} & \colhead{\multirow{2}{*}{$\Gamma_\mathrm{int}$}} & \colhead{$E_\mathrm{cut}$} \\
\colhead{Name} & \colhead{($\log_{10}{(B/\mathrm{G}})$)} & \colhead{($\mathrm{10^{-12}\,TeV^{-1}\,cm^{-2}\,s^{-1}}$)} &  & \colhead{($\mathrm{TeV}$)} \\
}
\tablecaption{Best-fit parameters for the combined LAT and \hess data in the presence of a halo.}
\label{tab:best-fit-par}
\startdata
\cutinhead{$t_\mathrm{max} = 10\,$yrs}
1ES0229+200 & $-16.0$ & 2.114 $\pm$ 0.223 & 1.632 $\pm$ 0.004 & 8.103 $\pm$ 3.022 \\
1ES0229+200 & $-15.5$ & 2.160 $\pm$ 0.066 & 1.633 $\pm$ 0.003 & 8.865 $\pm$ 2.414 \\
1ES0229+200 & $-15.0$ & 2.236 $\pm$ 6.016 & 1.639 $\pm$ 0.297 & 8.970 $\pm$ 10.878 \\
1ES0229+200 & $-14.5$ & 2.260 $\pm$ 0.073 & 1.635 $\pm$ 0.007 & 10.031 $\pm$ 3.545 \\
1ES0229+200 & $-14.0$ & 2.261 $\pm$ 0.085 & 1.639 $\pm$ 0.009 & 10.996 $\pm$ 0.500 \\
1ES0229+200 & $-13.5$ & 2.300 $\pm$ 0.105 & 1.638 $\pm$ 0.011 & 11.000 $\pm$ 0.080 \\
1ES0229+200 & $-13.0$ & 2.308 $\pm$ 0.092 & 1.638 $\pm$ 0.010 & 10.973 $\pm$ 1.487 \\
\hline
1ES0347$-$121 & $-16.0$ & 2.945 $\pm$ 0.063 & 1.633 $\pm$ 0.005 & 3.970 $\pm$ 1.244 \\
1ES0347$-$121 & $-15.5$ & 2.902 $\pm$ 0.119 & 1.633 $\pm$ 0.007 & 5.346 $\pm$ 2.173 \\
1ES0347$-$121 & $-15.0$ & 2.907 $\pm$ 0.104 & 1.655 $\pm$ 0.007 & 5.901 $\pm$ 2.799 \\
1ES0347$-$121 & $-14.5$ & 3.020 $\pm$ 0.093 & 1.633 $\pm$ 0.006 & 5.134 $\pm$ 2.421 \\
1ES0347$-$121 & $-14.0$ & 3.137 $\pm$ 0.087 & 1.631 $\pm$ 0.005 & 4.019 $\pm$ 1.721 \\
1ES0347$-$121 & $-13.5$ & 3.244 $\pm$ 0.066 & 1.633 $\pm$ 0.004 & 3.610 $\pm$ 1.410 \\
1ES0347$-$121 & $-13.0$ & 2.797 $\pm$ 0.148 & 1.659 $\pm$ 0.012 & 11.000 $\pm$ 0.014 \\
\hline
1ES1101$-$232 & $-16.0$ & 3.360 $\pm$ 0.111 & 1.632 $\pm$ 0.003 & 5.948 $\pm$ 1.609 \\
1ES1101$-$232 & $-15.5$ & 3.109 $\pm$ 0.025 & 1.627 $\pm$ 0.002 & 10.906 $\pm$ 1.529 \\
1ES1101$-$232 & $-15.0$ & 3.438 $\pm$ 0.056 & 1.633 $\pm$ 0.002 & 9.652 $\pm$ 3.680 \\
1ES1101$-$232 & $-14.5$ & 3.625 $\pm$ 0.174 & 1.633 $\pm$ 0.008 & 9.517 $\pm$ 3.018 \\
1ES1101$-$232 & $-14.0$ & 3.607 $\pm$ 0.184 & 1.627 $\pm$ 0.012 & 10.998 $\pm$ 0.315 \\
1ES1101$-$232 & $-13.5$ & 3.706 $\pm$ 3.540 & 1.633 $\pm$ 0.010 & 10.777 $\pm$ 88.543 \\
1ES1101$-$232 & $-13.0$ & 3.738 $\pm$ 0.540 & 1.632 $\pm$ 0.012 & 10.198 $\pm$ 14.941 \\
\hline
H2356$-$309 & $-16.0$ & 2.754 $\pm$ 0.075 & 1.717 $\pm$ 0.005 & 2.832 $\pm$ 0.573 \\
H2356$-$309 & $-15.5$ & 2.754 $\pm$ 0.690 & 1.726 $\pm$ 0.046 & 2.924 $\pm$ 0.895 \\
H2356$-$309 & $-15.0$ & 2.780 $\pm$ 0.591 & 1.727 $\pm$ 0.038 & 2.907 $\pm$ 1.988 \\
H2356$-$309 & $-14.5$ & 2.890 $\pm$ 0.024 & 1.710 $\pm$ 0.002 & 2.637 $\pm$ 0.503 \\
H2356$-$309 & $-14.0$ & 2.997 $\pm$ 0.049 & 1.707 $\pm$ 0.005 & 2.383 $\pm$ 0.437 \\
H2356$-$309 & $-13.5$ & 2.997 $\pm$ 0.046 & 1.697 $\pm$ 0.003 & 2.383 $\pm$ 0.436 \\
H2356$-$309 & $-13.0$ & 3.001 $\pm$ 0.058 & 1.696 $\pm$ 0.004 & 2.401 $\pm$ 0.450 \\
\hline
PKS0548$-$322 & $-16.0$ & 7.520 $\pm$ 0.071 & 1.774 $\pm$ 0.008 & 7.621 $\pm$ 0.077 \\
PKS0548$-$322 & $-15.5$ & 7.158 $\pm$ 0.161 & 1.803 $\pm$ 0.004 & 10.203 $\pm$ 4.209 \\
PKS0548$-$322 & $-15.0$ & 7.449 $\pm$ 0.310 & 1.794 $\pm$ 0.008 & 9.973 $\pm$ 5.004 \\
PKS0548$-$322 & $-14.5$ & 7.449 $\pm$ 0.294 & 1.789 $\pm$ 0.007 & 10.906 $\pm$ 3.623 \\
PKS0548$-$322 & $-14.0$ & 7.853 $\pm$ 0.313 & 1.780 $\pm$ 0.008 & 9.150 $\pm$ 5.184 \\
PKS0548$-$322 & $-13.5$ & 7.853 $\pm$ 0.306 & 1.780 $\pm$ 0.008 & 9.150 $\pm$ 5.177 \\
PKS0548$-$322 & $-13.0$ & 7.853 $\pm$ 0.307 & 1.780 $\pm$ 0.009 & 9.150 $\pm$ 5.211 \\
\cutinhead{$t_\mathrm{max} = 10^4\,$yrs}
1ES0229+200 & $-16.0$ & 2.257 $\pm$ 0.061 & 1.633 $\pm$ 0.001 & 3.171 $\pm$ 0.456 \\
1ES0229+200 & $-15.5$ & 2.274 $\pm$ 0.038 & 1.632 $\pm$ 0.002 & 4.106 $\pm$ 0.690 \\
1ES0229+200 & $-15.0$ & 2.142 $\pm$ 0.046 & 1.633 $\pm$ 0.001 & 6.321 $\pm$ 1.301 \\
1ES0229+200 & $-14.5$ & 2.146 $\pm$ 0.067 & 1.632 $\pm$ 0.004 & 8.070 $\pm$ 2.014 \\
1ES0229+200 & $-14.0$ & 2.206 $\pm$ 0.083 & 1.633 $\pm$ 0.005 & 8.673 $\pm$ 2.458 \\
1ES0229+200 & $-13.5$ & 2.217 $\pm$ 0.344 & 1.646 $\pm$ 0.013 & 9.921 $\pm$ 9.287 \\
1ES0229+200 & $-13.0$ & 2.227 $\pm$ 0.106 & 1.644 $\pm$ 0.011 & 10.979 $\pm$ 1.248 \\
\hline
1ES0347$-$121 & $-16.0$ & 4.398 $\pm$ 0.052 & 1.520 $\pm$ 0.003 & 1.000 $\pm$ 0.000 \\
1ES0347$-$121 & $-15.5$ & 3.618 $\pm$ 0.041 & 1.626 $\pm$ 0.002 & 1.269 $\pm$ 0.210 \\
1ES0347$-$121 & $-15.0$ & 2.997 $\pm$ 0.065 & 1.633 $\pm$ 0.002 & 2.976 $\pm$ 0.319 \\
1ES0347$-$121 & $-14.5$ & 2.948 $\pm$ 0.009 & 1.637 $\pm$ 0.005 & 4.149 $\pm$ 0.013 \\
1ES0347$-$121 & $-14.0$ & 2.965 $\pm$ 0.056 & 1.636 $\pm$ 0.004 & 5.046 $\pm$ 1.309 \\
1ES0347$-$121 & $-13.5$ & 2.964 $\pm$ 0.042 & 1.644 $\pm$ 0.003 & 5.222 $\pm$ 1.402 \\
1ES0347$-$121 & $-13.0$ & 2.960 $\pm$ 0.085 & 1.656 $\pm$ 0.006 & 5.022 $\pm$ 2.266 \\
\hline
1ES1101$-$232 & $-16.0$ & 3.534 $\pm$ 0.087 & 1.633 $\pm$ 0.002 & 1.747 $\pm$ 0.258 \\
1ES1101$-$232 & $-15.5$ & 3.366 $\pm$ 0.022 & 1.633 $\pm$ 0.001 & 2.897 $\pm$ 0.519 \\
1ES1101$-$232 & $-15.0$ & 3.364 $\pm$ 0.059 & 1.632 $\pm$ 0.003 & 4.254 $\pm$ 0.951 \\
1ES1101$-$232 & $-14.5$ & 3.378 $\pm$ 0.069 & 1.631 $\pm$ 0.004 & 6.024 $\pm$ 1.563 \\
1ES1101$-$232 & $-14.0$ & 3.197 $\pm$ 0.024 & 1.625 $\pm$ 0.001 & 10.368 $\pm$ 2.810 \\
1ES1101$-$232 & $-13.5$ & 3.365 $\pm$ 0.093 & 1.632 $\pm$ 0.005 & 10.964 $\pm$ 1.380 \\
1ES1101$-$232 & $-13.0$ & 3.538 $\pm$ 0.084 & 1.631 $\pm$ 0.005 & 10.991 $\pm$ 0.715 \\
\hline
H2356$-$309 & $-16.0$ & 3.555 $\pm$ 3.139 & 1.633 $\pm$ 0.002 & 1.255 $\pm$ 2.016 \\
H2356$-$309 & $-15.5$ & 3.696 $\pm$ 0.036 & 1.633 $\pm$ 0.002 & 1.328 $\pm$ 0.148 \\
H2356$-$309 & $-15.0$ & 3.494 $\pm$ 0.043 & 1.634 $\pm$ 0.002 & 1.640 $\pm$ 0.214 \\
H2356$-$309 & $-14.5$ & 2.789 $\pm$ 0.004 & 1.725 $\pm$ 0.006 & 2.704 $\pm$ 0.004 \\
H2356$-$309 & $-14.0$ & 2.795 $\pm$ 7.856 & 1.726 $\pm$ 0.159 & 2.770 $\pm$ 3.188 \\
H2356$-$309 & $-13.5$ & 2.841 $\pm$ 0.034 & 1.723 $\pm$ 0.002 & 2.703 $\pm$ 0.524 \\
H2356$-$309 & $-13.0$ & 2.900 $\pm$ 0.037 & 1.703 $\pm$ 0.003 & 2.638 $\pm$ 0.510 \\
\hline
PKS0548$-$322 & $-16.0$ & 7.793 $\pm$ 0.080 & 1.745 $\pm$ 0.001 & 5.245 $\pm$ 1.428 \\
PKS0548$-$322 & $-15.5$ & 7.877 $\pm$ 0.129 & 1.745 $\pm$ 0.002 & 5.870 $\pm$ 1.744 \\
PKS0548$-$322 & $-15.0$ & 7.680 $\pm$ 0.250 & 1.768 $\pm$ 0.008 & 6.844 $\pm$ 2.431 \\
PKS0548$-$322 & $-14.5$ & 7.610 $\pm$ 0.237 & 1.768 $\pm$ 0.006 & 8.011 $\pm$ 3.147 \\
PKS0548$-$322 & $-14.0$ & 7.539 $\pm$ 0.284 & 1.783 $\pm$ 0.007 & 9.001 $\pm$ 4.146 \\
PKS0548$-$322 & $-13.5$ & 7.543 $\pm$ 0.301 & 1.782 $\pm$ 0.009 & 10.057 $\pm$ 5.377 \\
PKS0548$-$322 & $-13.0$ & 7.554 $\pm$ 0.407 & 1.786 $\pm$ 0.011 & 10.222 $\pm$ 5.494 \\
\cutinhead{$t_\mathrm{max} = 10^7\,$yrs}
1ES0229+200 & $-16.0$ & 2.217 $\pm$ 0.042 & 1.633 $\pm$ 0.001 & 2.880 $\pm$ 0.408 \\
1ES0229+200 & $-15.5$ & 2.166 $\pm$ 0.030 & 1.633 $\pm$ 0.001 & 3.667 $\pm$ 0.552 \\
1ES0229+200 & $-15.0$ & 2.144 $\pm$ 0.044 & 1.633 $\pm$ 0.001 & 5.209 $\pm$ 0.959 \\
1ES0229+200 & $-14.5$ & 2.129 $\pm$ 0.165 & 1.633 $\pm$ 0.003 & 7.099 $\pm$ 2.412 \\
1ES0229+200 & $-14.0$ & 2.201 $\pm$ 0.104 & 1.632 $\pm$ 0.005 & 7.814 $\pm$ 2.125 \\
1ES0229+200 & $-13.5$ & 2.225 $\pm$ 0.124 & 1.635 $\pm$ 0.011 & 9.109 $\pm$ 3.505 \\
1ES0229+200 & $-13.0$ & 2.225 $\pm$ 0.080 & 1.635 $\pm$ 0.008 & 10.584 $\pm$ 3.262 \\
\hline
1ES0347$-$121 & $-16.0$ & 4.133 $\pm$ 0.418 & 1.520 $\pm$ 0.001 & 1.000 $\pm$ 0.000 \\
1ES0347$-$121 & $-15.5$ & 4.462 $\pm$ 0.060 & 1.520 $\pm$ 0.002 & 1.000 $\pm$ 0.005 \\
1ES0347$-$121 & $-15.0$ & 3.257 $\pm$ 0.055 & 1.632 $\pm$ 0.003 & 1.638 $\pm$ 0.324 \\
1ES0347$-$121 & $-14.5$ & 2.960 $\pm$ 0.317 & 1.632 $\pm$ 0.004 & 3.020 $\pm$ 0.482 \\
1ES0347$-$121 & $-14.0$ & 2.824 $\pm$ 0.117 & 1.632 $\pm$ 0.002 & 5.152 $\pm$ 2.231 \\
1ES0347$-$121 & $-13.5$ & 2.655 $\pm$ 1.179 & 1.672 $\pm$ 0.088 & 8.368 $\pm$ 6.833 \\
1ES0347$-$121 & $-13.0$ & 2.656 $\pm$ 0.241 & 1.672 $\pm$ 0.023 & 10.998 $\pm$ 0.100 \\
\hline
1ES1101$-$232 & $-16.0$ & 5.335 $\pm$ 0.435 & 1.520 $\pm$ 0.002 & 1.001 $\pm$ 0.012 \\
1ES1101$-$232 & $-15.5$ & 5.183 $\pm$ 0.081 & 1.520 $\pm$ 0.002 & 1.167 $\pm$ 0.139 \\
1ES1101$-$232 & $-15.0$ & 4.893 $\pm$ 0.042 & 1.520 $\pm$ 0.001 & 1.766 $\pm$ 0.262 \\
1ES1101$-$232 & $-14.5$ & 4.651 $\pm$ 0.149 & 1.553 $\pm$ 0.010 & 2.357 $\pm$ 0.224 \\
1ES1101$-$232 & $-14.0$ & 3.810 $\pm$ 0.044 & 1.599 $\pm$ 0.002 & 4.753 $\pm$ 1.205 \\
1ES1101$-$232 & $-13.5$ & 3.843 $\pm$ 0.080 & 1.600 $\pm$ 0.004 & 5.675 $\pm$ 1.756 \\
1ES1101$-$232 & $-13.0$ & 3.858 $\pm$ 0.070 & 1.604 $\pm$ 0.004 & 6.623 $\pm$ 2.599 \\
\hline
H2356$-$309 & $-16.0$ & 3.709 $\pm$ 0.025 & 1.632 $\pm$ 0.001 & 1.085 $\pm$ 0.100 \\
H2356$-$309 & $-15.5$ & 3.411 $\pm$ 0.096 & 1.632 $\pm$ 0.003 & 1.421 $\pm$ 0.176 \\
H2356$-$309 & $-15.0$ & 3.486 $\pm$ 0.042 & 1.632 $\pm$ 0.002 & 1.527 $\pm$ 0.186 \\
H2356$-$309 & $-14.5$ & 3.331 $\pm$ 0.046 & 1.663 $\pm$ 0.003 & 1.739 $\pm$ 0.239 \\
H2356$-$309 & $-14.0$ & 2.875 $\pm$ 0.107 & 1.706 $\pm$ 0.007 & 2.571 $\pm$ 0.526 \\
H2356$-$309 & $-13.5$ & 2.877 $\pm$ 0.042 & 1.703 $\pm$ 0.003 & 2.642 $\pm$ 0.501 \\
H2356$-$309 & $-13.0$ & 2.878 $\pm$ 0.040 & 1.709 $\pm$ 0.003 & 2.646 $\pm$ 0.510 \\
\hline
PKS0548$-$322 & $-16.0$ & 7.790 $\pm$ 0.097 & 1.745 $\pm$ 0.001 & 5.039 $\pm$ 1.201 \\
PKS0548$-$322 & $-15.5$ & 7.781 $\pm$ 0.084 & 1.745 $\pm$ 0.001 & 5.821 $\pm$ 1.684 \\
PKS0548$-$322 & $-15.0$ & 8.004 $\pm$ 0.249 & 1.745 $\pm$ 0.004 & 5.895 $\pm$ 1.845 \\
PKS0548$-$322 & $-14.5$ & 7.836 $\pm$ 0.969 & 1.765 $\pm$ 0.023 & 7.142 $\pm$ 3.571 \\
PKS0548$-$322 & $-14.0$ & 7.483 $\pm$ 0.336 & 1.769 $\pm$ 0.009 & 9.153 $\pm$ 4.287 \\
PKS0548$-$322 & $-13.5$ & 7.482 $\pm$ 0.297 & 1.783 $\pm$ 0.008 & 10.372 $\pm$ 5.814 \\
PKS0548$-$322 & $-13.0$ & 7.489 $\pm$ 0.412 & 1.792 $\pm$ 0.011 & 10.988 $\pm$ 2.337 \\
\enddata
\end{deluxetable}

\subsection{Likelihood profiles}
\label{app:likelihoods}

\begin{figure}
    \centering
    \gridline{
          \fig{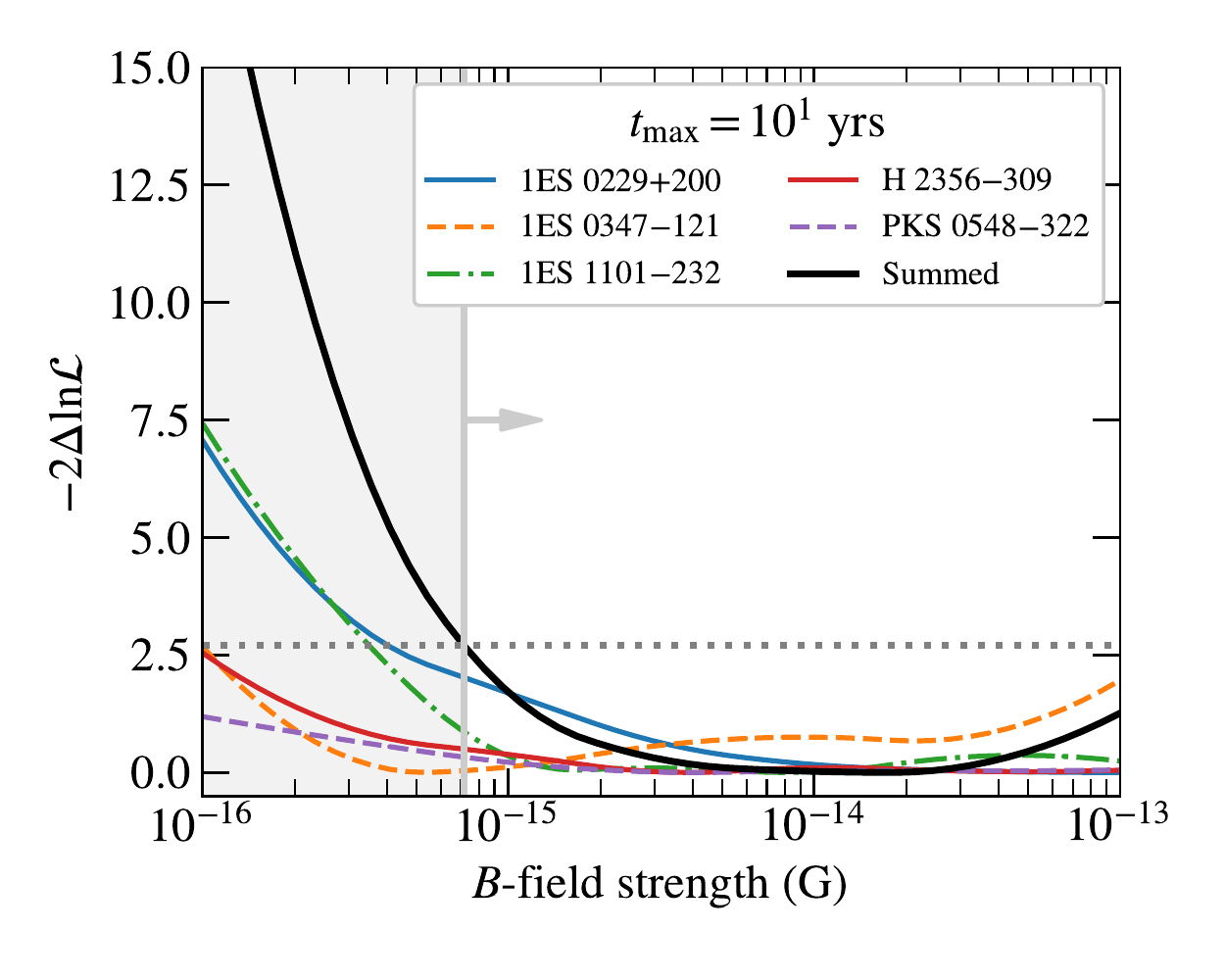}{0.4\textwidth}{(a)}
          \fig{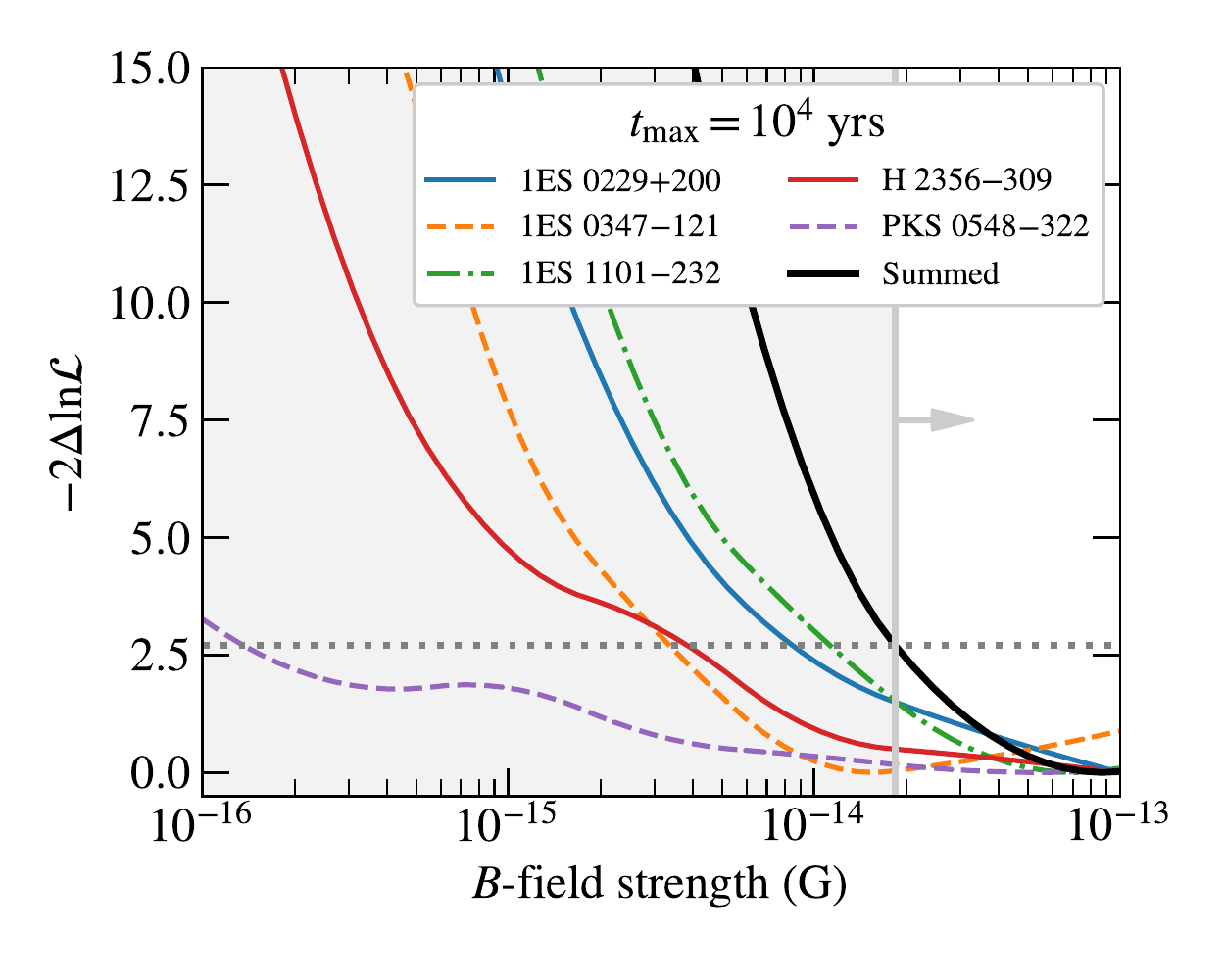}{0.4\textwidth}{(b)}
          }
    \gridline{
          \fig{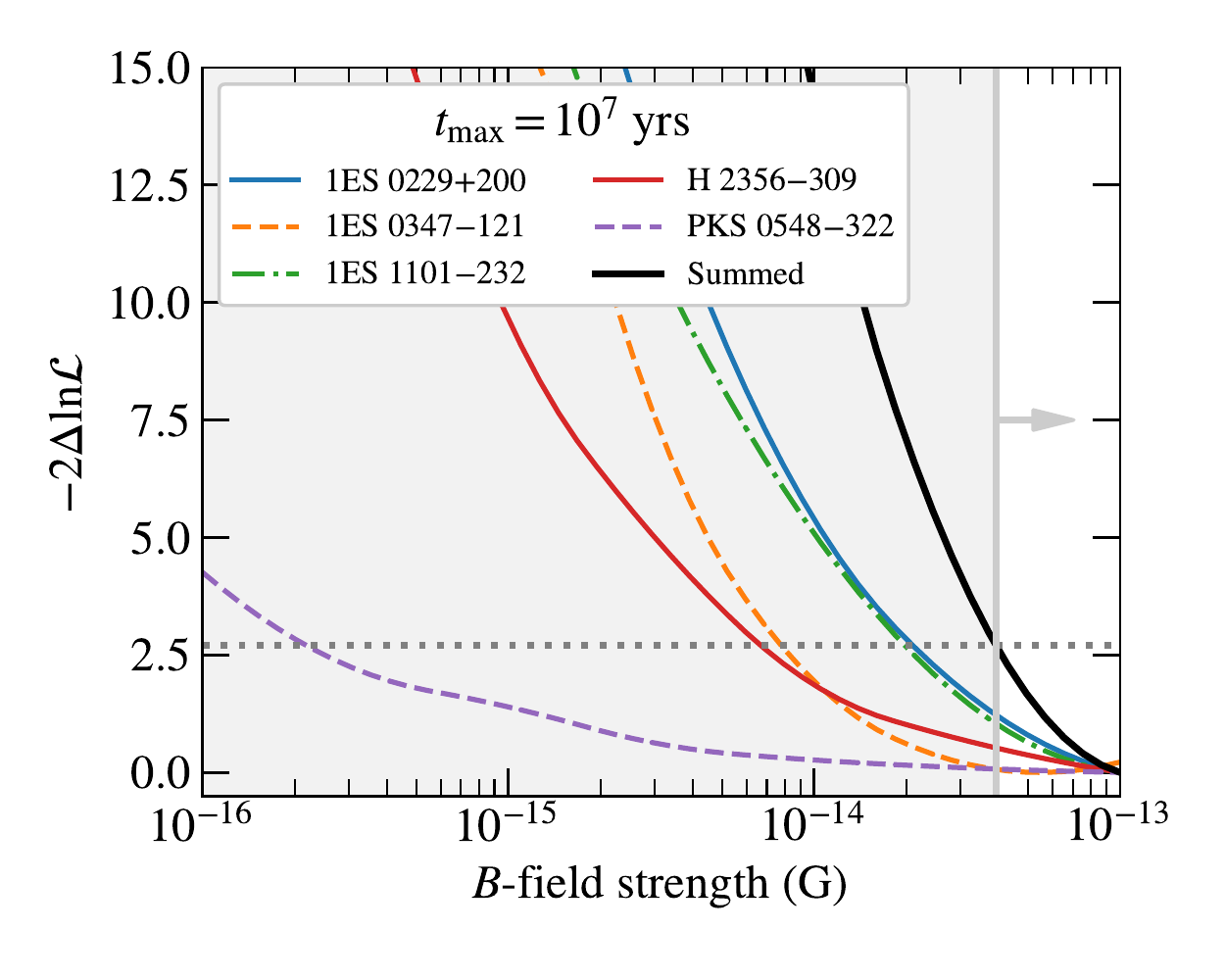}{0.4\textwidth}{(c)}
          }
    \caption{Likelihood profiles as function of the IGMF strength for (a) $t_\mathrm{max}=10\,\mathrm{yrs}$, (b) $t_\mathrm{max}=10^4\,\mathrm{yrs}$, and (c) $t_\mathrm{max}=10^7\,\mathrm{yrs}$. 
    Magnetic field values resulting in differences in the likelihood $-2\Delta\ln\mathcal{L} > 2.71$ (indicated by the gray dotted line) are ruled out at the 95\,\% confidence level, which is shown by the gray shaded region and arrow.
    The likelihood curves are smoothed using a quadratic spline interpolation.
    } 
    \label{fig:likelihoods}
\end{figure}

In Figure~\ref{fig:likelihoods}, we show the likelihood ratios $\lambda(B)$ combined for all sources as defined in Eq.~\eqref{eq:likelihood} for our choice of different blazar activity times (black solid lines). 
In addition, we show the likelihood ratios $-2\Delta\ln\mathcal{L}_i$ for the individual sources as colored lines. 
It is evident that shorter activity times lead to weaker constraints on the IGMF, as the likelihood curves are consistently shifted toward lower values of $B$.

\subsection{Additional systematic uncertainties}
\label{app:sys}
In the following, the effects of systematic uncertainties in addition to the unknown blazar activity time are discussed. 

Increasing the jet opening angle $\theta_\mathrm{jet}$ has a similar effect to increasing $t_\mathrm{max}$. 
Larger jet opening angles will lead to larger halos and larger cascade fluxes as fewer gamma rays are rejected in our parallel transport prescription (see Section~\ref{app:crpropa-rotations}).
This makes sense, as photons emitted along the edge of the jet could be deflected toward the observer.
However, given the observed spread of $\theta_\mathrm{jet}$~\citep[e.g.,][]{2017ApJ...846...98J}, 
the uncertainty is subdominant compared to the uncertain activity time. 
Similarly, the choice of the EBL model introduces a negligible uncertainty, as it typically affects the number of generated pairs of the order of $\sim 5\,\%$ in the energy and redshift range under consideration \citep{2016ApJ...827..147M}.

Another potentially important systematic uncertainty is the difference in the absolute energy scales of \hess and the LAT. 
For the LAT, the error on the energy scale is of the order of a few percent \citep{2012APh....35..346A},
whereas the reconstructed energies of \Grays observed by IACTs have an uncertainty of typically 10\,\% \citep{2000APh....12..207H}. We can allow for a relative shift in energy between the two instruments by adding a scaling parameter $s$ in our fitting procedure, such that the model fluxes fitted to \hess data are changed to  $\phi(E) \to \phi(E(1+s))$. We constrain $s$ to lie within $\pm0.2$. Repeating the fits for all sources, we find that for low magnetic fields $s$ tends to its lowest allowed value. 
This minimizes the cascade emission, as the \hess data are effectively shifted to lower fluxes. 
With this systematic term included, the limits on the magnetic field are weakened and $B < 5.3\times10^{-16}\,$G is ruled out for $t_\mathrm{max}=10\,$yrs.


\bibliography{main}{}
\bibliographystyle{aasjournal}



\end{document}

%% file: authors_list.tex
\author{F.~Aharonian}
\affiliation{Dublin Institute for Advanced Studies, 31 Fitzwilliam Place, Dublin 2, Ireland}
\affiliation{Max-Planck-Institut f\"ur Kernphysik, P.O. Box 103980, D 69029 Heidelberg, Germany}

\author{J.~Aschersleben}
\affiliation{Kapteyn Astronomical Institute, University of Groningen, Landleven 12, 9747 AD Groningen, The Netherlands}

\author[0000-0002-9326-6400]{M.~Backes}
\affiliation{University of Namibia, Department of Physics, Private Bag 13301, Windhoek 10005, Namibia}
\affiliation{Centre for Space Research, North-West University, Potchefstroom 2520, South Africa}

\author[0000-0002-5085-8828]{V.~Barbosa~Martins}
\affiliation{DESY, D-15738 Zeuthen, Germany}

\author[0000-0002-5797-3386]{R.~Batzofin}
\affiliation{Institut f\"ur Physik und Astronomie, Universit\"at Potsdam,  Karl-Liebknecht-Strasse 24/25, D 14476 Potsdam, Germany}

\author[0000-0002-2115-2930]{Y.~Becherini}
\affiliation{Université de Paris, CNRS, Astroparticule et Cosmologie, F-75013 Paris, France}
\affiliation{Department of Physics and Electrical Engineering, Linnaeus University,  351 95 V\"axj\"o, Sweden}

\author[0000-0002-2918-1824]{D.~Berge}
\affiliation{DESY, D-15738 Zeuthen, Germany}
\affiliation{Institut f\"ur Physik, Humboldt-Universit\"at zu Berlin, Newtonstr. 15, D 12489 Berlin, Germany}

\author{B.~Bi}
\affiliation{Institut f\"ur Astronomie und Astrophysik, Universit\"at T\"ubingen, Sand 1, D 72076 T\"ubingen, Germany}

\author{M.~Bouyahiaoui}
\affiliation{Max-Planck-Institut f\"ur Kernphysik, P.O. Box 103980, D 69029 Heidelberg, Germany}

\author[0000-0003-0268-5122]{M.~Breuhaus}
\affiliation{Max-Planck-Institut f\"ur Kernphysik, P.O. Box 103980, D 69029 Heidelberg, Germany}

% ...

\collaboration{300}{H.E.S.S. Collaboration}

\author[0000-0002-0738-7581]{M.~Meyer}
\altaffiliation{Now at CP3-Origins, University of Southern Denmark, Campusvej 55, 5230 Odense M, Denmark}
\affiliation{Universit\"at Hamburg, Institut f\"ur Experimentalphysik, Luruper Chaussee 149, D 22761 Hamburg, Germany}

\collaboration{300}{\fermiLAT Collaboration}

%% file: main.bbl
\begin{thebibliography}{}
\expandafter\ifx\csname natexlab\endcsname\relax\def\natexlab#1{#1}\fi
\providecommand{\url}[1]{\href{#1}{#1}}
\providecommand{\dodoi}[1]{doi:~\href{http://doi.org/#1}{\nolinkurl{#1}}}
\providecommand{\doeprint}[1]{\href{http://ascl.net/#1}{\nolinkurl{http://ascl.net/#1}}}
\providecommand{\doarXiv}[1]{\href{https://arxiv.org/abs/#1}{\nolinkurl{https://arxiv.org/abs/#1}}}

\bibitem[{{Abdalla} {et~al.}(2021){Abdalla}, {Abe}, {Acero}, {Acharyya},
  {Adam}, {Agudo}, {Aguirre-Santaella}, {Alfaro}, {Alfaro}, {Alispach},
  {Aloisio}, {Alves Batista}, {Amati}, {Amato}, {Ambrosi}, {Ang{\"u}ner},
  {Araudo}, {Armstrong}, {Arqueros}, {Arrabito}, {Asano}, {Ascas{\'\i}bar},
  {Ashley}, {Backes}, {Balazs}, {Balbo}, {Balmaverde}, {Baquero Larriva},
  {Barbosa Martins}, {Barkov}, {Baroncelli}, {Barres de Almeida}, {Barrio},
  {Batista}, {Becerra Gonz{\'a}lez}, {Becherini}, {Beck}, {Becker Tjus},
  {Belmont}, {Benbow}, {Bernardini}, {Berti}, {Berton}, {Bertucci}, {Beshley},
  {Bi}, {Biasuzzi}, {Biland}, {Bissaldi}, {Biteau}, {Blanch}, {Bocchino},
  {Boisson}, {Bolmont}, {Bonanno}, {Bonneau Arbeletche}, {Bonnoli}, {Bordas},
  {Bottacini}, {B{\"o}ttcher}, {Bozhilov}, {Bregeon}, {Brill}, {Brown},
  {Bruno}, {Bruno}, {Bulgarelli}, {Burton}, {Buscemi}, {Caccianiga}, {Cameron},
  {Capasso}, {Caprai}, {Caproni}, {Capuzzo-Dolcetta}, {Caraveo}, {Carosi},
  {Carosi}, {Casanova}, {Cascone}, {Cauz}, {Cerny}, {Cerruti}, {Chadwick},
  {Chaty}, {Chen}, {Chernyakova}, {Chiaro}, {Chiavassa}, {Chytka}, {Conforti},
  {Conte}, {Contreras}, {Coronado-Blazquez}, {Cortina}, {Costa}, {Costantini},
  {Covino}, {Cristofari}, {Cuevas}, {D'Ammando}, {Daniel}, {Davies}, {Dazzi},
  {De Angelis}, {de Bony de Lavergne}, {De Caprio}, {de C{\'a}ssia dos Anjos},
  {de Gouveia Dal Pino}, {De Lotto}, {De Martino}, {de Naurois}, {de O{\~n}a
  Wilhelmi}, {De Palma}, {de Souza}, {Delgado}, {Della Ceca}, {della Volpe},
  {Depaoli}, {Di Girolamo}, {Di Pierro}, {D{\'\i}az}, {D{\'\i}az-Bahamondes},
  {Diebold}, {Djannati-Ata{\"\i}}, {Dmytriiev}, {Dom{\'\i}nguez}, {Donini},
  {Dorner}, {Doro}, {Dournaux}, {Dwarkadas}, {Ebr}, {Eckner}, {Einecke},
  {Ekoume}, {Els{\"a}sser}, {Emery}, {Evoli}, {Fairbairn}, {Falceta-Goncalves},
  {Fegan}, {Feng}, {Ferrand}, {Fiandrini}, {Fiasson}, {Fioretti}, {Foffano},
  {Fonseca}, {Font}, {Fontaine}, {Franco}, {Freixas Coromina}, {Fukami},
  {Fukazawa}, {Fukui}, {Gaggero}, {Galanti}, {Gammaldi}, {Garcia},
  {Garczarczyk}, {Gascon}, {Gaug}, {Gent}, {Ghalumyan}, {Ghirlanda},
  {Gianotti}, {Giarrusso}, {Giavitto}, {Giglietto}, {Giordano}, {Glicenstein},
  {Goldoni}, {Gonz{\'a}lez}, {Gourgouliatos}, {Grabarczyk}, {Grandi}, {Granot},
  {Grasso}, {Green}, {Grube}, {Gueta}, {Gunji}, {Halim}, {Harvey}, {Hassan
  Collado}, {Hayashi}, {Heller}, {Hern{\'a}ndez Cadena}, {Hervet}, {Hinton},
  {Hiroshima}, {Hnatyk}, {Hnatyk}, {Hoffmann}, {Hofmann}, {Holder}, {Horan},
  {H{\"o}randel}, {Horvath}, {Hovatta}, {Hrabovsky}, {Hrupec}, {Hughes},
  {H{\"u}tten}, {Iarlori}, {Inada}, {Inoue}, {Insolia}, {Ionica}, {Iori},
  {Jacquemont}, {Jamrozy}, {Janecek}, {Jim{\'e}nez Mart{\'\i}nez}, {Jin},
  {Jung-Richardt}, {Jurysek}, {Kaaret}, {Karas}, {Karkar}, {Kawanaka},
  {Kerszberg}, {Kh{\'e}lifi}, {Kissmann}, {Kn{\"o}dlseder}, {Kobayashi},
  {Kohri}, {Komin}, {Kong}, {Kosack}, {Kubo}, {La Palombara}, {Lamanna},
  {Lang}, {Lapington}, {Laporte}, {Lefaucheur}, {Lemoine-Goumard}, {Lenain},
  {Leone}, {Leto}, {Leuschner}, {Lindfors}, {Lloyd}, {Lohse}, {Lombardi},
  {Longo}, {Lopez}, {L{\'o}pez}, {L{\'o}pez-Coto}, {Loporchio}, {Lucarelli},
  {Luque-Escamilla}, {Lyard}, {Maggio}, {Majczyna}, {Makariev}, {Mallamaci},
  {Mandat}, {Maneva}, {Manganaro}, {Manic{\`o}}, {Marcowith}, {Marculewicz},
  {Markoff}, {Marquez}, {Mart{\'\i}}, {Martinez}, {Mart{\'\i}nez},
  {Mart{\'\i}nez}, {Mart{\'\i}nez-Huerta}, {Maurin}, {Mazin}, {Mbarubucyeye},
  {Medina Miranda}, {Meyer}, {Micanovic}, {Miener}, {Minev}, {Miranda},
  {Mitchell}, {Mizuno}, {Mode}, {Moderski}, {Mohrmann}, {Molina}, {Montaruli},
  {Moralejo}, {Morales Merino}, {Morcuende-Parrilla}, {Morselli}, {Mukherjee},
  {Mundell}, {Murach}, {Muraishi}, {Nagai}, {Nakamori}, {Nemmen}, {Niemiec},
  {Nieto}, {Nievas}, {Nikolajuk}, {Nishijima}, {Noda}, {Nosek}, {Nozaki},
  {O'Brien}, {Ohira}, {Ohishi}, {Oka}, {Ong}, {Orienti}, {Orito}, {Orlandini},
  {Orlando}, {Osborne}, {Ostrowski}, {Oya}, {Pagliaro}, {Palatka}, {Paneque},
  {Pantaleo}, {Paredes}, {Parmiggiani}, {Patricelli}, {Pavleti{\'c}}, {Pe'er},
  {Pech}, {Pecimotika}, {Peresano}, {Persic}, {Petruk}, {Pfrang}, {Piatteli},
  {Pietropaolo}, {Pillera}, {Pilszyk}, {Pimentel}, {Pintore}, {Pita}, {Pohl},
  {Poireau}, {Polo}, {Prado}, {Prast}, {Principe}, {Produit}, {Prokoph},
  {Prouza}, {Przybilski}, {Pueschel}, {P{\"u}hlhofer}, {Pumo}, {Punch},
  {Queiroz}, {Quirrenbach}, {Rando}, {Razzaque}, {Rebert}, {Recchia},
  {Reichherzer}, {Reimer}, {Reimer}, {Renier}, {Reposeur}, {Rhode}, {Ribeiro},
  {Rib{\'o}}, {Richtler}, {Rico}, {Rieger}, {Rizi}, {Rodriguez}, {Rodriguez
  Fernandez}, {Rodriguez Ramirez}, {Rodr{\'\i}guez V{\'a}zquez}, {Romano},
  {Romeo}, {Roncadelli}, {Rosado}, {Rosales de Leon}, {Rowell}, {Rudak},
  {Rujopakarn}, {Russo}, {Sadeh}, {Saha}, {Saito}, {Salesa Greus}, {Sanchez},
  {S{\'a}nchez-Conde}, {Sangiorgi}, {Sano}, {Santander}, {Santos}, {Sanuy},
  {Sarkar}, {Saturni}, {Sawangwit}, {Scherer}, {Schleicher}, {Schovanek},
  {Schussler}, {Schwanke}, {Sciacca}, {Scuderi}, {Seglar Arroyo}, {Sergijenko},
  {Servillat}, {Seweryn}, {Shalchi}, {Sharma}, {Shellard}, {Siejkowski},
  {Sinha}, {Sliusar}, {Slowikowska}, {Sokolenko}, {Sol}, {Specovius},
  {Spencer}, {Spiga}, {Stamerra}, {Stani{\v{c}}}, {Starling}, {Stolarczyk},
  {Straumann}, {Stri{\v{s}}kovi{\'c}}, {Suda}, {{\'S}wierk}, {Tagliaferri},
  {Takahashi}, {Takahashi}, {Tavecchio}, {Taylor}, {Tejedor}, {Temnikov},
  {Terrier}, {Terzic}, {Testa}, {Tian}, {Tibaldo}, {Tonev}, {Torres},
  {Torresi}, {Tosti}, {Tothill}, {Tovmassian}, {Travnicek}, {Truzzi},
  {Tuossenel}, {Umana}, {Vacula}, {Vagelli}, {Valentino}, {Vallage},
  {Vallania}, {van Eldik}, {Varner}, {Vassiliev}, {V{\'a}zquez Acosta},
  {Vecchi}, {Veh}, {Vercellone}, {Vergani}, {Verguilov}, {Vettolani}, {Viana},
  {Vigorito}, {Vitale}, {Vorobiov}, {Vovk}, {Vuillaume}, {Wagner}, {Walter},
  {Watson}, {White}, {White}, {Wiemann}, {Wierzcholska}, {Will}, {Williams},
  {Wischnewski}, {Wolter}, {Yamazaki}, {Yanagita}, {Yang}, {Yoshikoshi},
  {Zacharias}, {Zaharijas}, {Zaric}, {Zavrtanik}, {Zavrtanik}, {Zdziarski},
  {Zech}, {Zechlin}, {Zhdanov}, \& {{\v{Z}}ivec}}]{2021JCAP...02..048A}
{Abdalla}, H., {Abe}, H., {Acero}, F., {et~al.} 2021, \jcap, 2021, 048,
  \dodoi{10.1088/1475-7516/2021/02/048}

\bibitem[{{Acciari} {et~al.}(2023){Acciari}, {Agudo}, {Aniello}, {Ansoldi},
  {Antonelli}, {Arbet Engels}, {Artero}, {Asano}, {Baack}, {Babi{\'c}},
  {Baquero}, {Barres de Almeida}, {Barrio}, {Batkovi{\'c}}, {Becerra
  Gonz{\'a}lez}, {Bednarek}, {Bernardini}, {Bernardos}, {Berti}, {Besenrieder},
  {Bhattacharyya}, {Bigongiari}, {Biland}, {Blanch}, {B{\"o}kenkamp},
  {Bonnoli}, {Bo{\v{s}}njak}, {Burelli}, {Busetto}, {Carosi}, {Ceribella},
  {Cerruti}, {Chai}, {Chilingarian}, {Cikota}, {Colombo}, {Contreras},
  {Cortina}, {Covino}, {D'Amico}, {D'Elia}, {da Vela}, {Dazzi}, {de Angelis},
  {de Lotto}, {Del Popolo}, {Delfino}, {Delgado}, {Delgado Mendez}, {Depaoli},
  {di Pierro}, {di Venere}, {Do Souto Espi{\~n}eira}, {Dominis Prester},
  {Donini}, {Dorner}, {Doro}, {Elsaesser}, {Fallah Ramazani}, {Fari{\~n}a},
  {Fattorini}, {Font}, {Fruck}, {Fukami}, {Fukazawa}, {Garc{\'\i}a L{\'o}pez},
  {Garczarczyk}, {Gasparyan}, {Gaug}, {Giglietto}, {Giordano}, {Gliwny},
  {Godinovi{\'c}}, {Green}, {Green}, {Hadasch}, {Hahn}, {Hassan}, {Heckmann},
  {Herrera}, {Hrupec}, {H{\"u}tten}, {Inada}, {Iotov}, {Ishio}, {Iwamura},
  {Jim{\'e}nez Mart{\'\i}nez}, {Jormanainen}, {Jouvin}, {Kerszberg},
  {Kobayashi}, {Kubo}, {Kushida}, {Lamastra}, {Lelas}, {Leone}, {Lindfors},
  {Linhoff}, {Liodakis}, {Lombardi}, {Longo}, {L{\'o}pez-Coto},
  {L{\'o}pez-Moya}, {L{\'o}pez-Oramas}, {Loporchio}, {Lorini}, {Machado de
  Oliveira Fraga}, {Maggio}, {Majumdar}, {Makariev}, {Mallamaci}, {Maneva},
  {Manganaro}, {Mannheim}, {Mariotti}, {Mart{\'\i}nez}, {Mas Aguilar}, {Mazin},
  {Menchiari}, {Mender}, {Mi{\'c}anovi{\'c}}, {Miceli}, {Miener}, {Miranda},
  {Mirzoyan}, {Molina}, {Mondal}, {Moralejo}, {Morcuende}, {Moreno}, {Moretti},
  {Nakamori}, {Nanci}, {Nava}, {Neustroev}, {Nievas Rosillo}, {Nigro},
  {Nilsson}, {Nishijima}, {Noda}, {Nozaki}, {Ohtani}, {Oka}, {Otero-Santos},
  {Paiano}, {Palatiello}, {Paneque}, {Paoletti}, {Paredes}, {Pavleti{\'c}},
  {Pe{\~n}il}, {Persic}, {Pihet}, {Prada Moroni}, {Prandini}, {Priyadarshi},
  {Puljak}, {Rhode}, {Rib{\'o}}, {Rico}, {Righi}, {Rugliancich}, {Sahakyan},
  {Saito}, {Sakurai}, {Satalecka}, {Saturni}, {Schleicher}, {Schmidt},
  {Schmuckermaier}, {Schubert}, {Schweizer}, {Sitarek}, {{\v{S}}nidari{\'c}},
  {Sobczynska}, {Spolon}, {Stamerra}, {Stri{\v{s}}kovi{\'c}}, {Strom},
  {Strzys}, {Suda}, {Suri{\'c}}, {Takahashi}, {Takeishi}, {Tavecchio},
  {Temnikov}, {Terzi{\'c}}, {Teshima}, {Tosti}, {Truzzi}, {Tutone}, {Ubach},
  {van Scherpenberg}, {Vanzo}, {Vazquez Acosta}, {Ventura}, {Verguilov},
  {Viale}, {Vigorito}, {Vitale}, {Vovk}, {Will}, {Wunderlich}, {Yamamoto},
  {Zari{\'c}}, \& {MAGIC Collaboration}}]{2023A&A...670A.145A}
{Acciari}, V.~A., {Agudo}, I., {Aniello}, T., {et~al.} 2023, \aap, 670, A145,
  \dodoi{10.1051/0004-6361/202244126}

\bibitem[{{Ackermann} {et~al.}(2012){Ackermann}, {Ajello}, {Allafort},
  {Atwood}, {Axelsson}, {Baldini}, {Barbiellini}, {Bastieri}, {Bechtol},
  {Bellazzini}, {Berenji}, {Bloom}, {Bonamente}, {Borgland}, {Bouvier},
  {Bregeon}, {Brez}, {Brigida}, {Bruel}, {Buehler}, {Buson}, {Caliandro},
  {Cameron}, {Caraveo}, {Casandjian}, {Cecchi}, {Charles}, {Chekhtman},
  {Chiang}, {Ciprini}, {Claus}, {Cohen-Tanugi}, {Cutini}, {de Palma}, {Dermer},
  {Digel}, {Do Couto E Silva}, {Drell}, {Drlica-Wagner}, {Dubois}, {Enoto},
  {Falletti}, {Favuzzi}, {Fegan}, {Focke}, {Fortin}, {Fukazawa}, {Funk},
  {Fusco}, {Gargano}, {Gehrels}, {Germani}, {Giglietto}, {Giordano},
  {Giroletti}, {Glanzman}, {Godfrey}, {Grenier}, {Grove}, {Guiriec}, {Hadasch},
  {Hayashida}, {Hays}, {Hughes}, {J{\'o}hannesson}, {Johnson}, {Johnson},
  {Kamae}, {Katagiri}, {Kataoka}, {Kn{\"o}dlseder}, {Kuss}, {Lande},
  {Latronico}, {Lee}, {Longo}, {Loparco}, {Lovellette}, {Lubrano}, {Madejski},
  {Mazziotta}, {McEnery}, {Michelson}, {Mizuno}, {Moiseev}, {Monte}, {Monzani},
  {Morselli}, {Moskalenko}, {Murgia}, {Nakamori}, {Naumann-Godo}, {Nolan},
  {Norris}, {Nuss}, {Ohsugi}, {Okumura}, {Omodei}, {Orlando}, {Ormes}, {Ozaki},
  {Paneque}, {Panetta}, {Parent}, {Pesce-Rollins}, {Pierbattista}, {Piron},
  {Rain{\`o}}, {Rando}, {Razzano}, {Reimer}, {Reimer}, {Reposeur}, {Ritz},
  {Rochester}, {Sgr{\`o}}, {Siskind}, {Smith}, {Spandre}, {Spinelli}, {Suson},
  {Takahashi}, {Tanaka}, {Thayer}, {Thayer}, {Thompson}, {Tibaldo}, {Tosti},
  {Troja}, {Usher}, {Vandenbroucke}, {Vasileiou}, {Vianello}, {Vilchez},
  {Vitale}, {Waite}, {Wang}, {Winer}, {Wood}, {Yang}, \&
  {Zimmer}}]{2012APh....35..346A}
{Ackermann}, M., {Ajello}, M., {Allafort}, A., {et~al.} 2012, Astroparticle
  Physics, 35, 346, \dodoi{10.1016/j.astropartphys.2011.10.007}

\bibitem[{{Ackermann} {et~al.}(2018){Ackermann}, {Ajello}, {Baldini}, {Ballet},
  {Barbiellini}, {Bastieri}, {Bellazzini}, {Bissaldi}, {Blandford}, {Bloom},
  {Bonino}, {Bottacini}, {Brandt}, {Bregeon}, {Bruel}, {Buehler}, {Cameron},
  {Caputo}, {Caraveo}, {Castro}, {Cavazzuti}, {Charles}, {Cheung}, {Chiaro},
  {Ciprini}, {Cohen-Tanugi}, {Costantin}, {Cutini}, {D'Ammando}, {de Palma},
  {Desai}, {Di Lalla}, {Di Mauro}, {Di Venere}, {Favuzzi}, {Finke},
  {Franckowiak}, {Fukazawa}, {Funk}, {Fusco}, {Gargano}, {Gasparrini},
  {Giglietto}, {Giordano}, {Giroletti}, {Green}, {Grenier}, {Guillemot},
  {Guiriec}, {Hays}, {Hewitt}, {Horan}, {J{\'o}hannesson}, {Kensei}, {Kuss},
  {Larsson}, {Latronico}, {Lemoine-Goumard}, {Li}, {Longo}, {Loparco},
  {Lovellette}, {Lubrano}, {Magill}, {Maldera}, {Manfreda}, {Mazziotta},
  {McEnery}, {Meyer}, {Mizuno}, {Monzani}, {Morselli}, {Moskalenko}, {Negro},
  {Nuss}, {Omodei}, {Orienti}, {Orlando}, {Ormes}, {Palatiello}, {Paliya},
  {Paneque}, {Perkins}, {Persic}, {Pesce-Rollins}, {Piron}, {Porter},
  {Principe}, {Rain{\`o}}, {Rando}, {Rani}, {Razzaque}, {Reimer}, {Reimer},
  {Reposeur}, {Sgr{\`o}}, {Siskind}, {Spandre}, {Spinelli}, {Suson}, {Tajima},
  {Thayer}, {Tibaldo}, {Torres}, {Tosti}, {Valverde}, {Venters}, {Vogel},
  {Wood}, {Wood}, {Zaharijas}, {Fermi-LAT Collaboration}, \&
  {Biteau}}]{2018ApJS..237...32A}
{Ackermann}, M., {Ajello}, M., {Baldini}, L., {et~al.} 2018, \apjs, 237, 32,
  \dodoi{10.3847/1538-4365/aacdf7}

\bibitem[{{Aharonian} {et~al.}(2006{\natexlab{a}}){Aharonian}, {Akhperjanian},
  {Bazer-Bachi}, {Beilicke}, {Benbow}, {Berge}, {Bernl{\"o}hr}, {Boisson},
  {Bolz}, {Borrel}, {Braun}, {Breitling}, {Brown}, {Chadwick}, {Chounet},
  {Cornils}, {Costamante}, {Degrange}, {Dickinson}, {Djannati-Ata{\"\i}},
  {Drury}, {Dubus}, {Emmanoulopoulos}, {Espigat}, {Feinstein}, {Fontaine},
  {Fuchs}, {Funk}, {Gallant}, {Giebels}, {Gillessen}, {Glicenstein}, {Goret},
  {Hadjichristidis}, {Hauser}, {Hauser}, {Heinzelmann}, {Henri}, {Hermann},
  {Hinton}, {Hofmann}, {Holleran}, {Horns}, {Jacholkowska}, {de Jager},
  {Kh{\'e}lifi}, {Klages}, {Komin}, {Konopelko}, {Latham}, {Le Gallou},
  {Lemi{\`e}re}, {Lemoine-Goumard}, {Leroy}, {Lohse}, {Martin},
  {Martineau-Huynh}, {Marcowith}, {Masterson}, {McComb}, {de Naurois}, {Nolan},
  {Noutsos}, {Orford}, {Osborne}, {Ouchrif}, {Panter}, {Pelletier}, {Pita},
  {P{\"u}hlhofer}, {Punch}, {Raubenheimer}, {Raue}, {Raux}, {Rayner}, {Reimer},
  {Reimer}, {Ripken}, {Rob}, {Rolland}, {Rowell}, {Sahakian}, {Saug{\'e}},
  {Schlenker}, {Schlickeiser}, {Schuster}, {Schwanke}, {Siewert}, {Sol},
  {Spangler}, {Steenkamp}, {Stegmann}, {Tavernet}, {Terrier}, {Th{\'e}oret},
  {Tluczykont}, {van Eldik}, {Vasileiadis}, {Venter}, {Vincent}, {V{\"o}lk}, \&
  {Wagner}}]{2006Natur.440.1018A}
{Aharonian}, F., {Akhperjanian}, A.~G., {Bazer-Bachi}, A.~R., {et~al.}
  2006{\natexlab{a}}, \nat, 440, 1018, \dodoi{10.1038/nature04680}

\bibitem[{{Aharonian} {et~al.}(2006{\natexlab{b}}){Aharonian}, {Akhperjanian},
  {Bazer-Bachi}, {Beilicke}, {Benbow}, {Berge}, {Bernl{\"o}hr}, {Boisson},
  {Bolz}, {Borrel}, {Braun}, {Breitling}, {Brown}, {B{\"u}hler},
  {B{\"u}sching}, {Carrigan}, {Chadwick}, {Chounet}, {Cornils}, {Costamante},
  {Degrange}, {Dickinson}, {Djannati-Ata{\"\i}}, {O'C. Drury}, {Dubus},
  {Egberts}, {Emmanoulopoulos}, {Espigat}, {Feinstein}, {Ferrero}, {Fiasson},
  {Fontaine}, {Funk}, {Funk}, {Gallant}, {Giebels}, {Glicenstein}, {Goret},
  {Hadjichristidis}, {Hauser}, {Hauser}, {Heinzelmann}, {Henri}, {Hermann},
  {Hinton}, {Hofmann}, {Holleran}, {Horns}, {Jacholkowska}, {de Jager},
  {Kh{\'e}lifi}, {Komin}, {Konopelko}, {Kosack}, {Latham}, {Le Gallou},
  {Lemi{\`e}re}, {Lemoine-Goumard}, {Lohse}, {Martin}, {Martineau-Huynh},
  {Marcowith}, {Masterson}, {McComb}, {de Naurois}, {Nedbal}, {Nolan},
  {Noutsos}, {Orford}, {Osborne}, {Ouchrif}, {Panter}, {Pelletier}, {Pita},
  {P{\"u}hlhofer}, {Punch}, {Raubenheimer}, {Raue}, {Rayner}, {Reimer},
  {Reimer}, {Ripken}, {Rob}, {Rolland}, {Rowell}, {Sahakian}, {Saug{\'e}},
  {Schlenker}, {Schlickeiser}, {Schwanke}, {Sol}, {Spangler}, {Spanier},
  {Steenkamp}, {Stegmann}, {Superina}, {Tavernet}, {Terrier}, {Th{\'e}oret},
  {Tluczykont}, {van Eldik}, {Vasileiadis}, {Venter}, {Vincent}, {V{\"o}lk},
  {Wagner}, \& {Ward}}]{2006A&A...457..899A}
---. 2006{\natexlab{b}}, \aap, 457, 899, \dodoi{10.1051/0004-6361:20065351}

\bibitem[{{Aharonian} {et~al.}(2007{\natexlab{a}}){Aharonian}, {Akhperjanian},
  {Barres de Almeida}, {Bazer-Bachi}, {Behera}, {Beilicke}, {Benbow},
  {Bernl{\"o}hr}, {Boisson}, {Bolz}, {Borrel}, {Braun}, {Brion}, {Brown},
  {B{\"u}hler}, {Bulik}, {B{\"u}sching}, {Boutelier}, {Carrigan}, {Chadwick},
  {Chounet}, {Clapson}, {Coignet}, {Cornils}, {Costamante}, {Dalton},
  {Degrange}, {Dickinson}, {Djannati-Ata{\"\i}}, {Domainko}, {O'C. Drury},
  {Dubois}, {Dubus}, {Dyks}, {Egberts}, {Emmanoulopoulos}, {Espigat},
  {Farnier}, {Feinstein}, {Fiasson}, {F{\"o}rster}, {Fontaine}, {Funk},
  {F{\"u}{\ss}ling}, {Gallant}, {Giebels}, {Glicenstein}, {Gl{\"u}ck}, {Goret},
  {Hadjichristidis}, {Hauser}, {Hauser}, {Heinzelmann}, {Henri}, {Hermann},
  {Hinton}, {Hoffmann}, {Hofmann}, {Holleran}, {Hoppe}, {Horns},
  {Jacholkowska}, {de Jager}, {Jung}, {Katarzy{\'n}ski}, {Kendziorra},
  {Kerschhaggl}, {Kh{\'e}lifi}, {Keogh}, {Komin}, {Kosack}, {Lamanna},
  {Latham}, {Lemi{\`e}re}, {Lemoine-Goumard}, {Lenain}, {Lohse}, {Martin},
  {Martineau-Huynh}, {Marcowith}, {Masterson}, {Maurin}, {Maurin}, {McComb},
  {Moderski}, {Moulin}, {de Naurois}, {Nedbal}, {Nolan}, {Ohm}, {Olive}, {de
  O{\~n}a Wilhelmi}, {Orford}, {Osborne}, {Ostrowski}, {Panter}, {Pedaletti},
  {Pelletier}, {Petrucci}, {Pita}, {P{\"u}hlhofer}, {Punch}, {Ranchon},
  {Raubenheimer}, {Raue}, {Rayner}, {Renaud}, {Ripken}, {Rob}, {Rolland},
  {Rosier-Lees}, {Rowell}, {Rudak}, {Ruppel}, {Sahakian}, {Santangelo},
  {Schlickeiser}, {Sch{\"o}ck}, {Schr{\"o}der}, {Schwanke}, {Schwarzburg},
  {Schwemmer}, {Shalchi}, {Sol}, {Spangler}, {Stawarz}, {Steenkamp},
  {Stegmann}, {Superina}, {Tam}, {Tavernet}, {Terrier}, {van Eldik},
  {Vasileiadis}, {Venter}, {Vialle}, {Vincent}, {Vivier}, {V{\"o}lk}, {Volpe},
  {Wagner}, {Ward}, {Zdziarski}, \& {Zech}}]{2007A&A...473L..25A}
{Aharonian}, F., {Akhperjanian}, A.~G., {Barres de Almeida}, U., {et~al.}
  2007{\natexlab{a}}, \aap, 473, L25, \dodoi{10.1051/0004-6361:20078412}

\bibitem[{{Aharonian} {et~al.}(2007{\natexlab{b}}){Aharonian}, {Akhperjanian},
  {Bazer-Bachi}, {Beilicke}, {Benbow}, {Berge}, {Bernl{\"o}hr}, {Boisson},
  {Bolz}, {Borrel}, {Braun}, {Brion}, {Brown}, {B{\"u}hler}, {B{\"u}sching},
  {Boutelier}, {Carrigan}, {Chadwick}, {Chounet}, {Coignet}, {Cornils},
  {Costamante}, {Degrange}, {Dickinson}, {Djannati-Ata{\"\i}}, {O'C. Drury},
  {Dubus}, {Egberts}, {Emmanoulopoulos}, {Espigat}, {Farnier}, {Feinstein},
  {Ferrero}, {Fiasson}, {Fontaine}, {Funk}, {Funk}, {F{\"u}{\ss}ling},
  {Gallant}, {Giebels}, {Glicenstein}, {Gl{\"u}ck}, {Goret}, {Hadjichristidis},
  {Hauser}, {Hauser}, {Heinzelmann}, {Henri}, {Hermann}, {Hinton}, {Hoffmann},
  {Hofmann}, {Holleran}, {Hoppe}, {Horns}, {Jacholkowska}, {de Jager},
  {Kendziorra}, {Kerschhaggl}, {Kh{\'e}lifi}, {Komin}, {Kosack}, {Lamanna},
  {Latham}, {Le Gallou}, {Lemi{\`e}re}, {Lemoine-Goumard}, {Lohse}, {Martin},
  {Martineau-Huynh}, {Marcowith}, {Masterson}, {Maurin}, {McComb}, {Moulin},
  {de Naurois}, {Nedbal}, {Nolan}, {Noutsos}, {Olive}, {Orford}, {Osborne},
  {Panter}, {Pelletier}, {Petrucci}, {Pita}, {P{\"u}hlhofer}, {Punch},
  {Ranchon}, {Raubenheimer}, {Raue}, {Rayner}, {Ripken}, {Rob}, {Rolland},
  {Rosier-Lees}, {Rowell}, {Sahakian}, {Santangelo}, {Saug{\'e}}, {Schlenker},
  {Schlickeiser}, {Schr{\"o}der}, {Schwanke}, {Schwarzburg}, {Schwemmer},
  {Shalchi}, {Sol}, {Spangler}, {Spanier}, {Steenkamp}, {Stegmann}, {Superina},
  {Tam}, {Tavernet}, {Terrier}, {Tluczykont}, {van Eldik}, {Vasileiadis},
  {Venter}, {Vialle}, {Vincent}, {V{\"o}lk}, {Wagner}, \&
  {Ward}}]{2007A&A...470..475A}
{Aharonian}, F., {Akhperjanian}, A.~G., {Bazer-Bachi}, A.~R., {et~al.}
  2007{\natexlab{b}}, \aap, 470, 475, \dodoi{10.1051/0004-6361:20077057}

\bibitem[{{Aharonian} {et~al.}(2010){Aharonian}, {Akhperjanian}, {Anton},
  {Barres de Almeida}, {Bazer-Bachi}, {Becherini}, {Behera}, {Benbow},
  {Bernl{\"o}hr}, {Bochow}, {Boisson}, {Bolmont}, {Borrel}, {Brucker}, {Brun},
  {Brun}, {B{\"u}hler}, {Bulik}, {B{\"u}sching}, {Boutelier}, {Chadwick},
  {Charbonnier}, {Chaves}, {Cheesebrough}, {Chounet}, {Clapson}, {Coignet},
  {Dalton}, {Daniel}, {Davids}, {Degrange}, {Deil}, {Dickinson},
  {Djannati-Ata{\"\i}}, {Domainko}, {O'C. Drury}, {Dubois}, {Dubus}, {Dyks},
  {Dyrda}, {Egberts}, {Emmanoulopoulos}, {Espigat}, {Farnier}, {Feinstein},
  {Fiasson}, {F{\"o}rster}, {Fontaine}, {F{\"u}{\ss}ling}, {Gabici}, {Gallant},
  {G{\'e}rard}, {Gerbig}, {Giebels}, {Glicenstein}, {Gl{\"u}ck}, {Goret},
  {G{\"o}ring}, {Hauser}, {Hauser}, {Heinz}, {Heinzelmann}, {Henri}, {Hermann},
  {Hinton}, {Hoffmann}, {Hofmann}, {Holleran}, {Hoppe}, {Horns},
  {Jacholkowska}, {de Jager}, {Jahn}, {Jung}, {Katarzy{\'n}ski}, {Katz},
  {Kaufmann}, {Kendziorra}, {Kerschhaggl}, {Khangulyan}, {Kh{\'e}lifi},
  {Keogh}, {Klu{\'z}niak}, {Kneiske}, {Komin}, {Kosack}, {Lamanna}, {Lenain},
  {Lohse}, {Marandon}, {Martin}, {Martineau-Huynh}, {Marcowith}, {Masbou},
  {Maurin}, {McComb}, {Medina}, {Moderski}, {Moulin}, {Naumann-Godo}, {de
  Naurois}, {Nedbal}, {Nekrassov}, {Nicholas}, {Niemiec}, {Nolan}, {Ohm},
  {Olive}, {de O{\~n}a Wilhelmi}, {Orford}, {Ostrowski}, {Panter}, {Paz
  Arribas}, {Pedaletti}, {Pelletier}, {Petrucci}, {Pita}, {P{\"u}hlhofer},
  {Punch}, {Quirrenbach}, {Raubenheimer}, {Raue}, {Rayner}, {Renaud}, {Rieger},
  {Ripken}, {Rob}, {Rosier-Lees}, {Rowell}, {Rudak}, {Rulten}, {Ruppel},
  {Sahakian}, {Santangelo}, {Schlickeiser}, {Sch{\"o}ck}, {Schr{\"o}der},
  {Schwanke}, {Schwarzburg}, {Schwemmer}, {Shalchi}, {Sikora}, {Skilton},
  {Sol}, {Spangler}, {Stawarz}, {Steenkamp}, {Stegmann}, {Stinzing},
  {Superina}, {Szostek}, {Tam}, {Tavernet}, {Terrier}, {Tibolla}, {Tluczykont},
  {van Eldik}, {Vasileiadis}, {Venter}, {Venter}, {Vialle}, {Vincent},
  {Vivier}, {V{\"o}lk}, {Volpe}, {Wagner}, {Ward}, {Zdziarski}, \&
  {Zech}}]{2010A&A...521A..69A}
{Aharonian}, F., {Akhperjanian}, A.~G., {Anton}, G., {et~al.} 2010, \aap, 521,
  A69, \dodoi{10.1051/0004-6361/200912363}

\bibitem[{{Aharonian} {et~al.}(1994){Aharonian}, {Coppi}, \&
  {Voelk}}]{1994ApJ...423L...5A}
{Aharonian}, F.~A., {Coppi}, P.~S., \& {Voelk}, H.~J. 1994, \apjl, 423, L5,
  \dodoi{10.1086/187222}

\bibitem[{{Aleksi{\'c}} {et~al.}(2010){Aleksi{\'c}}, {Antonelli}, {Antoranz},
  {Backes}, {Baixeras}, {Barrio}, {Bastieri}, {Becerra Gonz{\'a}lez},
  {Bednarek}, {Berdyugin}, {Berger}, {Bernardini}, {Biland}, {Blanch}, {Bock},
  {Bonnoli}, {Bordas}, {Borla Tridon}, {Bosch-Ramon}, {Bose}, {Braun}, {Bretz},
  {Britzger}, {Camara}, {Carmona}, {Carosi}, {Colin}, {Commichau}, {Contreras},
  {Cortina}, {Costado}, {Covino}, {Dazzi}, {de Angelis}, {de Cea Del Pozo}, {de
  Los Reyes}, {de Lotto}, {de Maria}, {de Sabata}, {Delgado Mendez}, {Doert},
  {Dom{\'\i}nguez}, {Dominis Prester}, {Dorner}, {Doro}, {Elsaesser},
  {Errando}, {Ferenc}, {Fonseca}, {Font}, {Garc{\'\i}a L{\'o}pez},
  {Garczarczyk}, {Gaug}, {Godinovic}, {Hadasch}, {Herrero}, {Hildebrand},
  {H{\"o}hne-M{\"o}nch}, {Hose}, {Hrupec}, {Hsu}, {Jogler}, {Klepser},
  {Kr{\"a}henb{\"u}hl}, {Kranich}, {La Barbera}, {Laille}, {Leonardo},
  {Lindfors}, {Lombardi}, {Longo}, {L{\'o}pez}, {Lorenz}, {Majumdar}, {Maneva},
  {Mankuzhiyil}, {Mannheim}, {Maraschi}, {Mariotti}, {Mart{\'\i}nez}, {Mazin},
  {Meucci}, {Miranda}, {Mirzoyan}, {Miyamoto}, {Mold{\'o}n}, {Moles},
  {Moralejo}, {Nieto}, {Nilsson}, {Ninkovic}, {Orito}, {Oya}, {Paiano},
  {Paoletti}, {Paredes}, {Partini}, {Pasanen}, {Pascoli}, {Pauss}, {Pegna},
  {Perez-Torres}, {Persic}, {Peruzzo}, {Prada}, {Prandini}, {Puchades},
  {Puljak}, {Reichardt}, {Rhode}, {Rib{\'o}}, {Rico}, {Rissi}, {R{\"u}gamer},
  {Saggion}, {Saito}, {Salvati}, {S{\'a}nchez-Conde}, {Satalecka}, {Scalzotto},
  {Scapin}, {Schultz}, {Schweizer}, {Shayduk}, {Shore}, {Sierpowska-Bartosik},
  {Sillanp{\"a}{\"a}}, {Sitarek}, {Sobczynska}, {Spanier}, {Spiro}, {Stamerra},
  {Steinke}, {Struebig}, {Suric}, {Takalo}, {Tavecchio}, {Temnikov}, {Terzic},
  {Tescaro}, {Teshima}, {Torres}, {Vankov}, {Wagner}, {Weitzel}, {Zabalza},
  {Zandanel}, {Zanin}, {Neronov}, \& {Semikoz}}]{2010A&A...524A..77A}
{Aleksi{\'c}}, J., {Antonelli}, L.~A., {Antoranz}, P., {et~al.} 2010, \aap,
  524, A77, \dodoi{10.1051/0004-6361/201014747}

\bibitem[{{Alves Batista} \& {Saveliev}(2021)}]{2021Univ....7..223A}
{Alves Batista}, R., \& {Saveliev}, A. 2021, Universe, 7, 223,
  \dodoi{10.3390/universe7070223}

\bibitem[{{Alves Batista} {et~al.}(2016{\natexlab{a}}){Alves Batista},
  {Saveliev}, {Sigl}, \& {Vachaspati}}]{2016PhRvD..94h3005A}
{Alves Batista}, R., {Saveliev}, A., {Sigl}, G., \& {Vachaspati}, T.
  2016{\natexlab{a}}, \prd, 94, 083005, \dodoi{10.1103/PhysRevD.94.083005}

\bibitem[{{Alves Batista} {et~al.}(2016{\natexlab{b}}){Alves Batista},
  {Dundovic}, {Erdmann}, {Kampert}, {Kuempel}, {M{\"u}ller}, {Sigl}, {van
  Vliet}, {Walz}, \& {Winchen}}]{2016JCAP...05..038A}
{Alves Batista}, R., {Dundovic}, A., {Erdmann}, M., {et~al.}
  2016{\natexlab{b}}, \jcap, 2016, 038, \dodoi{10.1088/1475-7516/2016/05/038}

\bibitem[{{Archambault} {et~al.}(2017){Archambault}, {Archer}, {Benbow},
  {Buchovecky}, {Bugaev}, {Cerruti}, {Connolly}, {Cui}, {Falcone},
  {Fern{\'a}ndez Alonso}, {Finley}, {Fleischhack}, {Fortson}, {Furniss},
  {Griffin}, {H{\"u}tten}, {Hervet}, {Holder}, {Humensky}, {Johnson}, {Kaaret},
  {Kar}, {Kieda}, {Krause}, {Krennrich}, {Lang}, {Lin}, {Maier}, {McArthur},
  {Moriarty}, {Nieto}, {O'Brien}, {Ong}, {Otte}, {Pohl}, {Popkow}, {Pueschel},
  {Quinn}, {Ragan}, {Reynolds}, {Richards}, {Roache}, {Rovero}, {Sadeh},
  {Shahinyan}, {Staszak}, {Telezhinsky}, {Tyler}, {Wakely}, {Weinstein},
  {Weisgarber}, {Wilcox}, {Wilhelm}, {Williams}, \&
  {Zitzer}}]{2017ApJ...835..288A}
{Archambault}, S., {Archer}, A., {Benbow}, W., {et~al.} 2017, \apj, 835, 288,
  \dodoi{10.3847/1538-4357/835/2/288}

\bibitem[{{Astropy Collaboration} {et~al.}(2013){Astropy Collaboration},
  {Robitaille}, {Tollerud}, {Greenfield}, {Droettboom}, {Bray}, {Aldcroft},
  {Davis}, {Ginsburg}, {Price-Whelan}, {Kerzendorf}, {Conley}, {Crighton},
  {Barbary}, {Muna}, {Ferguson}, {Grollier}, {Parikh}, {Nair}, {Unther},
  {Deil}, {Woillez}, {Conseil}, {Kramer}, {Turner}, {Singer}, {Fox}, {Weaver},
  {Zabalza}, {Edwards}, {Azalee Bostroem}, {Burke}, {Casey}, {Crawford},
  {Dencheva}, {Ely}, {Jenness}, {Labrie}, {Lim}, {Pierfederici}, {Pontzen},
  {Ptak}, {Refsdal}, {Servillat}, \& {Streicher}}]{2013A&A...558A..33A}
{Astropy Collaboration}, {Robitaille}, T.~P., {Tollerud}, E.~J., {et~al.} 2013,
  \aap, 558, A33, \dodoi{10.1051/0004-6361/201322068}

\bibitem[{{Atwood} {et~al.}(2013){Atwood}, {Albert}, {Baldini}, {Tinivella},
  {Bregeon}, {Pesce-Rollins}, {Sgr{\`o}}, {Bruel}, {Charles}, {Drlica-Wagner},
  {Franckowiak}, {Jogler}, {Rochester}, {Usher}, {Wood}, {Cohen-Tanugi}, \&
  {Zimmer}}]{2013arXiv1303.3514A}
{Atwood}, W., {Albert}, A., {Baldini}, L., {et~al.} 2013, arXiv e-prints,
  arXiv:1303.3514.
\newblock \doarXiv{1303.3514}

\bibitem[{{Atwood} {et~al.}(2009){Atwood}, {Abdo}, {Ackermann}, {Althouse},
  {Anderson}, {Axelsson}, {Baldini}, {Ballet}, {Band}, {Barbiellini},
  {Bartelt}, {Bastieri}, {Baughman}, {Bechtol}, {B{\'e}d{\'e}r{\`e}de},
  {Bellardi}, {Bellazzini}, {Berenji}, {Bignami}, {Bisello}, {Bissaldi},
  {Blandford}, {Bloom}, {Bogart}, {Bonamente}, {Bonnell}, {Borgland},
  {Bouvier}, {Bregeon}, {Brez}, {Brigida}, {Bruel}, {Burnett}, {Busetto},
  {Caliandro}, {Cameron}, {Caraveo}, {Carius}, {Carlson}, {Casandjian},
  {Cavazzuti}, {Ceccanti}, {Cecchi}, {Charles}, {Chekhtman}, {Cheung},
  {Chiang}, {Chipaux}, {Cillis}, {Ciprini}, {Claus}, {Cohen-Tanugi},
  {Condamoor}, {Conrad}, {Corbet}, {Corucci}, {Costamante}, {Cutini}, {Davis},
  {Decotigny}, {DeKlotz}, {Dermer}, {de Angelis}, {Digel}, {do Couto e Silva},
  {Drell}, {Dubois}, {Dumora}, {Edmonds}, {Fabiani}, {Farnier}, {Favuzzi},
  {Flath}, {Fleury}, {Focke}, {Funk}, {Fusco}, {Gargano}, {Gasparrini},
  {Gehrels}, {Gentit}, {Germani}, {Giebels}, {Giglietto}, {Giommi}, {Giordano},
  {Glanzman}, {Godfrey}, {Grenier}, {Grondin}, {Grove}, {Guillemot}, {Guiriec},
  {Haller}, {Harding}, {Hart}, {Hays}, {Healey}, {Hirayama}, {Hjalmarsdotter},
  {Horn}, {Hughes}, {J{\'o}hannesson}, {Johansson}, {Johnson}, {Johnson},
  {Johnson}, {Johnson}, {Kamae}, {Katagiri}, {Kataoka}, {Kavelaars}, {Kawai},
  {Kelly}, {Kerr}, {Klamra}, {Kn{\"o}dlseder}, {Kocian}, {Komin}, {Kuehn},
  {Kuss}, {Landriu}, {Latronico}, {Lee}, {Lee}, {Lemoine-Goumard}, {Lionetto},
  {Longo}, {Loparco}, {Lott}, {Lovellette}, {Lubrano}, {Madejski}, {Makeev},
  {Marangelli}, {Massai}, {Mazziotta}, {McEnery}, {Menon}, {Meurer},
  {Michelson}, {Minuti}, {Mirizzi}, {Mitthumsiri}, {Mizuno}, {Moiseev},
  {Monte}, {Monzani}, {Moretti}, {Morselli}, {Moskalenko}, {Murgia},
  {Nakamori}, {Nishino}, {Nolan}, {Norris}, {Nuss}, {Ohno}, {Ohsugi}, {Omodei},
  {Orlando}, {Ormes}, {Paccagnella}, {Paneque}, {Panetta}, {Parent}, {Pearce},
  {Pepe}, {Perazzo}, {Pesce-Rollins}, {Picozza}, {Pieri}, {Pinchera}, {Piron},
  {Porter}, {Poupard}, {Rain{\`o}}, {Rando}, {Rapposelli}, {Razzano}, {Reimer},
  {Reimer}, {Reposeur}, {Reyes}, {Ritz}, {Rochester}, {Rodriguez}, {Romani},
  {Roth}, {Russell}, {Ryde}, {Sabatini}, {Sadrozinski}, {Sanchez}, {Sander},
  {Sapozhnikov}, {Parkinson}, {Scargle}, {Schalk}, {Scolieri}, {Sgr{\`o}},
  {Share}, {Shaw}, {Shimokawabe}, {Shrader}, {Sierpowska-Bartosik}, {Siskind},
  {Smith}, {Smith}, {Spandre}, {Spinelli}, {Starck}, {Stephens}, {Strickman},
  {Strong}, {Suson}, {Tajima}, {Takahashi}, {Takahashi}, {Tanaka}, {Tenze},
  {Tether}, {Thayer}, {Thayer}, {Thompson}, {Tibaldo}, {Tibolla}, {Torres},
  {Tosti}, {Tramacere}, {Turri}, {Usher}, {Vilchez}, {Vitale}, {Wang},
  {Watters}, {Winer}, {Wood}, {Ylinen}, \& {Ziegler}}]{2009ApJ...697.1071A}
{Atwood}, W.~B., {Abdo}, A.~A., {Ackermann}, M., {et~al.} 2009, \apj, 697,
  1071, \dodoi{10.1088/0004-637X/697/2/1071}

\bibitem[{{Ballet} {et~al.}(2020){Ballet}, {Burnett}, {Digel}, \&
  {Lott}}]{2020arXiv200511208B}
{Ballet}, J., {Burnett}, T.~H., {Digel}, S.~W., \& {Lott}, B. 2020, arXiv
  e-prints, arXiv:2005.11208.
\newblock \doarXiv{2005.11208}

\bibitem[{{Broderick} {et~al.}(2012){Broderick}, {Chang}, \&
  {Pfrommer}}]{2012ApJ...752...22B}
{Broderick}, A.~E., {Chang}, P., \& {Pfrommer}, C. 2012, \apj, 752, 22,
  \dodoi{10.1088/0004-637X/752/1/22}

\bibitem[{{Caprini} \& {Gabici}(2015)}]{2015PhRvD..91l3514C}
{Caprini}, C., \& {Gabici}, S. 2015, \prd, 91, 123514,
  \dodoi{10.1103/PhysRevD.91.123514}

\bibitem[{{de Naurois} \& {Rolland}(2009)}]{2009APh....32..231D}
{de Naurois}, M., \& {Rolland}, L. 2009, Astroparticle Physics, 32, 231,
  \dodoi{10.1016/j.astropartphys.2009.09.001}

\bibitem[{{Deil} {et~al.}(2017){Deil}, {Zanin}, {Lefaucheur}, {Boisson},
  {Khelifi}, {Terrier}, {Wood}, {Mohrmann}, {Chakraborty}, {Watson},
  {Lopez-Coto}, {Klepser}, {Cerruti}, {Lenain}, {Acero}, {Djannati-Ata{\"\i}},
  {Pita}, {Bosnjak}, {Trichard}, {Vuillaume}, {Donath}, {Consortium}, {King},
  {Jouvin}, {Owen}, {Sipocz}, {Lennarz}, {Voruganti}, {Spir-Jacob}, {Ruiz}, \&
  {Arribas}}]{gammapy:2017}
{Deil}, C., {Zanin}, R., {Lefaucheur}, J., {et~al.} 2017, in International
  Cosmic Ray Conference, Vol. 301, 35th International Cosmic Ray Conference
  (ICRC2017), 766.
\newblock \doarXiv{1709.01751}

\bibitem[{{Dermer} {et~al.}(2011){Dermer}, {Cavadini}, {Razzaque}, {Finke},
  {Chiang}, \& {Lott}}]{2011ApJ...733L..21D}
{Dermer}, C.~D., {Cavadini}, M., {Razzaque}, S., {et~al.} 2011, \apjl, 733,
  L21, \dodoi{10.1088/2041-8205/733/2/L21}

\bibitem[{{Dom{\'\i}nguez} {et~al.}(2011){Dom{\'\i}nguez}, {Primack},
  {Rosario}, {Prada}, {Gilmore}, {Faber}, {Koo}, {Somerville},
  {P{\'e}rez-Torres}, {P{\'e}rez-Gonz{\'a}lez}, {Huang}, {Davis},
  {Guhathakurta}, {Barmby}, {Conselice}, {Lozano}, {Newman}, \&
  {Cooper}}]{2011MNRAS.410.2556D}
{Dom{\'\i}nguez}, A., {Primack}, J.~R., {Rosario}, D.~J., {et~al.} 2011,
  \mnras, 410, 2556, \dodoi{10.1111/j.1365-2966.2010.17631.x}

\bibitem[{{Durrer} \& {Neronov}(2013)}]{2013A&ARv..21...62D}
{Durrer}, R., \& {Neronov}, A. 2013, \aapr, 21, 62,
  \dodoi{10.1007/s00159-013-0062-7}

\bibitem[{{Ebeling} {et~al.}(2006){Ebeling}, {White}, \&
  {Rangarajan}}]{2006MNRAS.368...65E}
{Ebeling}, H., {White}, D.~A., \& {Rangarajan}, F.~V.~N. 2006, \mnras, 368, 65,
  \dodoi{10.1111/j.1365-2966.2006.10135.x}

\bibitem[{{Fermi Science Support Development Team}(2019)}]{2019ascl.soft05011F}
{Fermi Science Support Development Team}. 2019, {Fermitools: Fermi Science
  Tools}, Astrophysics Source Code Library, record ascl:1905.011.
\newblock \doeprint{1905.011}

\bibitem[{{Fosbury} \& {Disney}(1976)}]{1976ApJ...207L..75F}
{Fosbury}, R.~A.~E., \& {Disney}, M.~J. 1976, \apjl, 207, L75,
  \dodoi{10.1086/182183}

\bibitem[{{H.~E.~S.~S. Collaboration} {et~al.}(2010){H.~E.~S.~S.
  Collaboration}, {Abramowski}, {Acero}, {Aharonian}, {Akhperjanian}, {Anton},
  {Barres de Almeida}, {Bazer-Bachi}, {Becherini}, {Behera}, {Benbow},
  {Bernl{\"o}hr}, {Bochow}, {Boisson}, {Bolmont}, {Borrel}, {Brucker}, {Brun},
  {Brun}, {B{\"u}hler}, {Bulik}, {B{\"u}sching}, {Boutelier}, {Chadwick},
  {Charbonnier}, {Chaves}, {Cheesebrough}, {Conrad}, {Chounet}, {Clapson},
  {Coignet}, {Costamante}, {Dalton}, {Daniel}, {Davids}, {Degrange}, {Deil},
  {Dickinson}, {Djannati-Ata{\"\i}}, {Domainko}, {O'C. Drury}, {Dubois},
  {Dubus}, {Dyks}, {Dyrda}, {Egberts}, {Eger}, {Espigat}, {Fallon}, {Farnier},
  {Fegan}, {Feinstein}, {Fernandes}, {Fiasson}, {F{\"o}rster}, {Fontaine},
  {F{\"u}{\ss}ling}, {Gabici}, {Gallant}, {G{\'e}rard}, {Gerbig}, {Giebels},
  {Glicenstein}, {Gl{\"u}ck}, {Goret}, {G{\"o}ring}, {Hampf}, {Hauser},
  {Heinz}, {Heinzelmann}, {Henri}, {Hermann}, {Hinton}, {Hoffmann}, {Hofmann},
  {Hofverberg}, {Holleran}, {Hoppe}, {Horns}, {Jacholkowska}, {de Jager},
  {Jahn}, {Jung}, {Katarzy{\'n}ski}, {Katz}, {Kaufmann}, {Kerschhaggl},
  {Khangulyan}, {Kh{\'e}lifi}, {Keogh}, {Klochkov}, {Klu{\v{z}}niak},
  {Kneiske}, {Komin}, {Kosack}, {Kossakowski}, {Lamanna}, {Lenain}, {Lohse},
  {Lu}, {Marandon}, {Marcowith}, {Masbou}, {Maurin}, {McComb}, {Medina},
  {M{\'e}hault}, {Moderski}, {Moulin}, {Naumann-Godo}, {de Naurois}, {Nedbal},
  {Nekrassov}, {Nguyen}, {Nicholas}, {Niemiec}, {Nolan}, {Ohm}, {Olive}, {de
  O{\~n}a Wilhelmi}, {Opitz}, {Orford}, {Ostrowski}, {Panter}, {Paz Arribas},
  {Pedaletti}, {Pelletier}, {Petrucci}, {Pita}, {P{\"u}hlhofer}, {Punch},
  {Quirrenbach}, {Raubenheimer}, {Raue}, {Rayner}, {Reimer}, {Renaud}, {de Los
  Reyes}, {Rieger}, {Ripken}, {Rob}, {Rosier-Lees}, {Rowell}, {Rudak},
  {Rulten}, {Ruppel}, {Ryde}, {Sahakian}, {Santangelo}, {Schlickeiser},
  {Sch{\"o}ck}, {Sch{\"o}nwald}, {Schwanke}, {Schwarzburg}, {Schwemmer},
  {Shalchi}, {Sushch}, {Sikora}, {Skilton}, {Sol}, {Stawarz}, {Steenkamp},
  {Stegmann}, {Stinzing}, {Szostek}, {Tam}, {Tavernet}, {Terrier}, {Tibolla},
  {Tluczykont}, {Valerius}, {van Eldik}, {Vasileiadis}, {Venter}, {Venter},
  {Vialle}, {Viana}, {Vincent}, {Vivier}, {V{\"o}lk}, {Volpe}, {Vorobiov},
  {Wagner}, {Ward}, {Zdziarski}, {Zech}, \& {Zechlin}}]{2010A&A...516A..56H}
{H.~E.~S.~S. Collaboration}, {Abramowski}, A., {Acero}, F., {et~al.} 2010,
  \aap, 516, A56, \dodoi{10.1051/0004-6361/201014321}

\bibitem[{{H.~E.~S.~S. Collaboration} {et~al.}(2014){H.~E.~S.~S.
  Collaboration}, {Abramowski}, {Aharonian}, {Ait Benkhali}, {Akhperjanian},
  {Ang{\"u}ner}, {Anton}, {Backes}, {Balenderan}, {Balzer}, {Barnacka},
  {Becherini}, {Becker Tjus}, {Bernl{\"o}hr}, {Birsin}, {Bissaldi}, {Biteau},
  {B{\"o}ttcher}, {Boisson}, {Bolmont}, {Bordas}, {Brucker}, {Brun}, {Brun},
  {Bulik}, {Carrigan}, {Casanova}, {Chadwick}, {Chalme-Calvet}, {Chaves},
  {Cheesebrough}, {Chr{\'e}tien}, {Colafrancesco}, {Cologna}, {Conrad},
  {Couturier}, {Cui}, {Dalton}, {Daniel}, {Davids}, {Degrange}, {Deil},
  {deWilt}, {Dickinson}, {Djannati-At{\"a}{\i}}, {Domainko}, {Drury}, {Dubus},
  {Dutson}, {Dyks}, {Dyrda}, {Edwards}, {Egberts}, {Eger}, {Espigat},
  {Farnier}, {Fegan}, {Feinstein}, {Fernandes}, {Fernandez}, {Fiasson},
  {Fontaine}, {F{\"o}rster}, {F{\"u}{\ss}ling}, {Gajdus}, {Gallant},
  {Garrigoux}, {Giavitto}, {Giebels}, {Glicenstein}, {Grondin},
  {Grudzi{\'n}ska}, {H{\"a}ffner}, {Hahn}, {Harris}, {Heinzelmann}, {Henri},
  {Hermann}, {Hervet}, {Hillert}, {Hinton}, {Hofmann}, {Hofverberg}, {Holler},
  {Horns}, {Jacholkowska}, {Jahn}, {Jamrozy}, {Janiak}, {Jankowsky}, {Jung},
  {Kastendieck}, {Katarzy{\'n}ski}, {Katz}, {Kaufmann}, {Kh{\'e}lifi},
  {Kieffer}, {Klepser}, {Klochkov}, {Klu{\'z}niak}, {Kneiske}, {Kolitzus},
  {Komin}, {Kosack}, {Krakau}, {Krayzel}, {Kr{\"u}ger}, {Laffon}, {Lamanna},
  {Lefaucheur}, {Lemie`re}, {Lemoine-Goumard}, {Lenain}, {Lohse}, {Lopatin},
  {Lu}, {Marandon}, {Marcowith}, {Marx}, {Maurin}, {Maxted}, {Mayer}, {McComb},
  {M{\'e}hault}, {Meintjes}, {Menzler}, {Meyer}, {Moderski}, {Mohamed},
  {Moulin}, {Murach}, {Naumann}, {de Naurois}, {Niemiec}, {Nolan}, {Oakes},
  {Odaka}, {Ohm}, {de O{\~n}a Wilhelmi}, {Opitz}, {Ostrowski}, {Oya}, {Panter},
  {Parsons}, {Paz Arribas}, {Pekeur}, {Pelletier}, {Perez}, {Petrucci},
  {Peyaud}, {Pita}, {Poon}, {P{\"u}hlhofer}, {Punch}, {Quirrenbach}, {Raab},
  {Raue}, {Reichardt}, {Reimer}, {Reimer}, {Renaud}, {de los Reyes}, {Rieger},
  {Rob}, {Romoli}, {Rosier-Lees}, {Rowell}, {Rudak}, {Rulten}, {Sahakian},
  {Sanchez}, {Santangelo}, {Schlickeiser}, {Sch{\"u}ssler}, {Schulz},
  {Schwanke}, {Schwarzburg}, {Schwemmer}, {Sol}, {Spengler}, {Spies},
  {Stawarz}, {Steenkamp}, {Stegmann}, {Stinzing}, {Stycz}, {Sushch},
  {Tavernet}, {Tavernier}, {Taylor}, {Terrier}, {Tluczykont}, {Trichard},
  {Valerius}, {van Eldik}, {van Soelen}, {Vasileiadis}, {Venter}, {Viana},
  {Vincent}, {V{\"o}lk}, {Volpe}, {Vorster}, {Vuillaume}, {Wagner}, {Wagner},
  {Wagner}, {Ward}, {Weidinger}, {Weitzel}, {White}, {Wierzcholska},
  {Willmann}, {W{\"o}rnlein}, {Wouters}, {Yang}, {Zabalza}, {Zacharias},
  {Zdziarski}, {Zech}, {Zechlin}, \& {Malyshev}}]{2014A&A...562A.145H}
{H.~E.~S.~S. Collaboration}, {Abramowski}, A., {Aharonian}, F., {et~al.} 2014,
  \aap, 562, A145, \dodoi{10.1051/0004-6361/201322510}

\bibitem[{Harris {et~al.}(2020)Harris, Millman, van~der Walt, Gommers,
  Virtanen, Cournapeau, Wieser, Taylor, Berg, Smith, Kern, Picus, Hoyer, van
  Kerkwijk, Brett, Haldane, del R{\'{i}}o, Wiebe, Peterson,
  G{\'{e}}rard-Marchant, Sheppard, Reddy, Weckesser, Abbasi, Gohlke, \&
  Oliphant}]{harris2020array}
Harris, C.~R., Millman, K.~J., van~der Walt, S.~J., {et~al.} 2020, Nature, 585,
  357, \dodoi{10.1038/s41586-020-2649-2}

\bibitem[{{Hofmann} {et~al.}(2000){Hofmann}, {Lampeitl}, {Konopelko}, \&
  {Krawczynski}}]{2000APh....12..207H}
{Hofmann}, W., {Lampeitl}, H., {Konopelko}, A., \& {Krawczynski}, H. 2000,
  Astroparticle Physics, 12, 207, \dodoi{10.1016/S0927-6505(99)00109-7}

\bibitem[{{Jones} {et~al.}(2009){Jones}, {Read}, {Saunders}, {Colless},
  {Jarrett}, {Parker}, {Fairall}, {Mauch}, {Sadler}, {Watson}, {Burton},
  {Campbell}, {Cass}, {Croom}, {Dawe}, {Fiegert}, {Frankcombe}, {Hartley},
  {Huchra}, {James}, {Kirby}, {Lahav}, {Lucey}, {Mamon}, {Moore}, {Peterson},
  {Prior}, {Proust}, {Russell}, {Safouris}, {Wakamatsu}, {Westra}, \&
  {Williams}}]{2009MNRAS.399..683J}
{Jones}, D.~H., {Read}, M.~A., {Saunders}, W., {et~al.} 2009, \mnras, 399, 683,
  \dodoi{10.1111/j.1365-2966.2009.15338.x}

\bibitem[{{Jorstad} {et~al.}(2017){Jorstad}, {Marscher}, {Morozova},
  {Troitsky}, {Agudo}, {Casadio}, {Foord}, {G{\'o}mez}, {MacDonald}, {Molina},
  {L{\"a}hteenm{\"a}ki}, {Tammi}, \& {Tornikoski}}]{2017ApJ...846...98J}
{Jorstad}, S.~G., {Marscher}, A.~P., {Morozova}, D.~A., {et~al.} 2017, \apj,
  846, 98, \dodoi{10.3847/1538-4357/aa8407}

\bibitem[{{Kobayashi}(2014)}]{2014JCAP...05..040K}
{Kobayashi}, T. 2014, \jcap, 2014, 040, \dodoi{10.1088/1475-7516/2014/05/040}

\bibitem[{{Lott} {et~al.}(2020){Lott}, {Gasparrini}, \&
  {Ciprini}}]{2020arXiv201008406L}
{Lott}, B., {Gasparrini}, D., \& {Ciprini}, S. 2020, arXiv e-prints,
  arXiv:2010.08406.
\newblock \doarXiv{2010.08406}

\bibitem[{{Meyer} {et~al.}(2016){Meyer}, {Conrad}, \&
  {Dickinson}}]{2016ApJ...827..147M}
{Meyer}, M., {Conrad}, J., \& {Dickinson}, H. 2016, \apj, 827, 147,
  \dodoi{10.3847/0004-637X/827/2/147}

\bibitem[{{Miniati} {et~al.}(2018){Miniati}, {Gregori}, {Reville}, \&
  {Sarkar}}]{2018PhRvL.121b1301M}
{Miniati}, F., {Gregori}, G., {Reville}, B., \& {Sarkar}, S. 2018, \prl, 121,
  021301, \dodoi{10.1103/PhysRevLett.121.021301}

\bibitem[{{Neronov} \& {Semikoz}(2009)}]{2009PhRvD..80l3012N}
{Neronov}, A., \& {Semikoz}, D.~V. 2009, \prd, 80, 123012,
  \dodoi{10.1103/PhysRevD.80.123012}

\bibitem[{{Neronov} \& {Vovk}(2010)}]{2010Sci...328...73N}
{Neronov}, A., \& {Vovk}, I. 2010, Science, 328, 73,
  \dodoi{10.1126/science.1184192}

\bibitem[{{Parma} {et~al.}(2002){Parma}, {Murgia}, {de Ruiter}, \&
  {Fanti}}]{2002NewAR..46..313P}
{Parma}, P., {Murgia}, M., {de Ruiter}, H.~R., \& {Fanti}, R. 2002, \nar, 46,
  313, \dodoi{10.1016/S1387-6473(01)00201-9}

\bibitem[{{Parsons} \& {Hinton}(2014)}]{2014APh....56...26P}
{Parsons}, R.~D., \& {Hinton}, J.~A. 2014, Astroparticle Physics, 56, 26,
  \dodoi{10.1016/j.astropartphys.2014.03.002}

\bibitem[{{Pe{\~n}a-Herazo} {et~al.}(2021){Pe{\~n}a-Herazo}, {Massaro}, {Gu},
  {Paggi}, {Landoni}, {D'Abrusco}, {Ricci}, {Masetti}, \&
  {Chavushyan}}]{2021AJ....161..196P}
{Pe{\~n}a-Herazo}, H.~A., {Massaro}, F., {Gu}, M., {et~al.} 2021, \aj, 161,
  196, \dodoi{10.3847/1538-3881/abe41d}

\bibitem[{{Plaga}(1995)}]{1995Natur.374..430P}
{Plaga}, R. 1995, \nat, 374, 430, \dodoi{10.1038/374430a0}

\bibitem[{{Pshirkov} {et~al.}(2016){Pshirkov}, {Tinyakov}, \&
  {Urban}}]{2016PhRvL.116s1302P}
{Pshirkov}, M.~S., {Tinyakov}, P.~G., \& {Urban}, F.~R. 2016, \prl, 116,
  191302, \dodoi{10.1103/PhysRevLett.116.191302}

\bibitem[{{Remillard} {et~al.}(1989){Remillard}, {Tuohy}, {Brissenden},
  {Buckley}, {Schwartz}, {Feigelson}, \& {Tapia}}]{1989ApJ...345..140R}
{Remillard}, R.~A., {Tuohy}, I.~R., {Brissenden}, R.~J.~V., {et~al.} 1989,
  \apj, 345, 140, \dodoi{10.1086/167888}

\bibitem[{Virtanen {et~al.}(2020)Virtanen, Gommers, Oliphant, Haberland, Reddy,
  Cournapeau, Burovski, Peterson, Weckesser, Bright, {van der Walt}, Brett,
  Wilson, Millman, Mayorov, Nelson, Jones, Kern, Larson, Carey, Polat, Feng,
  Moore, {VanderPlas}, Laxalde, Perktold, Cimrman, Henriksen, Quintero, Harris,
  Archibald, Ribeiro, Pedregosa, {van Mulbregt}, \& {SciPy 1.0
  Contributors}}]{2020SciPy-NMeth}
Virtanen, P., Gommers, R., Oliphant, T.~E., {et~al.} 2020, Nature Methods, 17,
  261, \dodoi{10.1038/s41592-019-0686-2}

\bibitem[{{Woo} {et~al.}(2005){Woo}, {Urry}, {van der Marel}, {Lira}, \&
  {Maza}}]{2005ApJ...631..762W}
{Woo}, J.-H., {Urry}, C.~M., {van der Marel}, R.~P., {Lira}, P., \& {Maza}, J.
  2005, \apj, 631, 762, \dodoi{10.1086/432681}

\bibitem[{{Wood} {et~al.}(2017){Wood}, {Caputo}, {Charles}, {Di Mauro},
  {Magill}, {Perkins}, \& {Fermi-LAT Collaboration}}]{2017ICRC...35..824W}
{Wood}, M., {Caputo}, R., {Charles}, E., {et~al.} 2017, in International Cosmic
  Ray Conference, Vol. 301, 35th International Cosmic Ray Conference
  (ICRC2017), 824.
\newblock \doarXiv{1707.09551}

\end{thebibliography}
